# The ENIGMA of CDMA REVISITED

K. Alishahi, S. Dashmiz, *Student Member, IEEE*, P. Pad, *Student Member, IEEE*, F. Marvasti, *Senior Member, IEEE*, M. H. Shafinia, *Student Member, IEEE*, and M. Mansouri, *Student Member, IEEE*

**Abstract**

In this paper, we explore the mystery of synchronous CDMA as applied to wireless and optical communication systems under very general settings for the user symbols and the signature matrix entries. The channel is modeled with real/complex additive noise of arbitrary distribution. Two problems are addressed. The first problem concerns whether overloaded error free codes exist in the absence of additive noise under these general settings, and if so whether there are any practical optimum decoding algorithms. The second one is about the bounds for the sum channel capacity when user data and signature codes employ any real or complex alphabets (finite or infinite). In response to the first problem, we have developed practical Maximum Likelihood (ML) decoding algorithms for overloaded CDMA systems for a large class of alphabets. In response to the second problem, a general theorem has been developed in which the sum capacity lower bounds with respect to the number of users and spreading gain and Signal-to-Noise Ratio (SNR) can be derived as special cases for a given CDMA system. To show the power and utility of the main theorem, a number of sum capacity bounds for special cases are simulated. An important conclusion of this paper is that the lower and upper bounds of the sum capacity for small/medium size CDMA systems depend on both the input and the signature symbols; this is contrary to the asymptotic results for large scale systems reported in the literature (also confirmed in this paper) where the signature symbols and statistics disappear in the asymptotic sum capacity. Moreover, these questions are investigated for the case when not all users are active. Furthermore, upper and asymptotic bounds are derived and numerically evaluated and compared to other derivations.

K. Alishahi is affiliated with Advanced Communication Research Institute (ACRI) and Dept. of Mathematical Sciences, Sharif University of Technology.

S. Dashmiz, P. Pad, F. Marvasti, M. H. Shafinia and S. M. Mansouri are affiliated with ACRI and Electrical Engineering Department, Sharif University of Technology marvasti@sharif.edu.

# I. INTRODUCTION

Code Division Multiple Access (CDMA) has become the standard for the third generation of Global System for Mobile (GSM) communications and American cellular systems [1]. Also, optical CDMA systems have become an alternative multiple access for fiber optics and optical wireless systems [2]–[4]. The reasons, to name a few, are its simplicity, high overloading factor and soft hand-off. On the other hand, the theoretical developments in CDMA sum capacity have been limited to users with Gaussian inputs where Welch Bound Equality (WBE) codes achieve the theoretical capacity [5]–[9]. For binary CDMA, only asymptotic results were known [10]–[13] prior to our recent papers [14], [15] (to be discussed in the next paragraph). For overloaded CDMA, only WBE sequences were known to minimize the user interference using autocorrelator detectors and iterative methods [16]–[28]. Simple ML decoding for binary overloaded CDMA and its superiority to WBE sequences were shown in [14], [29]–[31]. ML decoding of general user inputs and signature matrices has not been reported in the literature, nor has the evaluation of the sum capacity for such general CDMA systems. The revisitation and the solutions of such CDMA problems are the main topics of the current paper.

For binary synchronous CDMA systems, a class of errorless codes for overloaded CDMA systems were introduced [14], [32]; also in [15], we derived a relatively tight family of bounds for the sum capacity of binary CDMA systems with binary signatures. The present paper is a continuation of the previous ones; the constraints of binary multi-user and binary signatures are now relaxed. The errorless codes (COW[1] and COO[2]) developed in [32] and [14] are also extended to nonbinary cases. We have also extended the results to active user detection in a CDMA system. Below, we will give a brief summary of the previous results:

In [14], [29], [32], a class of overloaded errorless codes for binary multi-users and binary/ternary signatures for wireless ($\pm 1$ for COW) and optical ($0, 1$ for COO) applications were developed. Mow [32] presented a unifying approach to find one-to-one $m \times n$ binary and ternary matrices for binary inputs for multiuser applications. He also applied constructive theorems developed by the previous authors [33]–[39] to enlarge such matrices. This paper also discusses asymptotic behavior of such matrices. In [14], we have also developed injective binary matrices for binary inputs independently. In the same paper, we have also suggested ML decoding using tensor products.

In addition, in [14], we developed lower and upper bounds for the sum capacity of a binary synchronous

[1]Codes for Overloaded Wireless systems

[2]Codes for Overloaded Optical systems

CDMA under noiseless environment conditions and the assumption that users are jointly dependent. These sum capacity bounds suggest that there exists a linear region with respect to the number of users far beyond the orthogonal Walsh codes ($n = m$ where $m$ is the chip rate), i.e., errorless overloaded ($n \gg m$) codes do exist. The bounds also suggest that there is a threshold beyond which overloaded errorless codes do not exist. In the same paper, we developed practical COW/COO codes ($n = 104$, $m = 64$) with simple Maximum Likelihood (ML) decoders that outperformed Welch Bound Equality (WBE) codes both in terms of complexity and performance. WBE codes are optimum (capacity achieving) for multi-users of Gaussian real inputs but not for binary users.

In [15], we extended the bounds to a noisy channel with arbitrary distribution. We also changed the assumption of joint probability of multi-users to a more realistic scenario of independence among users. For the noiseless case, the same bounds were derived as the ones in [14] despite the independence assumption. Simulation results again showed that there is a linear region beyond Hadamard codes and there is a degraded region for a highly overloaded CDMA. The asymptotic results for the sum capacity simplified the equations and thus the computer simulations, and we managed to compare our results to that of Tanaka [10]. The main contribution of [15] are tight bounds with closed form derivations that, unlike the previous papers, depend not on the limiting cases but rather on the number of users $n$, spreading gain $m$, and the noise distribution that need not be Gaussian.

In the present paper, we will extend the binary results to multi-users and signature matrices with entries from finite/infinite, real/complex alphabets. Extensions of overloaded codes (COW and COO) to nonbinary CDMA and active user detection are also considered. An important achievement in this regard is the development of practical overloaded codes with optimal low complexity decoding algorithms.

Bounds for both noisy and noiseless channels are considered. Our main achievement in this regard is the development of a general theorem for the lower bounds where any CDMA system with arbitrary user data and signature matrix symbols can be modeled as a special case. We further extend the sum capacity bounds to k-active users where not all users transmit data.

Some of the main results and theorems will be reviewed in the next section on preliminaries; the structure of the paper is as shown below[3]:

CONTENTS

[3]The table of contents is for the convenience of the reviewers and will not be in the final galley proof.

In the section on preliminaries, four topics are reviewed: 1- power constraint considerations, 2- the definition of sum capacity and the main known results, 3- review of the COW/COO matrices, and 4- review of the sum capacity bounds for the binary CDMA. Section III deals with the generalization of the COW/COO codes to the nonbinary finite real/complex CDMA systems (GCO). Practical algorithms for the reconstruction of GCO matrices and bounds on the overloading factor are introduced in this section. The performances of GCO matrices, as compared to the WBE codes, and practical ML decoders are covered in this section. Active user detection for CDMA systems is considered as a special case of GCO matrices and is discussed at the end of this section.

Section IV covers the derivation of lower bounds for the sum capacity of an arbitrary CDMA system with no channel noise; a general theorem is developed and special cases of finite $(q, l)$-ary CDMA systems are developed. Many examples such as binary/ternary and binary/quaternary CDMA systems are derived with simulation results. A presentation of upper bounds and asymptotic cases is the final parts of this section.

Section V is similar to the previous section except additive channel noise with arbitrary distribution is also considered; the noiseless case of Section IV can be derived from the noisy case in the limiting case. In this section, we will develop a general theorem for the lower bound that is valid for any type of CDMA system; from this theorem, we shall develop in the subsequent two sections, the special cases for various types of input and signature symbols. In SectionVI, we will derive the lower bound from our main theorem for the special cases of finite $(q,l)$-ary and finite/real-complex CDMA systems. As special cases, some examples are simulated for the binary/ternary, ternary/ternary, and binary/real CDMA systems with additive Gaussian noise. Upper bounds and asymptotic limits are also derived in the same section.

Section VII is related to the bounds for the case when the number of active users is either deterministic or random. For the deterministic case, special cases of binary/real, real/real, and binary/binary are also formulated and simulated. For the random case, similar results are derived. The asymptotic bound for the random case is also considered.

Finally, a summary of the main results, concluding remarks, and future studies are covered in Section VIII. All the proofs of the theorems and examples are provided in the appendices.

## II. Preliminaries

Consider a CDMA system with $n$ users and $m$ chips as $Y=\frac{1}{\sqrt{m}}\mathbf{A}X+N$, where $X$ is the $n\times 1$ user data vector, $N$ is the $m\times 1$ noise vector, $\mathbf{A}$ is the $m\times n$ signature matrix, and $Y$ is the $m\times 1$ received signal (these vectors and matrices could be real or complex).

### *A. Power Constraint Issues*

Since the total power of users and noise in the above model are, respectively, equal to $\text{tr}\left(\mathbb{E}\left(\frac{1}{m}\mathbf{A}XX^*\mathbf{A}^*\right)\right)$ and $\mathbb{E}\left(N^*N\right)$ (the symbol * stands for the Hermitian transpose), the multi-user SNR is defined as

$$\text{SNR}=\frac{\text{tr}\left(\mathbb{E}\left(\frac{1}{m}\mathbf{A}XX^*\mathbf{A}^*\right)\right)}{\mathbb{E}\left(N^*N\right)}. \tag{1}$$

For simplicity, we will assume that $N=[N_1,N_2,\cdots,N_m]^T$, where $N_j$'s are i.i.d. random variables with $f(\cdot)$ as common Probability Distribution Function (pdf) of variance $\sigma_f^2$. This assumption implies that $\mathbb{E}\left(N^*N\right)=m\sigma_f^2$ and hence the SNR definition can be written as

$$\text{SNR}=\frac{\text{tr}\left(\mathbb{E}\left(\frac{1}{m}\mathbf{A}XX^*\mathbf{A}^*\right)\right)}{m\sigma_f^2}. \tag{2}$$

If $\frac{m}{n}\text{SNR}\leq\eta$, the above equation can be written as

$$\text{tr}\left(\mathbb{E}\left(\frac{1}{m}\mathbf{A}XX^*\mathbf{A}^*\right)\right)\leq n\eta\sigma_f^2. \tag{3}$$

For a given signature matrix $\frac{1}{\sqrt{m}}\mathbf{A}$ and $\eta$, the sum capacity is defined as

$$C\left(\mathbf{A},\eta\right)=\max\left\{\mathbb{I}\left(X;Y\right)|X\sim p_1\left(x_1\right)\times p_2\left(x_2\right)\times\cdots\times p_n\left(x_n\right)\ \text{such that (3) is satisfied}\right\}. \tag{4}$$

However, we would like to maximize $C(\mathbf{A}, \eta)$ over a class of admissible signature matrices $\frac{1}{\sqrt{m}}\mathbf{A}$ and input vectors $X$ as follows:

Let $\mathcal{S}$ and $\mathcal{I}$ be two given sets of real or complex numbers which could be finite or infinite. Assume that

$$\mathbf{A} = r\mathbf{B} \quad r \in \mathbb{R}^+, \quad \mathbf{B} \in \mathcal{M}_{m\times n}(\mathcal{S}) \tag{5}$$

and

$$X \in \mathcal{M}_{n\times 1}(\mathcal{I}), \tag{6}$$

where $r$ is the scale factor that guarantees the power condition in (3). With the assumption that $X_i$'s are i.i.d. with distribution $p(\cdot)$, we have $\mathbb{E}(XX^*) = \sigma_p^2\mathbf{I}_n + {\mu_p}^2\mathbf{J}_n$, where $\mu_p$ and $\sigma_p^2$ denote the mean and variance; $\mathbf{I}_n$ and $\mathbf{J}_n$ are, respectively, the $n \times n$ identity matrix and the matrix with all entries equal to one. Therefore, we have

$$\text{tr}\left(\mathbb{E}\left(\frac{1}{m}\mathbf{A}XX^*\mathbf{A}^*\right)\right) = \frac{r^2}{m}\big({\sigma_p}^2\text{tr}(BB^*) + {\mu_p}^2\text{tr}(B\mathbf{J}B^*)\big). \tag{7}$$

Hence the SNR condition (3) can be written in the following form

$$\frac{r^2}{m}\big(\sigma_p^2\text{tr}(BB^*) + {\mu_p}^2\text{tr}(B\mathbf{J}B^*)\big) \leq n\eta\sigma_f^2. \tag{8}$$

### *B. Definition of Sum Channel Capacity and the Main Known Relevant Results*

*Definition 1 (The Sum Channel Capacity):*

Define $C(m, n, \mathcal{I}, \mathcal{S}, \eta)$ to be the maximum value of $C(\mathbf{A}, \eta)$ in (4) over all signature matrices $\frac{1}{\sqrt{m}}\mathbf{A}$, where (5), (6) and (8) are satisfied. For noiseless systems, $\eta = \infty$, we use the notation $C(m, n, \mathcal{I}, \mathcal{S})$.

*Remark 1:* References [10], [19], [40] have used the average mutual information over all signature matrices as the sum capacity. Such interpretation is actually less than the sum capacity as defined in the above. In the present paper, our upper bounds are naturally the upper bounds for the average mutual information. Due to the techniques we use, the lower bounds derived in the present paper have smaller values than the average mutual information.

Given the above definition, we are now ready to review the main relevant results from the CDMA literature:

*Theorem 1 (Main Known Result for the Sum Channel Capacity):*

When the input vector $X$ is Gaussian, the signature matrix $\frac{\mathbf{A}}{\sqrt{m}}$ is arbitrary, and the additive noise is white Gaussian, the sum channel capacity can be shown to be [5]

$$C = \frac{1}{2}\log\left(\det\left(\mathbf{I} + \frac{2\eta}{m}\mathbf{A}\mathbf{A}^T\right)\right), \quad \text{for Real System} \tag{9}$$

$$C = \log\left(\det\left(\mathbf{I} + \frac{\eta}{m}\mathbf{A}\mathbf{A}^*\right)\right), \quad \text{for Complex System,} \tag{10}$$

where $\eta$ is the normalized SNR as defined in (3).

*Example 1 (The Sum Channel Capacity for WBE Matrices):*
Here we consider real WBE signature matrices since they maximize the capacity in (9), [7]. For underloaded case $(n < m)$ and for the normalized WBE matrices, (9) becomes

$$C = \frac{n}{2}\log(1 + 2\eta), \tag{11}$$

and likewise for the overloaded case $(n > m)$, (9) becomes

$$C = \frac{m}{2}\log\left(1 + \frac{2n}{m}\eta\right). \tag{12}$$

*Theorem 2:* Tanaka's Asymptotic Bound for Binary User Input
For binary bipolar input and binary matrix, Tanaka derived an asymptotic bound using the replica theory in statistical physics [10].

$$C_\theta = \frac{1}{2\beta}\log\left(1 + \frac{\beta(1-\theta)}{\sigma^2}\right) + g(\lambda, \theta)\log(e), \tag{13}$$

in which

$$g(\lambda, \theta) = \frac{\lambda}{2}(1 + \theta) - \int \ln\left(\cosh\left(\sqrt{\lambda}Z + \lambda\right)\right)D_Z, \tag{14}$$

where $D_Z$ is the standard normal measure and

$$\lambda = \frac{1}{\sigma^2 + \beta(1-\theta)}, \tag{15}$$

$$\theta = \int \tanh\left(\sqrt{\lambda}Z + \lambda\right)D_Z, \tag{16}$$

where $\beta = \frac{n}{m}$ and $\sigma^2$ is the variance of the noise.

We have shown in [15] that (13) approaches our asymptotic upper bound when $\beta$ increases. The replica theory is a non-rigorous mathematical analysis. A rigorous proof of Tanaka's results is given in [11], [40], [41]. The extension of Tanaka's bound to arbitrary input symbols is given in [12]; for complex input symbols, the following asymptotic result can be derived:

*Theorem 3:* Guo-Verdu Asymptotic Bound for Arbitrary User Input [12]

$$C_{joint}(\beta) = \beta\mathbb{E}\{\mathbb{I}\left(\eta' \text{ s}\right)\} + \left(\eta' - 1\right)\log e - \log\eta', \tag{17}$$

in which

$$\mathbb{I}\left(\eta' \ \mathrm{s}\right) = \mathbb{D}\left(p_{Z|X,\mathrm{s};\eta'} \| p_{Z|\mathrm{s};\eta'} | p_X\right), \tag{18}$$

where $\beta = n/m$, s is the single user SNR, $\eta'$ is the multiuser efficiency determined from recursive relationships that depend on the type of detectors and $Z = \sqrt{\mathrm{s}}X + \frac{N}{\sqrt{\eta'}}$.

It is also shown in [40] that the asymptotic bound does not depend on the distribution of the signature matrix. In this paper, we will show that for the noiseless case, the asymptotic lower bound depends on the input and signature symbols; however, for the noisy case, the asymptotic lower bound does not depend on the signature matrix symbols.

Since most of the results of [14] and [15] are needed in the present paper, a summary of the results are also given here in the preliminaries.

### *C. Summary of [14]: A Class of Errorless Codes for Overloaded Synchronous Wireless and Optical CDMA Systems*

[14] is on the construction and simple ML decoding of errorless binary codes in the absence of noise and without near-far effects for a synchronous CDMA system. The main results of this paper that will be needed in the present paper are the following theorems:

*Theorem 4 (Enlarging COW Matrices):*
Assume that $\mathbf{C}_{cow}$ is an $m \times n$ COW matrix and $\mathbf{P}$ is an invertible $k \times k$ $\{\pm 1\}$ matrix, then $\mathbf{D}_{cow} = \mathbf{P} \otimes \mathbf{C}_{cow}$ is a $km \times kn$ COW matrix, where $\otimes$ denotes the Kronecker product.
Moreover the decoding of a system with the code matrix $\mathbf{D}_{cow}$ can be reduced to $k$ decoding systems with the code matrix $\mathbf{C}_{cow}$, where the decoding of $\mathbf{C}_{cow}$ can be implemented by $2^{n-m}$ Euclidean distance measurements using ML decoding [30].

*Theorem 5 (Highly Overloaded COW Matrices):*
Assume $\mathbf{C}_{cow}$ is an $m \times n$ COW matrix and $\mathbf{H}_2$ is a $2 \times 2$ Hadamard matrix. We can add $\lceil (m-1)\log_3 2 \rceil$ columns to $\mathbf{H}_2 \otimes \mathbf{C}_{cow}$ to obtain another COW matrix.

*Theorem 6 (Upper Bound for the Overloading Factor):*
For a COW/COO matrix, the total number of users $n$ is upper bounded by

$$n \leq -m \left( \sum_{i=0}^{n} \frac{\binom{n}{i}}{2^n} \log_2 \frac{\binom{n}{i}}{2^n} \right), \tag{19}$$

where $\binom{n}{i} = \frac{n!}{i!(n-i)!}$.

### D. Summary of [15]:Bounds on the Sum Capacity of Synchronous Binary CDMA Channels

*Theorem 7 (Noiseless Lower Bound):*

For any $m$ and $n$, the sum capacity is lower bounded by

$$C\left(m, n, \{\pm 1\}, \{\pm 1\}\right) \geq n - \log \sum_{j=0}^{\lfloor \frac{n}{2} \rfloor} \binom{n}{2j} \left( \frac{\binom{2j}{j}}{2^{2j}} \right)^m . \tag{20}$$

*Theorem 8 (Noisy Lower Bound):*

In Additive White Gaussian Noise (AWGN) channel with the real noise of variance $\sigma^2$, $\eta = \frac{1}{2\sigma^2}$ is equivalent to the normalized SNR, $E_b/N_0$. For any $m$ and $n$ and any positive real number $\gamma$, the sum capacity, as described in Definition 1, is lower bounded by:

$$C\left(m, n, \{\pm 1\}, \{\pm 1\}, \eta\right) \geq n - m\gamma \log \sqrt{e} - \log \sum_{k=0}^{n} \binom{n}{k} \left( \sum_{j=0}^{k} \frac{\binom{k}{j}}{2^k} \frac{e^{-\left(\frac{4\eta\gamma}{1+\gamma}\right)\left(\frac{2j-k}{\sqrt{m}}\right)^2}}{\sqrt{\gamma+1}} \right)^m . \tag{21}$$

*Theorem 9 (Conjectured Noisy Upper Bound):*

If the noise pdf $f(\cdot)$ is symmetric, we have

$$C\left(m, n, \{\pm 1\}, \{\pm 1\}, \eta\right) \leq \min\left(n, m\left(\mathbb{H}\left(\tilde{f}\right) - \mathbb{H}\left(f\right)\right)\right), \tag{22}$$

where

$$\tilde{f}\left(x\right) = \sum_{j=0}^{n} \frac{\binom{n}{j}}{2^n} f\left(x - \frac{2j-n}{\sqrt{m}}\right). \tag{23}$$

*Theorem 10 (Asymptotic Noiseless Lower Bound):*

For large scale systems when $\frac{n}{m \log n}$ is kept constant, the bound in (20) becomes

$$\lim_{\substack{n/(m\log n)\to\imath \\ m,n\to\infty}} \frac{1}{n} C\left(m, n, \{\pm 1\}, \{\pm 1\}\right) = \min\left(1, \frac{1}{2\imath}\right). \tag{24}$$

*Theorem 11 (Asymptotic Noisy Lower Bound):*

Likewise, for large scale systems when $\beta = \frac{n}{m}$ is kept constant, the bound in (21) becomes

$$\lim_{\substack{n/m\to\beta\\ m,n\to\infty}} \frac{1}{n} C\left(m,n,\{\pm 1\},\{\pm 1\},\eta\right)$$
$$\geq \sup_{\gamma}\left(1-\sup_{t\in[0,1]}\left[H\left(t\right)-\frac{1}{2\beta}\log\frac{1+2\eta\gamma\sigma^{2}+4t\beta}{e^{\gamma}}\right]\right), \tag{25}$$

where $H(t) = -t\log(t) - (1-t)\log(1-t)$.

*Theorem 12 (Asymptotic Noisy Upper Bound):*
Similarly, the bound in (22) becomes

$$\lim_{\substack{n/m\to\beta\\ m,n\to\infty}} \frac{1}{n} C\left(m,n,\{\pm 1\},\{\pm 1\},\eta\right) \leq \min\left(1, \frac{1}{2\beta}\log\left(1+2\eta\beta\right)\right). \tag{26}$$

*Note 1:* Numerical Improvements of the Bounds for Overloaded Binary CDMA
The bounds developed from Theorems 8-12 for the binary CDMA systems are less accurate for under-loaded (small $\beta$) and small scales $(m,n)$. However, for $\beta < 1$, the Hadamard signature matrix is equivalent to the single user Binary Phase Shift Keying (BPSK) and its actual capacity is known analytically and thus can be simulated. Fig. 1 shows the actual normalized sum capacity based on both hard and softdesition (soft decision is identical to hard for $\eta = 12\ dB$) for $\beta \leq 1, m = 8$. The COW/COO overloaded codes behave similar to the Hadamard codes that are for the fully/underloaded CDMA systems. If the noise level is less than a certain amount, the probability of error can be determined based on hard decision ML decoding, and the actual normalized sum capacity can be simulated. The same figure shows the capacity of a specific COW matrix for $\beta$ in the range $(1, 13/8)$ for a signal-to-noise ratio $\eta$ of $12\ dB$. Clearly, we can have better numerical bounds than that of Theorem 8.

Below, we will generalize the COW/COO codes of subsection II-C to nonbinary cases. The generalization of such codes for active users are also given in the same section. The rest of the sections are generalizations of the binary sum capacity bounds given in subsection II-D for finite/infinite, and real/complex symbols including the near-far effects.

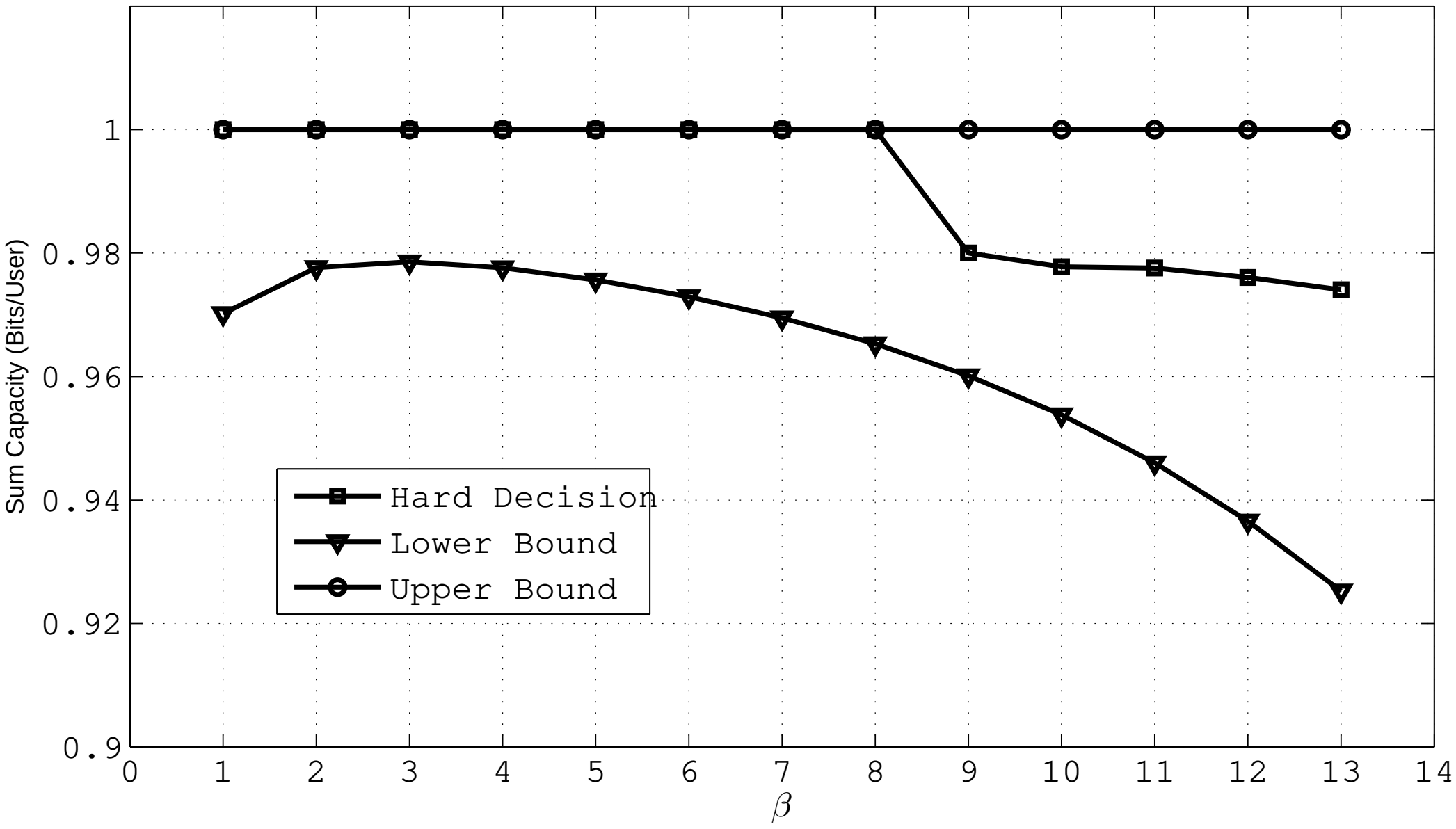

Fig. 1. Normalized upper and lower bounds for the sum capacity derived from (21) and (22) and the actual noormalized hard decision capacity. For $m = 8$, $\eta = 12$ (dB) and $\beta > 1$, the computer simulated sum capacity for the overloaded COW matrix is a better lower bound than the one predicted analytically from (21).

## III. Generalized COW/COO (GCO) Matrices

In this section, we extend the class of COW/COO codes discussed in Subsection II-C when the alphabets of the signatures and the user inputs are not limited to binary sets. In the following, we will discuss the construction of GCO matrices and the ML decoding of such codes. The concepts of fairness and active user detection will be examined.

### *A. Constructing GCO Matrices*

Suppose that the inputs are from a given set $\mathcal{I}$ and the signature entries are from a given set $\mathcal{S}$. For errorless communications, we need to find $m \times n$ signature matrices that are one-to-one over the set $\mathcal{I}^n$. We call such matrices GCO$\,(m, n, \mathcal{I}, \mathcal{S})$. The main problem is to find values of $m$ and $n$ such that GCO$\,(m, n, \mathcal{I}, \mathcal{S})$ matrices exist. In the following, we will develop a general theorem related to GCO matrices. Consequently, we will give corollaries and examples to construct and decode GCO matrices for overloaded nonbinary CDMA systems. In this section, we define $\tilde{\mathbf{I}}_m$ to be the first $m-1$ columns of the $m \times m$ identity matrix $\mathbf{I}_m$, $\tilde{\mathbf{J}}_m$ to be the first $m-1$ columns of the $m \times m$ all one matrix $\mathbf{J}_m$, $\tilde{\mathbf{K}}_m$ to be $2\tilde{\mathbf{I}}_m - \tilde{\mathbf{J}}_m$ and $\tilde{\mathbf{0}}_m$ to be the first $m-1$ columns of the $m \times m$ zero matrix $\mathbf{0}_m$. We can generalize

Theorem 4 for enlarging COW matrices to GCO matrices as shown in the following theorem:

*Theorem 13 (Constructing Larger GCO Matrices from Smaller Ones):*

Assume that $\mathbf{A}$ is a GCO$\,(m,n,\mathcal{I},\mathcal{S})$ matrix and $\mathbf{P}$ is a $w\times w$ invertible matrix with entries belonging to the set $\mathcal{R}$. Then, $\mathbf{P}\otimes\mathbf{A}$ is a GCO$\,(mw,nw,\mathcal{I},\mathcal{S}\cdot\mathcal{R})$ matrix where $\mathcal{S}\cdot\mathcal{R}$ is the set of all products of the elements of $\mathcal{S}$ and $\mathcal{R}$.

The proof is very similar to that of Theorem 4 in the preliminaries section, which is given in [29].

A generalization of Theorem 5 to GCO matrices is given below:

*Theorem 14 (A General Method for Constructing GCO Matrices):*

For integer sets of $\mathcal{I},\mathcal{S}$, assume that $\mathbf{C}$ is a GCO$\,(m,n,\mathcal{I},\mathcal{S})$, $\mathbf{H}_w$ is a $w\times w$ Hadamard matrix and $k$ is an integer number. Also, suppose that $\pm k\mathcal{S}\subset\mathcal{S}'$ and $d$ is the largest non-zero integer such that $\frac{\mathcal{I}-\mathcal{I}}{d}\subset\mathbb{Z}$. If $\mathbf{M}_1,\mathbf{M}_2,\ldots,\mathbf{M}_w$ are matrices with entries from $\mathcal{S}'$ such that $\sum_{i=1}^{w}\mathbf{M}_i=t\mathbf{I}_m$, where $t$ is any integer, then

$$\mathbf{B}=\left[k\mathbf{H}_w\otimes\mathbf{C}\;\begin{matrix}\mathbf{M}_1\\\mathbf{M}_2\\\vdots\\\mathbf{M}_w\end{matrix}\right]\tag{27}$$

is a GCO$\,(wm,wn+m,\mathcal{I},\mathcal{S}')$ matrix if

$$\left\{\frac{uwk}{t}:u\in\mathbb{Z}\right\}\cap\frac{\mathcal{I}-\mathcal{I}}{d}=\{0\}.\tag{28}$$

moreover, when $\mathcal{S}$ consists of only odd numbers and $\sum_{i=1}^{w}\mathbf{M}_i=t\tilde{\mathbf{I}}_m$, then $\mathbf{B}$ is GCO$\,(wm,wn+m-1,\mathcal{I},\mathcal{S}')$ matrix if

$$\left\{\frac{2uwk}{t}:u\in\mathbb{Z}\right\}\cap\frac{\mathcal{I}-\mathcal{I}}{d}=\{0\}.\tag{29}$$

The proof is given in Appendix A.

*Corollary 1 (Constructing Binary GCO Matrices):*

Assume that $\mathbf{A}$ is a GCO$\,(m,n,\{0,\pm1,\ldots,\pm q\},\{\pm1\})$ matrix, $\mathbf{H}_{4r}$ is a $4r\times4r$ Hadamard matrix and $4r\geq2q+1$. Also, assume that $\mathbf{M}_1=\tilde{\mathbf{J}}_m$, $\mathbf{M}_2=\tilde{\mathbf{K}}_m$, $\mathbf{M}_i=\tilde{\mathbf{J}}_m$ for $i=3,4,\ldots,2r+1$ and $\mathbf{M}_i=-\tilde{\mathbf{J}}_m$ for $i=2r+2,2r+3,\ldots,4r$. Now,

$$\left[\mathbf{H}_{4r} \otimes \mathbf{A} \begin{matrix} \mathbf{M}_1 \\ \mathbf{M}_2 \\ \vdots \\ \mathbf{M}_{4r} \end{matrix}\right] \tag{30}$$

is a GCO $(4rm, 4rn + m - 1, \{0, \pm 1, \ldots, \pm q\}, \{\pm 1\})$ matrix.

The proof is given in Appendix B.

*Example 2 ((3,2)-ary GCO):* Table I is a GCO $(16, 22, \{0, \pm 1\}, \{\pm 1\})$ matrix which is about $38\%$ over-loaded [29]. Using the above theorem, we get a GCO $(64, 103, \{0, \pm 1\}, \{\pm 1\})$ matrix which is about $60\%$ over-loaded. Notice that the first $16$ columns of the above matrix is a $16 \times 16$ Hadamard matrix.

TABLE I

$\mathbf{A}_{16\times 22}$ WHERE $+$ DENOTES $+1$ AND $-$ DENOTES $-1$.

$$\begin{bmatrix}
+&+&+&+&+&+&+&+&+&+&+&+&+&+&+&+&+&+&+&+&+&+\\
+&-&+&-&+&-&+&-&+&-&+&-&+&-&+&-&-&-&+&-&+&+\\
+&+&-&-&+&+&-&-&+&+&-&-&+&+&-&-&+&-&-&-&+&-\\
+&-&-&+&+&-&-&+&+&-&-&+&+&-&-&+&+&+&-&-&+&+\\
+&+&+&+&-&-&-&-&+&+&+&+&-&-&-&-&-&-&-&-&-&-\\
+&-&+&-&-&+&-&+&+&-&+&-&-&+&-&+&+&-&+&-&+&+\\
+&+&-&-&-&-&+&+&+&+&-&-&-&-&+&+&-&+&+&+&+&-\\
+&-&-&+&-&+&+&-&+&-&-&+&-&+&+&-&-&-&-&+&+&-\\
+&+&+&+&+&+&+&+&-&-&-&-&-&-&-&-&-&-&+&-&-&-\\
+&-&+&-&+&-&+&-&-&+&-&+&-&+&-&+&+&-&+&-&-&+\\
+&+&-&-&+&+&-&-&-&-&+&+&-&-&+&+&+&+&+&-&-&+\\
+&-&-&+&+&-&-&+&-&+&+&-&-&+&+&-&-&+&+&+&+&+\\
+&+&+&+&-&-&-&-&-&-&-&-&+&+&+&+&+&+&-&-&-&-\\
+&-&+&-&-&+&-&+&-&+&-&+&+&-&+&-&+&-&+&-&+&+\\
+&+&-&-&-&-&+&+&-&-&+&+&+&+&-&-&+&-&-&+&+&-\\
+&-&-&+&-&+&+&-&-&+&+&-&+&-&-&+&+&+&+&-&-&+
\end{bmatrix}$$

*Example 3 ((7,2)-ary GCO):* Note that $\mathbf{H}_8$ is a GCO $(8, 8, \{0, \pm 1, \pm 2, \pm 3\}, \{\pm 1\})$ matrix. Using the above theorem, we get a GCO $(64, 71, \{0, \pm 1, \pm 2, \pm 3\}, \{\pm 1\})$ matrix.

*Example 4 ((2,2)-ary GCO):* Using theorem 14, one can start with a $2 \times 2$ Hadamard matrix and construct a GCO $(64, 193, \{\pm 1\}, \{\pm 1\})$ matrix which is about $200\%$ over-loaded. This example is similar to that of [32], [42].

*Example 5 ((2,3)-ary GCO):* Suppose $\mathbf{A}$ is a GCO $(m, n, \{\pm 1\}, \{\pm 1\})$; from Theorem 14, by setting $k = 1$, $w = 1$, $\mathcal{S} = \{\pm 1\}$, $\mathcal{S}' = \{0, \pm 1\}$ and $\mathbf{M}_1 = \tilde{\mathbf{I}}_m$, we get $t = 1$ and $d = 2$. One can easily check that

$$\left\{\frac{2uwk}{t} : u \in \mathbb{Z}\right\} \cap \frac{\mathcal{I} - \mathcal{I}}{d} = \{0\}. \tag{31}$$

Thus $[\mathbf{A}\quad \mathbf{M}_1]$ is a GCO $(m, m+n-1, \{\pm 1\}, \{0, \pm 1\})$ matrix. If we choose $\mathbf{A}$ to be the GCO $(64, 193, \{\pm 1\}, \{\pm 1\})$ matrix generated from Example 4, we get a GCO $(64, 256, \{\pm 1\}, \{0, \pm 1\})$ matrix, which is $300\%$ overload. This example is also similar to that of [32].

*Corollary 2 (Constructing Nonbinary GCO Matrices):*
Assume that $\mathbf{A}$ is a GCO $(m, n, \{0, \pm 1, \ldots, \pm q\}, \{0, \pm 1, \ldots, \pm l\})$ matrix and $\mathbf{H}_w$ is a $w \times w$ Hadamard matrix. Also, assume that $\mathbf{M}_1 = \mathbf{I}_m$ and $\mathbf{M}_i = \mathbf{0}_m$ for $i = 2, 3, \ldots, w$. Now,

$$\left[ k\mathbf{H}_w \otimes \mathbf{A} \begin{matrix} \mathbf{M}_1 \\ \mathbf{M}_2 \\ \vdots \\ \mathbf{M}_w \end{matrix} \right] \tag{32}$$

is a GCO $(mw, nw+m, \{0, \pm 1, \ldots, \pm q\}, \{0, \pm 1, \ldots, \pm lk\})$ matrix, if $wk > 2q$.

The proof is similar to that of Corollary 1.

*Example 6 ((3,3)-ary GCO):* It is easy to verify that the matrix shown in Table II is a GCO $(4, 6, \{0, \pm 1\}, \{0, \pm 1\})$ matrix. Applying the previous theorem two times, we obtain a GCO $(64, 128, \{0, \pm 1\}, \{0, \pm 1\})$ matrix.

TABLE II

$\mathbf{A}_{4\times 6}$ WHERE " $+$ " DENOTES $+1$ AND " $-$ " DENOTES $-1$.

$$\begin{bmatrix} + & + & + & + & + & + \\ + & - & - & + & 0 & - \\ 0 & + & - & + & - & + \\ + & 0 & + & - & - & - \end{bmatrix}$$

*Example 7 ((3,5)-ary GCO):* From the previous example, we conclude that there exists a GCO $(64, 128, \{0, \pm 1\}, \{0, \pm 1\})$ matrix. Applying the previous theorem , we obtain a GCO $(128, 320, \{0, \pm 1\}, \{0, \pm 1, \pm 2\})$ matrix.

*Theorem 15 (Generation of Overloaded Complex GCO Matrices for Arbitrary Input Integers):* Let $\mathcal{I}$ be a finite set of integers and assume that $\mathbf{A}_i$ is a GCO $(m, n, \mathcal{I}, \{\pm 1\})$ matrix for $i = 0, 1, 2, \ldots, k$, and $\{1, \xi_1, \xi_2, \ldots, \xi_k\}$ are $k+1$ AIN[4] (real or complex) numbers. Also, assume that $\mathbf{B}_i = \left[\mathbf{H}_w \otimes \mathbf{A}_i \;\; \mathbf{M}^i\right]$ where $\mathbf{M}^i = \begin{pmatrix} \mathbf{M}_1{}^i \\ \vdots \\ \mathbf{M}_w{}^i \end{pmatrix}$ are GCO $(m, n, \mathcal{I}, \{0, \pm 1\})$ matrices generated as demonstrated in Theorem 14 for $i = 1, 2, \ldots, k$. Suppose $\mathbf{D}_i = \mathbf{H}_w \otimes \mathbf{A}_i$ for $i = 0, 1, 2, \ldots, k$ and $\tilde{\mathbf{M}}^i$ is obtained by replacing each 0 entry of $\xi_i \mathbf{M}^i$ by $+1$ or $-1$ arbitrarily. Then,
$\mathbf{B} = [\mathbf{D}_0 \quad \xi_1 \mathbf{D}_1 \quad \xi_2 \mathbf{D}_2 \quad \ldots \quad \xi_k \mathbf{D}_k \quad \tilde{\mathbf{M}}^1 \ldots \tilde{\mathbf{M}}^k]$ is a
GCO $(mw, (k+1)nw + (m-1)\,k, \{\pm 1\}, \{\pm 1, \pm\xi_1, \ldots, \pm\xi_k\})$ matrix.
The proof is given in Appendix C.

*Corollary 3 (Binary Input with Complex Matrix GCO):* Assume that $\mathbf{A}_i$'s are GCO $(m, n, \{\pm 1\}, \{\pm 1\})$ for $i = 0, 1, 2, \ldots, k-1$, and $\xi$ is a complex number with minimal polynomial of degree $k$. Also, assume that $\mathbf{M}_i = -\tilde{\mathbf{J}}_m + \left(1 + \xi^i\right) \tilde{\mathbf{I}}_m$ for $i = 1, \ldots, k-1$. Then, $[\mathbf{A}_0 \quad \xi \mathbf{A}_1 \quad \ldots \quad \xi^{k-1} \mathbf{A}_{k-1} \quad \mathbf{M}_1 \quad \ldots \quad \mathbf{M}_{k-1}]$ is a
GCO $\left(m, kn + (m-1)\,(k-1)\,, \{\pm 1\}, \{\pm 1, \pm\xi, \ldots, \pm\xi^{k-1}\}\right)$ matrix.
The proof is straightforward from Example 5 and Theorem 15.

*Example 8 (Binary Input with Quaternary Complex Matrix):* For the special case of $\xi = e^{\frac{j\pi}{k}}$, $m = 64$, $n = 193$, $k = 2$, we get a GCO$(64, 449, \{\pm 1\}, \{\pm 1, \pm j\})$. But in [43] they achieved an overloaded GCO$(64, 128, \{\pm 1\}, \{\pm 1, \pm j\})$, which is much smaller than the matrix derived from the above theorem. Now that we have given examples of constructing GCO codes, we can discuss the upper bound for the number of users and practical and optimum decoding of such codes.

### *B. Upper Bound for the Overloading Factor*

The following theorem provides a general upper-bound for the overloading factor for GCO matrices.

[4]We call a set of numbers Algebraically Independent Numbers (AIN), if linear combinations of the numbers with integer coefficient cannot become zero.

*Theorem 16 (A General Upper Bound for GCO$(m,n,\mathcal{I},\mathcal{S})$ Matrices):*
If there exists a GCO$(m,n,\mathcal{I},\mathcal{S})$, then

$$n\log_2|\mathcal{I}| \le m \max_{(a_1,\dots,a_n)\in\mathcal{S}^n} \mathbb{H}(Y_1), \tag{33}$$

where $\mathbb{P}(Y_1=l)$ is equal to the number of solutions of equation $\sum_{j=1}^{n} a_j X_j = l$ in $\mathcal{I}$, divided by $|\mathcal{I}|^n$. The Proof is given in Appendix D.

*Example 9 (Binary and Ternary Cases):*
For $\mathcal{I}=\mathcal{S}=\{\pm1\}$, the upper bound (33) becomes identical to that given in (19). On the other hand, for the special case $\mathcal{I}=\{0,\pm1\}$ and $\mathcal{S}=\{\pm1\}$, we get the following upper bound [29].

$$n \le -m\left(\sum_{i=0}^{n} \frac{g(n,k)}{3^n}\log_2\frac{g(n,k)}{3^n}\right), \tag{34}$$

where

$$g(n,k) = \sum_{r=0}^{\lfloor n-k\rfloor} \binom{n}{r}\binom{n-r}{k+r}. \tag{35}$$

The above upper bound is simulated in Fig. 2. This figure shows that we cannot have errorless communications beyond $230$ ternary users (equivalent to $230\log_2 3 \approx 365$ bits) with $64$ chips. It is interesting to compare this number to the binary case in (19) where the upper bound for the number of users is $268$.

## C. ML Decoding for a Class of GCO Codes

The direct ML decoding of GCO codes is computationally very expensive for moderate values of $m$ and $n$. In this subsection, we present two lemmas for decreasing the computational complexity of the ML decoding for a class of GCO codes.

*Lemma 1 (Decoding Method for Large Matrices):*
Suppose $\mathbf{D}_{wm\times wn} = \mathbf{P}_{w\times w}\otimes\mathbf{A}_{m\times n}$ is constructed as discussed in Theorem 13. The decoding problem of a system with the code matrix $\mathbf{D}$ can be reduced to $k$ decoding problems of a system with the code matrix $\mathbf{A}$.

*Proof*: The proof is similar to that of Theorem 4 in the preliminaries.

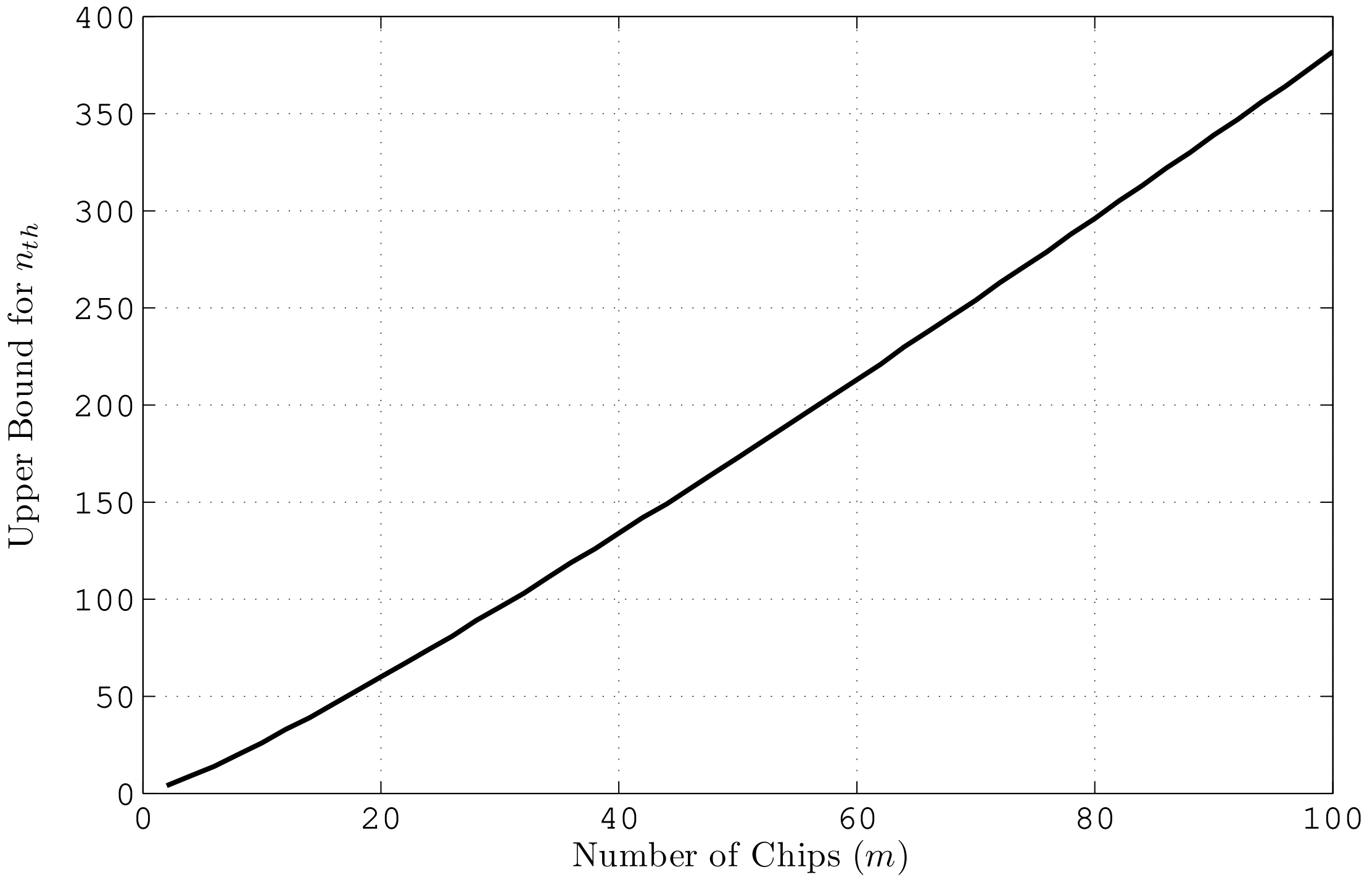

Fig. 2. Upper bound on the maximum number of overloaded users ($n_{th}$) vs. the number of chips for a $(3,2)$-ary CDMA system in the absence of additive noise.

*Lemma 2 (Optimum Decoding Algorithm):*

If $\mathbf{A}$ is a full rank GCO $(m, n, \mathcal{I}, \mathcal{S})$, then the decoding problem for a system with the code matrix $\mathbf{A}$ can be performed through $|\mathcal{I}|^{n-m}$ Euclidean distance measurements.

*Proof*: The proof is similar to that of Theorem 4 in the preliminaries; however, instead of using the $sign$ function, a vector is mapped to the nearest $\mathcal{I}$-vector.

Lemmas 1 and 2 lead to significant decrease of decoding complexity. Since each vector is mapped to the nearest $\mathcal{I}$ vector, it is not hard to show that if $m$ columns of $\mathbf{A}$ make a scaled unitary matrix, then the proposed method for decoding is an ML decoder.

### *D. Simulation Results for GCO Matrices*

Figure 3 shows the performance of a GCO matrix $(3,2)$ generated from the kronecker product of the matrix represented in Table I by a $4 \times 4$ Hadamard matrix in a noisy environment. This GCO matrix has dimension $64 \times 88$; it also confirms that GCO codes, similar to COW/COO codes for the binary case, are superior to WBE sequences. The ML decoding has been performed by using Lemmas 1 and 2.

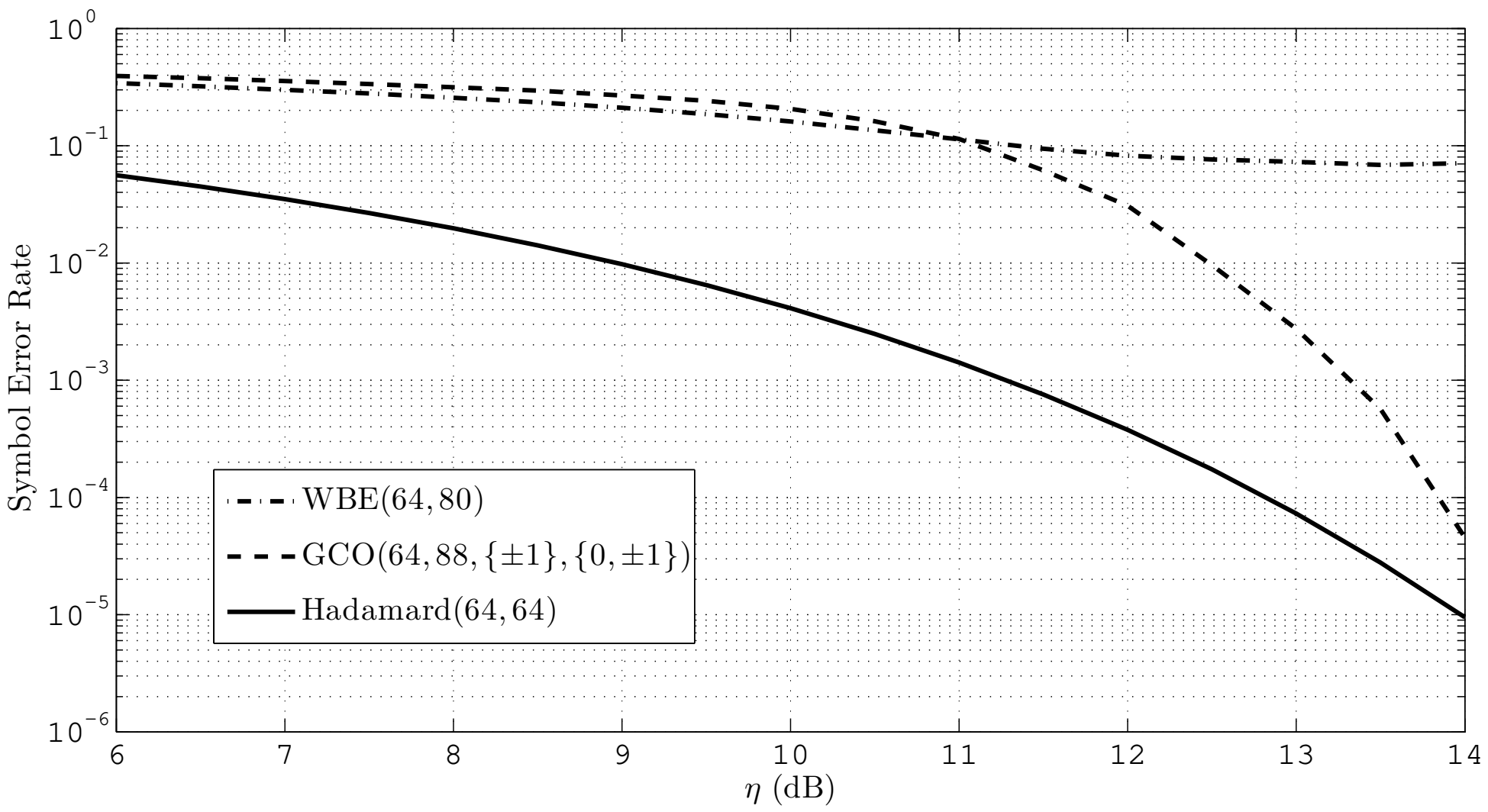

Fig. 3. Symbol-Error-Rate versus normalized SNR ($\eta$) for an overloaded CDMA system with $64$ chips and $88$ users for a (3,2)-ary GCO system with signature matrix $\mathbf{H}_4 \otimes \mathbf{A}_{16\times 22}$ (Table I).

*Note 2:* Fairness

The COW/COO codes, for a given signature matrix, do not have identical performance in the presence of additive noise. The performance of ($m = 64$,$n = 96$) COW codes is shown in Fig. 4. On the same figure, one can see (64,96) Binary WBE (BWBE) codes [44] for comparison. Although the BWBE code looks fairer, its performance is not acceptable. On the other hand, the Hadamard codes for underloaded CDMA and the Optical Orthogonal Codes (OOC) are fair, but OOC codes have poor performance for the overloaded case.

The extension of the concept of fairness to the general case is also simulated in Fig. 5 for a GCO $(64, 88, \{\pm 1\}, \{0, \pm 1\})$. This figure also shows that the GCO codes are not as fair as WBE codes but their performances are better than that of WBE codes.

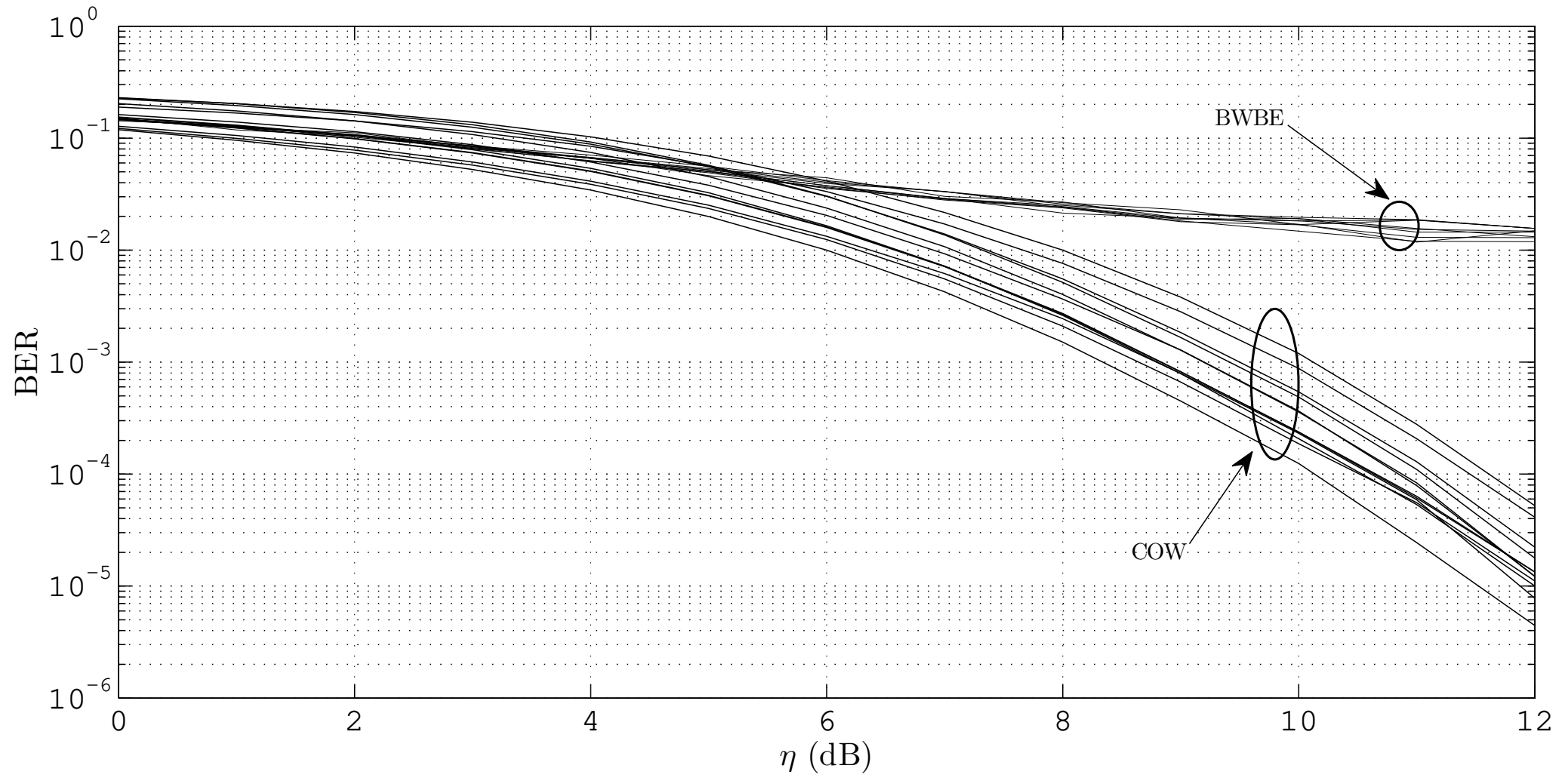

Fig. 4. Bit Error Rate (BER) versus $\eta$ for different code signatures for specific COW and BWBE matrices. The COW codes are not as fair as BWBE codes but their performances are better than that of BWBE codes.

### *E. Codes for Overloaded Wireless CDMA with Detection of Active Users (COWDA)*

In a practical CDMA system, not all users are always active. This can be modeled by $\pm 1$ for binary user data and $0$ when nothing is sent[5]. This becomes problematic for Multi User Detection (MUD); to obtain errorless communication, we need binary signature matrices that are one-to-one over ternary input vectors. This class of codes are named Codes for Overloaded Wireless CDMA with Detection of Active (COWDA) users. This class is identical to GCO$\,(m, n, \{0, \pm 1\}, \{\pm 1\})$ as discussed in Section III.

*1) Decoding Algorithm and Simulation Results:* The decoding algorithm is identical to that of Sub-section III-C. This decoding has the added advantage that it works with an unknown number of active users [29].

To show the behavior of COWDA codes with respect to additive noise, we have simulated a CDMA system with $64$ chips and $88$ users in the presence of AWGN. The BER versus $\eta$ is depicted in Fig. 6. The simulation result for the COWDA code is obtained by using input vectors $X$ with equiprobable entries in the set $\{0, \pm 1\}$. This is not a typical case and does not occur in practice because, in actual systems, when a user is active, it stays active for a period of time. Thus, entries at different times are not independent and do not randomly alternate between $\pm 1$ and $0$. Therefore, the BER curve of COWDA is an upper bound for practical situations.

[5]Or more generally when user data consists of $\{0,\ \mathcal{I}\}$.

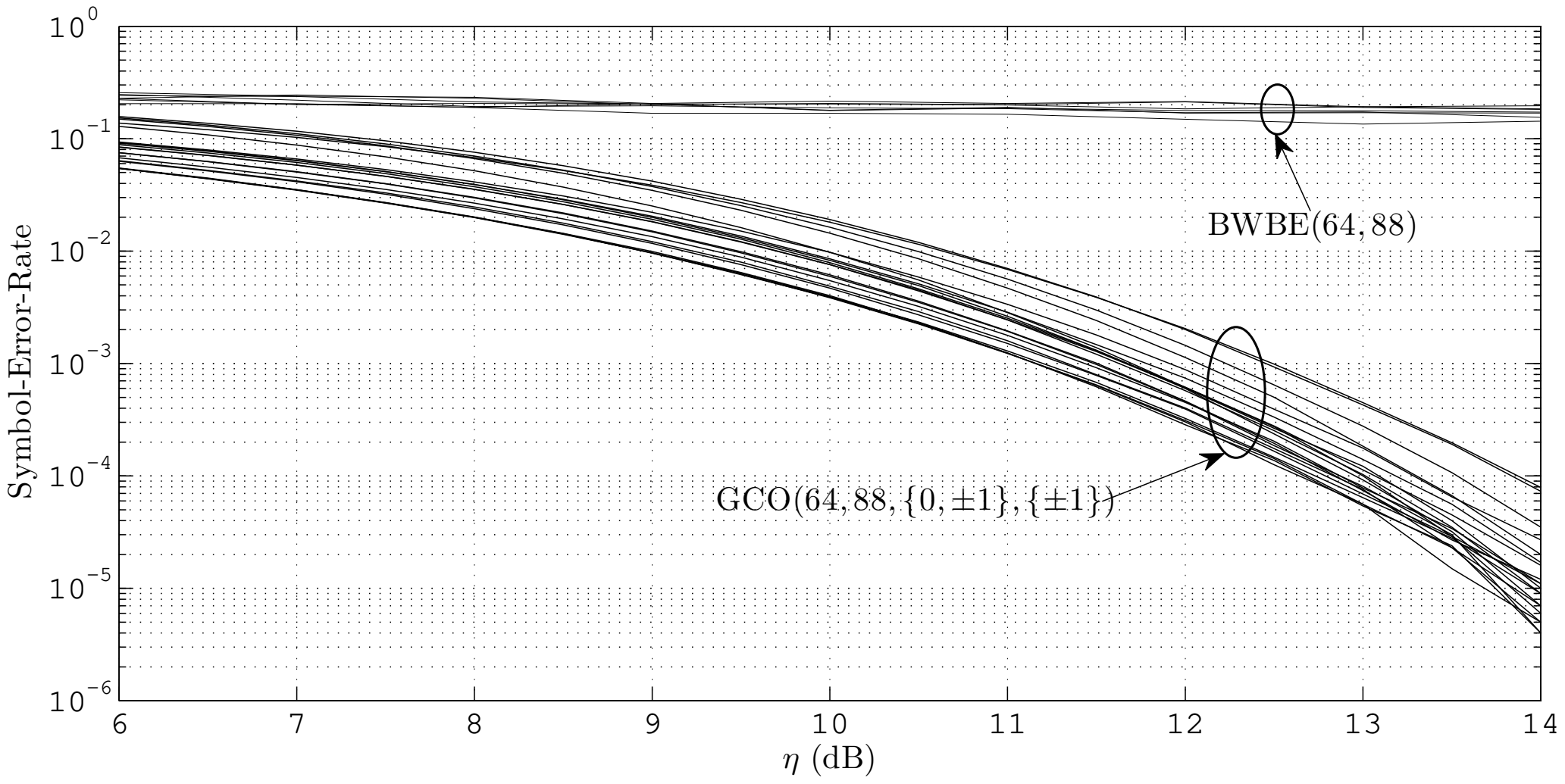

Fig. 5. Symbol Error Rate versus $\eta$ for different code signatures for the GCO $(64, 88, \{\pm 1\}, \{0, \pm 1\})$ and WBE matrices. The GCO codes are not as fair as WBE codes but their performances are better than that of WBE codes.

We have compared COWDA/ML decoder from the BER point of view with COW/ML decoder and BWBE/iterative decoder as depicted in Fig. 6; this figure shows the degradation of COWDA relative to COW/ML where there is no need for additional process of active user detection. As shown in Fig. 6, BWBE codes cannot reach error probabilities below a certain threshold even by increasing the power. Clearly, the BER of COWDA tends to zero as $\eta$ increases. This means that we can have a CDMA system with a desired low BER using COWDA matrices. For variable-active users, the sum capacity bounds will be discussed in Section VII

## IV. Sum Capacity Bounds for General User Inputs and Matrices- the Noiseless Case

Let $\mathcal{I}$ and $\mathcal{S}$ be two sets of (real or complex) numbers. In this section, we consider the noiseless CDMA channel $Y = \frac{1}{\sqrt{m}}\mathbf{A}X$ where the entries of the user input vector $X$ and the signature matrix $\mathbf{A}$ belong to $\mathcal{I}$ and $\mathcal{S}$, respectively, i.e., $X \in \mathcal{M}_{n\times 1}(\mathcal{I})$ and $\mathbf{A} \in \mathcal{M}_{m\times n}(\mathcal{S})$.[6]

To state our essential lemma about the expected mutual information, we should fix some notations:

*Definition 2 (The Probability of Difference of Two i.i.d. Random Variables):*

[6]For the noiseless case, the power constraint in (3) is irrelevant and therefore, $r$ in (5) can be set to 1.

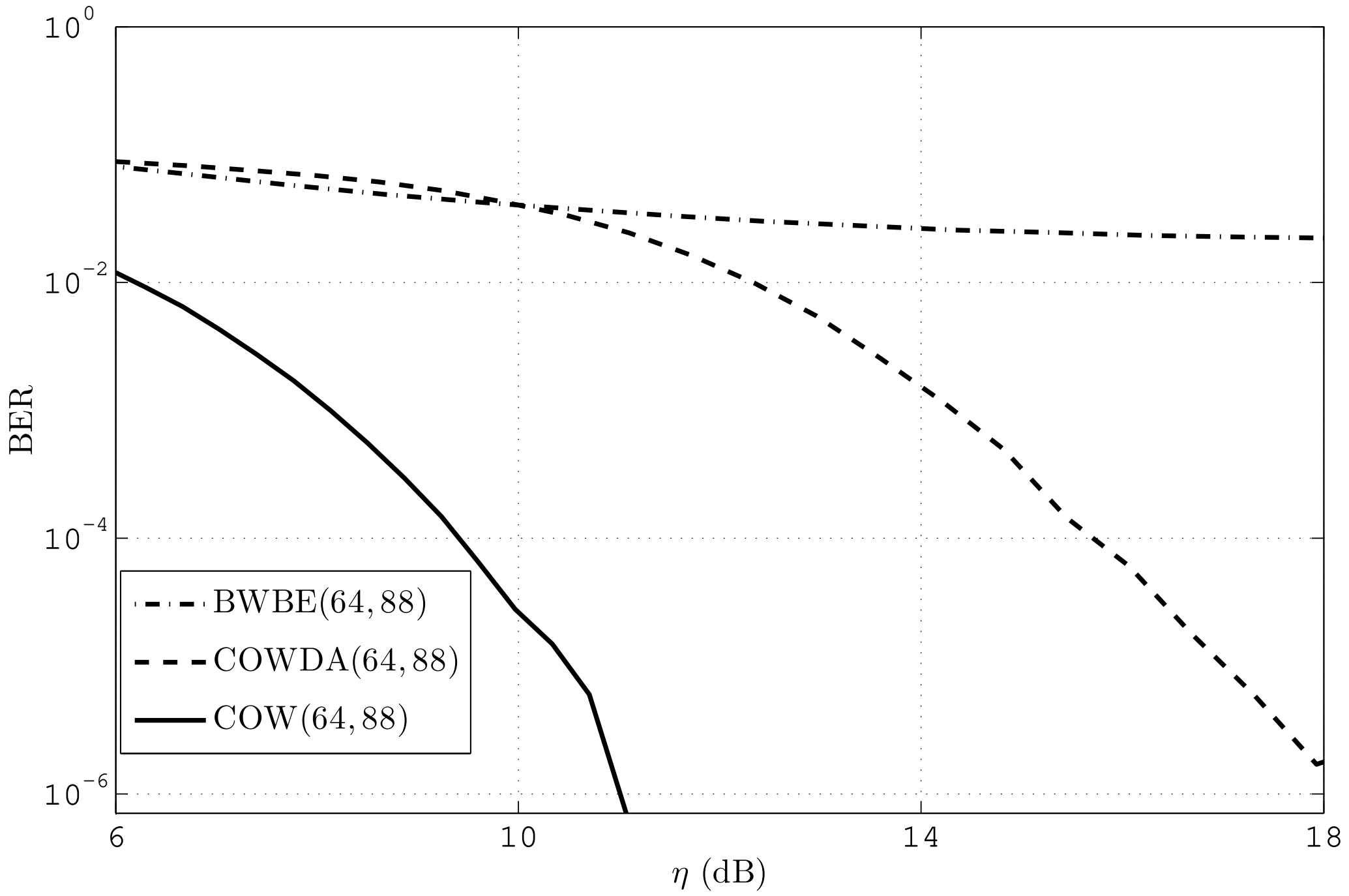

Fig. 6. BER versus $\eta$ for a COWDA matrix with 64 chips and 88 users. The COW performance is for the case when all users are active.

$\tilde{\mathcal{I}}$, the difference set of $\mathcal{I}$, is defined as:

$$\tilde{\mathcal{I}} = \mathcal{I} - \mathcal{I} = \{\imath - \imath' | \imath, \imath' \in \mathcal{I}\}. \tag{36}$$

For a probability law $p(\cdot)$ on $\mathcal{I}$, we define $\tilde{p}(\cdot)$ to be the probability law on $\tilde{\mathcal{I}}$ which is the pdf of the difference of two independent random variables with the same probability distribution $p(\cdot)$.

Finally, let $\pi(\cdot)$ be a probability distribution on $\mathcal{S}$. The probability measure $\mathbb{P}^{\pi}$ is the probability measure on $\mathcal{M}_{m\times n}(\mathcal{S})$ induced by choosing entries of the random matrix independently and with the same distribution $\pi(\cdot)$.

## *A. Sum Capacity Lower Bound*

In order to obtain lower bounds for the sum capacity for various scenarios, we prove a general theorem. Special cases are derived from this theorem.

*Theorem 17 (A General Lower Bound for the Noiseless Case):*
In the absence of additive noise, for any $\mathcal{I}$ and $\mathcal{S}$, we have

$$C\left(m,n,\mathcal{I},\mathcal{S}\right) \geq \sup_{p,\pi}\left\{-\log \mathbb{E}_{\tilde{X}}\left(\mathbb{P}\Big(a^T\tilde{X}=0\Big)^m\right)\right\}, \tag{37}$$

where $a \in \mathcal{S}^n$ and $\tilde{X} \in \tilde{\mathcal{I}}^n$ with i.i.d. entries with distributions $\pi(\cdot)$ and $\tilde{p}(\cdot)$, respectively.
The proof is given in Appendix E.

For the case when the input and the signature matrix alphabets are finite, the above theorem can be simplified as follows:

*Corollary 4 (Finite Input/Finite Signature Matrix):* Suppose that $\tilde{\mathcal{I}} = \{\tilde{\imath}_1, \ldots, \tilde{\imath}_{\tilde{q}}\}$. Let $k_1, \ldots, k_{\tilde{q}}$ be nonnegative integers that satisfy $k_1 + \cdots + k_{\tilde{q}} = n$. Define $\mathcal{B}\left(k_1, \ldots, k_{\tilde{q}}\right)$ to be the set of all $n \times 1$ $\tilde{X}$ vectors such that the number of occurrences of $\tilde{\imath}_1, \ldots, \tilde{\imath}_{\tilde{q}}$ in $\tilde{X}$ are $k_1, \ldots, k_{\tilde{q}}$, respectively. It is easy to show that if $\tilde{X}_1, \tilde{X}_2 \in \mathcal{B}\left(k_1, \ldots, k_{\tilde{q}}\right)$, then $\mathbb{P}\left(a^T\tilde{X}_1 = 0\right) = \mathbb{P}\left(a^T\tilde{X}_2 = 0\right) = \Phi\left(k_1, \ldots, k_{\tilde{q}}, \mathcal{S}\right)$. Thus, (37) becomes

$$C\left(m,n,\mathcal{I},\mathcal{S}\right) \geq -\log \sum_{k_1+\cdots+k_{\tilde{q}}=n} \binom{n}{k_1, \cdots, k_{\tilde{q}}} \tilde{p}\left(\tilde{\imath}_1\right)^{k_1} \cdots \tilde{p}\left(\tilde{\imath}_{\tilde{q}}\right)^{k_{\tilde{q}}} \Phi\left(k_1, \ldots, k_{\tilde{q}}, \mathcal{S}\right)^m. \tag{38}$$

## B. Sum Capacity Upper Bound

Below a general conjectured upper bound for the noiseless case is given.

*Theorem 18 (A General Upper Bound for the Noiseless Case):*
In the absence of additive noise, if $\mathcal{I} = \{\imath_1, \ldots, \imath_q\}$ with distribution $p(\imath_i) = p_i$ and $\mathcal{S} = \{s_1, \ldots, s_l\}$, we have the following conjectured upper bound:

$$C\left(m,n,\mathcal{I},\mathcal{S}\right) \leq \max_{\substack{\sum_{i=1}^{l} u_i = n \\ p(\cdot)}} \left\{\min\left(n\mathbb{H}(\mathcal{I}), m\mathbb{H}(\tilde{f})\right)\right\}, \tag{39}$$

in which

$$\tilde{f}(z) = \sum_{\substack{\sum_{j=1}^{q} v_{ij} = u_i \\ 1 \leq i \leq l}} \left(\prod_{k=1}^{l} \binom{u_k}{v_{k1}, \ldots, v_{kq}}\right)\left(\prod_{k=1}^{q} p_k^{\sum_{\alpha=1}^{l} v_{\alpha k}}\right) \delta\Big(z - \frac{1}{\sqrt{m}}\Big(\sum_{k=1}^{l} s_k \sum_{\alpha=1}^{q} v_{k\alpha}\imath_\alpha\Big)\Big), \tag{40}$$

where $\delta$ is the Dirac function.
Also, when $s_i = e^{\frac{2\pi i}{l}\sqrt{-1}}$ and $l$ divides $n$, we conjecture that $u_1 = u_2 = \cdots = u_l = \frac{n}{l}$.

This conjecture is a special case of the conjectured upper bound for the noisy Theorem 23 when the noise variance is zero without restricting the signature to have unity magnitude. An upper bound that does not depend on the symbol alphabets but is simpler to evaluate is given in following theorem. However the bound is not as tight as the above upper bound.

*Theorem 19 (Noiseless Upper Bound):*
If $|\mathcal{I}| = q$ and $|\mathcal{S}| = l$, then

$$C(m,n,\mathcal{I},\mathcal{S}) \leq m \cdot \max_{\sum_{i=1}^{l} u_i} \sum_{j=1}^{l} \mathbb{H}\left(\text{Mult}\left(u_j, \frac{1}{q}, \ldots, \frac{1}{q}\right)\right), \tag{41}$$

where $\text{Mult}\left(t, \theta_1, \ldots, \theta_q\right)$ denotes the multinomial distribution with probability $\binom{t}{x_1 \ldots x_q}\theta_1^{x_1} \ldots \theta_q^{x_q}$ at $(x_1, \ldots, x_q)$ when $x_1, \ldots, x_q$ are nonnegative integers and $x_1 + \cdots + x_q = t$.

*Remark 2:* It seems reasonable to believe that the maximum value in the above bound is attained at $u_1 = \cdots = u_l = \frac{n}{l}$ and hence the bound may be simplified as

$$C\left(m,n,\mathcal{I},\mathcal{S}\right) \leq ml\mathbb{H}\left(\text{Mult}\left(\frac{n}{l}, \frac{1}{q}, \ldots, \frac{1}{q}\right)\right). \tag{42}$$

The proof is given in Appendix F.

## C. Asymptotic Bounds

*Theorem 20 (Asymptotic Noiseless Lower Bound):*
Let $p$ and $\pi$ be probability distributions on $\mathcal{I}$ and $\mathcal{S}$ with $\mu_\pi = 0$ and $\tilde{p}$ be the distribution induced on $\tilde{\mathcal{I}}$ as before. Then

$$\lim_{\substack{n,m\to\infty \\ \frac{n}{m\log n}\to\zeta}} \frac{1}{n} C\left(m,n,\mathcal{I},\mathcal{S}\right) \geq \min_{J\subseteq\tilde{\mathcal{I}}} \left\{\frac{\text{rank}\left(J\cdot\mathcal{S}\right)}{2\zeta} - \log\tilde{p}\left(J\right)\right\}, \tag{43}$$

where $J \cdot \mathcal{S} = \{js | j \in J, s \in \mathcal{S}\}$, $\tilde{p}\left(J\right) = \sum_{j\in J}\tilde{p}\left(j\right)$ and for a set of numbers $\Lambda$, $\text{rank}\left(\Lambda\right)$ denotes the dimension of $\Lambda$ as a set of vectors over the field of rational numbers $\mathbb{Q}$.
For the proof see Appendix G.

*Example 10:* Let $\mathcal{I} = \mathcal{S} = \{\pm 1\}$ and $\pi$, $p$ be uniform distributions on $\mathcal{I}$ and $\mathcal{S}$. Therefore, $\tilde{\mathcal{I}} = \{-2, 0, 2\}$ and $\tilde{p}\left(-2\right) = \tilde{p}\left(2\right) = \frac{1}{4}$ and $\tilde{p}\left(0\right) = \frac{1}{2}$. Now the maximum $J \subseteq \tilde{I}$ with $\text{rank}\left(J \cdot \mathcal{S}\right) = 1$ is

$J = \tilde{\mathcal{I}}$ with $\tilde{p}(J) = 1$ and therefore the bound in theorem becomes

$$\min\left\{\frac{0}{2\zeta} - \log\frac{1}{2}, \frac{1}{2\zeta} - \log(1)\right\} = \min\left\{1, \frac{1}{2\zeta}\right\}. \quad (44)$$

This is the same result as the binary case given in (24) of Preliminaries.

*Example 11:* Let $\mathcal{I} = \{\pm 1\}$, $\mathcal{S} = \{\pm 1, \pm j\}$ and $\pi$, $p$ be uniform distributions on $\mathcal{I}$ and $\mathcal{S}$. Therefore $\tilde{\mathcal{I}} = \{-2, 0, 2\}$ and $\tilde{p}(-2) = \tilde{p}(2) = \frac{1}{4}$ and $\tilde{p}(0) = \frac{1}{2}$. Now the maximum $J \subseteq \tilde{I}$ with $\text{rank}(J \cdot \mathcal{S}) = 2$ is $J = \tilde{\mathcal{I}}$ with $\tilde{p}(J) = 1$ and therefore the bound in theorem becomes

$$\min\left\{\frac{0}{2\zeta} - \log\frac{1}{2}, \frac{2}{2\zeta} - \log(1)\right\} = \min\left\{1, \frac{1}{\zeta}\right\}. \quad (45)$$

The above asymptotic bound and finite lower bounds for various values of $m$ are depicted in Fig. 7. This figure shows that, unlike the asymptotic results for the noisy case that will be discussed in the next

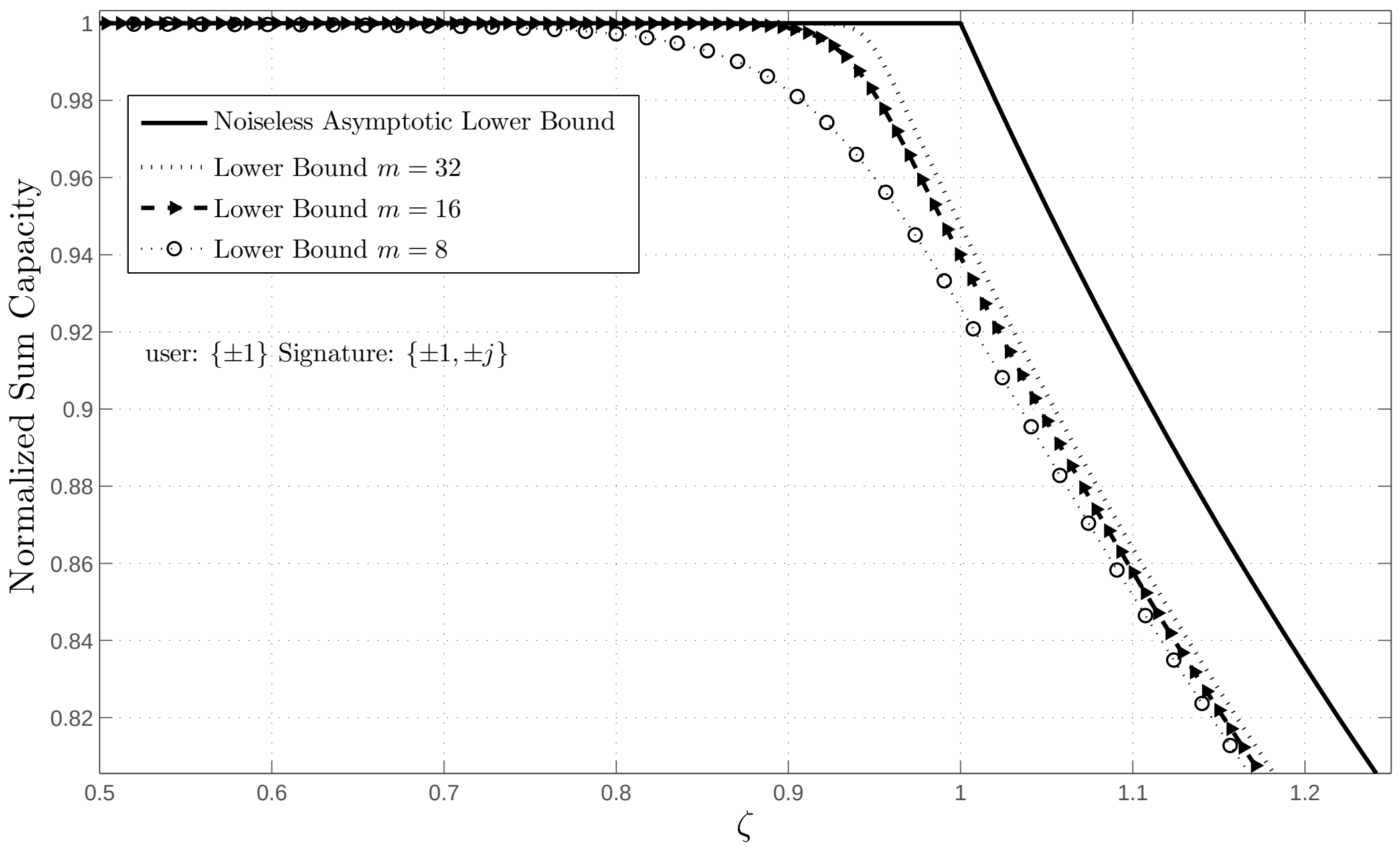

Fig. 7. The asymptotic noiseless lower bound is compared to the normalized finite scaled CDMA systems. This figure shows that small to medium scaled systems cannot be accurately estimated by the asymptotic lower bound for high values of $\zeta$.

section, medium scaled systems cannot be accurately estimated by the asymptotic lower bounds.

### *D. Examples*

Below, we will give many examples and simulation results that are special cases of Corollary 4. The derivation of the following formulas are all given in Appendix H.

*Example 12:* Binary Wireless CDMA, $\mathcal{I} = \mathcal{S} = \{\pm 1\}$, $p(1) = p_1$, $\pi(1) = \pi_1$

$$C(m,n,\mathcal{I},\mathcal{S}) \geq \sup_{p_1,\pi_1} \{-\log \sum_{k_1+k_2+k_3=n} \binom{n}{k_1,k_2,k_3} (p_1(1-p_1))^{k_1+k_2} \left(p_1^2 + (1-p_1)^2\right)^{k_3}$$

$$\times \left( \sum_{\alpha=0}^{k_1} \binom{k_1}{\alpha} \binom{k_2}{\alpha + \frac{k_2-k_1}{2}} {\pi_1}^{2\alpha + \frac{k_2-k_1}{2}} (1-\pi_1)^{(\frac{3k_1+k_2}{2} - 2\alpha)} \right)^m \}. \tag{46}$$

When $\pi_1 = p_1 = 1/2$, the above equation becomes identical to (20) in the preliminaries section; the simulation of (46) for the uniform distribution is given in Fig. 8.

*Example 13:* Binary Optical CDMA, $\mathcal{I} = \mathcal{S} = \{0, 1\}$, $p(0) = p_0$, $\pi(0) = \pi_0$

$$C(m,n,\mathcal{I},\mathcal{S}) \geq \sup_{p_0,\pi_0} \{-\log \sum_{k_1+k_2+k_3=n} \binom{n}{k_1,k_2,k_3} (p_0(1-p_0))^{k_1+k_2} \left(p_0^2 + (1-p_0)^2\right)^{k_3} \tag{47}$$

$$\times \left( \sum_{\alpha=0}^{k_1} \binom{k_1}{\alpha} \binom{k_2}{\alpha} (1-\pi_0)^{2\alpha} \pi_0^{k_1+k_2-2\alpha} \right)^m \}.$$

Fig. 8 also shows the above lower bound when $p_0 = \pi_0 = 1/2$.

*Example 14:* Binary Complex Signature (Uniform Distribution), $\mathcal{I} = \{\pm 1\}$, $\mathcal{S} = \{1, j\}$

$$C(m,n,\mathcal{I},\mathcal{S}) \geq -\log \left( \sum_{k=0}^{\lfloor n/2 \rfloor} \binom{n}{k,k,n-2k} \left(\frac{1}{2}\right)^{n+2k} \left( \frac{\binom{2k}{k}}{2^{2k}} \right)^m \right). \tag{48}$$

The above bound is also plotted in Fig. 8. This figure shows that the lower bound for the complex signature matrix is slightly better than the optical case and the optical case is better than the binary wireless. We shall see in Subsection VI-C.1 that under additive noisy environment, this statement may not be true.

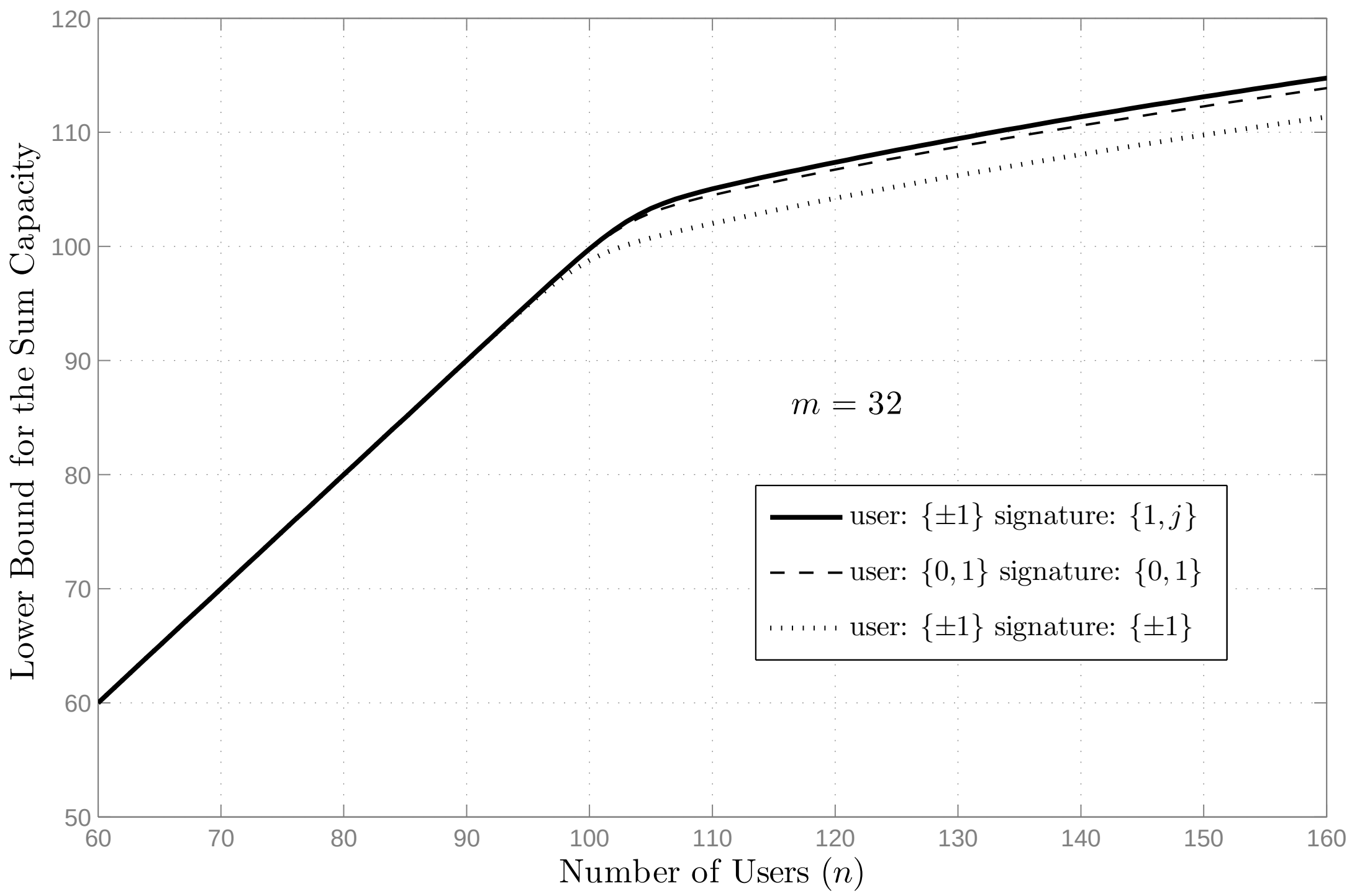

Fig. 8. The sum capacity lower bound versus the number of users for binary input and binary signature matrices when all the probabilities are equal to $1/2$ for $m = 32$.

*Example 15:* Binary/Ternary, $\mathcal{I} = \{\pm 1\}$, $\mathcal{S} = \{0, \pm 1\}$, $\pi(0) = \pi_0$ and $\pi(+1) = \pi(-1) = \pi_1$

$$C(m, n, \mathcal{I}, \mathcal{S}) \geq \sup_{\pi_0, \pi_1} \left\{ -\log \sum_{k=0}^{n} \binom{n}{k} \left(\frac{1}{2}\right)^n \left( \sum_{\alpha=0}^{k} \binom{k}{\alpha, \alpha, k-2\alpha} \pi_0^{k-2\alpha} \pi_1^{2\alpha} \right)^m \right\}. \tag{49}$$

See Fig. 9 for the simulation result of the above equation. A slightly different bound for this system is given in [45].

*Example 16:* Binary/Ternary with Uniform Distribution, $\mathcal{I} = \{\pm 1\}$, $\mathcal{S}$ is a set of ternary AIN's

$$C(m, n, \mathcal{I}, \mathcal{S}) \geq -\log \sum_{k=0}^{n} \binom{n}{k, k, n-2k} \left(\frac{1}{2}\right)^{n+2k} \left( \frac{1}{3^{2k}} \sum_{\alpha+\beta+\delta=k} \binom{k}{\alpha, \beta, \delta}^2 \right)^m. \tag{50}$$

The above bound is plotted in Fig. 9.

*Example 17:* Binary/Ternary CDMA system (Uniform Distribution), $\mathcal{I} = \{\pm 1\}$, $\mathcal{S} = \{1, e^{\pm j\frac{2\pi}{3}}\}$

$$C(m,n,\mathcal{I},\mathcal{S}) \geq -\log \sum_{k_1+k_2+k_3=n} \binom{n}{k_1,k_2,k_3} \left(\frac{1}{2}\right)^{n+k_1+k_2}$$
$$\times \left(\frac{1}{3^{k_1+k_2}} \sum_{\alpha+\beta+\delta=k_1} \binom{k_1}{\alpha,\beta,\delta} \binom{k_2}{\alpha-\frac{k_1-k_2}{3}, \beta-\frac{k_1-k_2}{3}, \delta-\frac{k_1-k_2}{3}} \right)^m . \tag{51}$$

The above bound is simulated in Fig. 9. This figure shows that the lower bound for the AIN ternary signature matrix is slightly better than the third roots of unity, which is significantly better than the $\{0, \pm 1\}$ case.

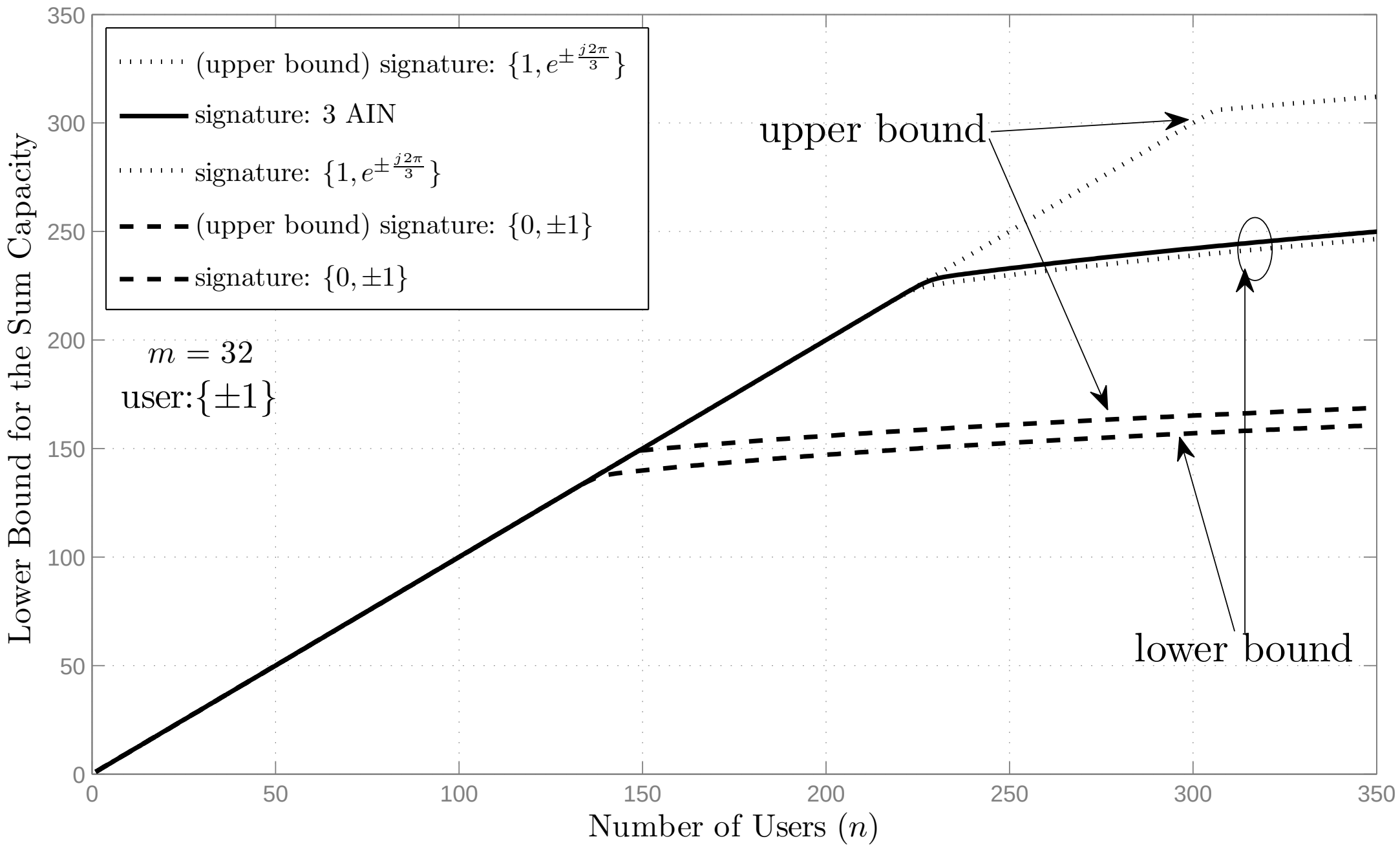

Fig. 9. The sum capacity lower and upper bounds versus the number of users for binary input and ternary signature matrices.

*Example 18:* Ternary/Binary CDMA System, $\mathcal{I} = \{0, \pm 1\}, \mathcal{S} = \{\pm 1\}$, $p\,(0) = p_0$, $p(+1) = p(-1) = p_1$ and $\pi(+1) = \frac{1}{2}$

$$C(m,n,\mathcal{I},\mathcal{S}) \geq \sup_{p(\cdot)}\{-\log \sum_{k_1+k_2+k_3=n} \binom{n}{k_1,k_2,k_3} \left(2p_1^2\right)^{k_1} (4p_0p_1)^{k_2} \left({p_0}^2 + 2p_1^2\right)^{k_3}$$
$$\times \left( \sum_{\alpha=0}^{k_1} \binom{k_1}{\alpha} \binom{k_2}{\frac{k_2}{2}+k_1-2\alpha} \left(\frac{1}{2}\right)^{k_1+k_2} \right)^m \}. \tag{52}$$

The above bound is simulated in Fig. 10 and is compared to the ternary/ternary case of the next example.

*Example 19:* Ternary Wireless System, $\mathcal{I} = \mathcal{S} = \{0, \pm 1\}$, $p\,(0) = p_0$, $p(+1) = p(-1) = p_1$, $\pi(0) = \pi_0$ and $\pi(+1) = \pi(-1) = \pi_1$

$$C(m,n,\mathcal{I},\mathcal{S}) \geq \sup_{p(\cdot),\pi(\cdot)}\{-\log \sum_{k_1+k_2+k_3=n} \binom{n}{k_1,k_2,k_3} \left(2p_1^2\right)^{k_1} (4p_0p_1)^{k_2} \left(p_0^2 + 2p_1^2\right)^{k_3}$$
$$\times \left( \sum_{\alpha+\beta+\delta=k_1} \sum_{\alpha'=0}^{k_2} \binom{k_1}{\alpha,\beta,\delta} \binom{k_2}{\alpha',\beta',k_2-\alpha'-\beta'} \pi_1^{\alpha+\beta+\alpha'+\beta'} \pi_0^{k_1+k_2-\alpha-\beta-\alpha'-\beta'} \right)^m \}, \tag{53}$$

in which $\beta' = 2(\alpha - \beta) + \alpha'$.

The above bound is simulated in Fig. 10. This figure shows the upper and lower bounds for the ternary/ternary system, which is relatively tight. It also shows the lower bound of ternary/binary system of Example 18, which is lower than the ternary/ternary system.

*Example 20:* Binary/Quaternary CDMA System when $\mathcal{I} = \{\pm 1\}$ and $\mathcal{S}$ is a set of four AIN's

$$C(m,n,\mathcal{I},\mathcal{S}) \geq -\log \sum_{k=0}^{\lfloor \frac{n}{2} \rfloor} \binom{n}{k,k,n-2k} \left(\frac{1}{2}\right)^{n+2k} \left( \frac{1}{4^{2k}} \sum_{\alpha=0}^{k} \binom{k}{m}^2 \binom{2m}{m} \binom{2k-2m}{k-m} \right)^m. \tag{54}$$

The above formula is simulated in Fig. 11 and will be compared to the next example with complex signature but the same cardinality.

*Example 21:* Binary/Quaternary System for $\mathcal{I} = \{\pm 1\}$, $\mathcal{S} = \{\pm 1, \pm j\}$

$$C(m,n,\mathcal{I},\mathcal{S}) \geq n - \log \sum_{k=0}^{\lfloor \frac{n}{2} \rfloor} \binom{n}{2k} \left( \frac{\binom{2k}{k}}{2^{2k}} \right)^{2m}. \tag{55}$$

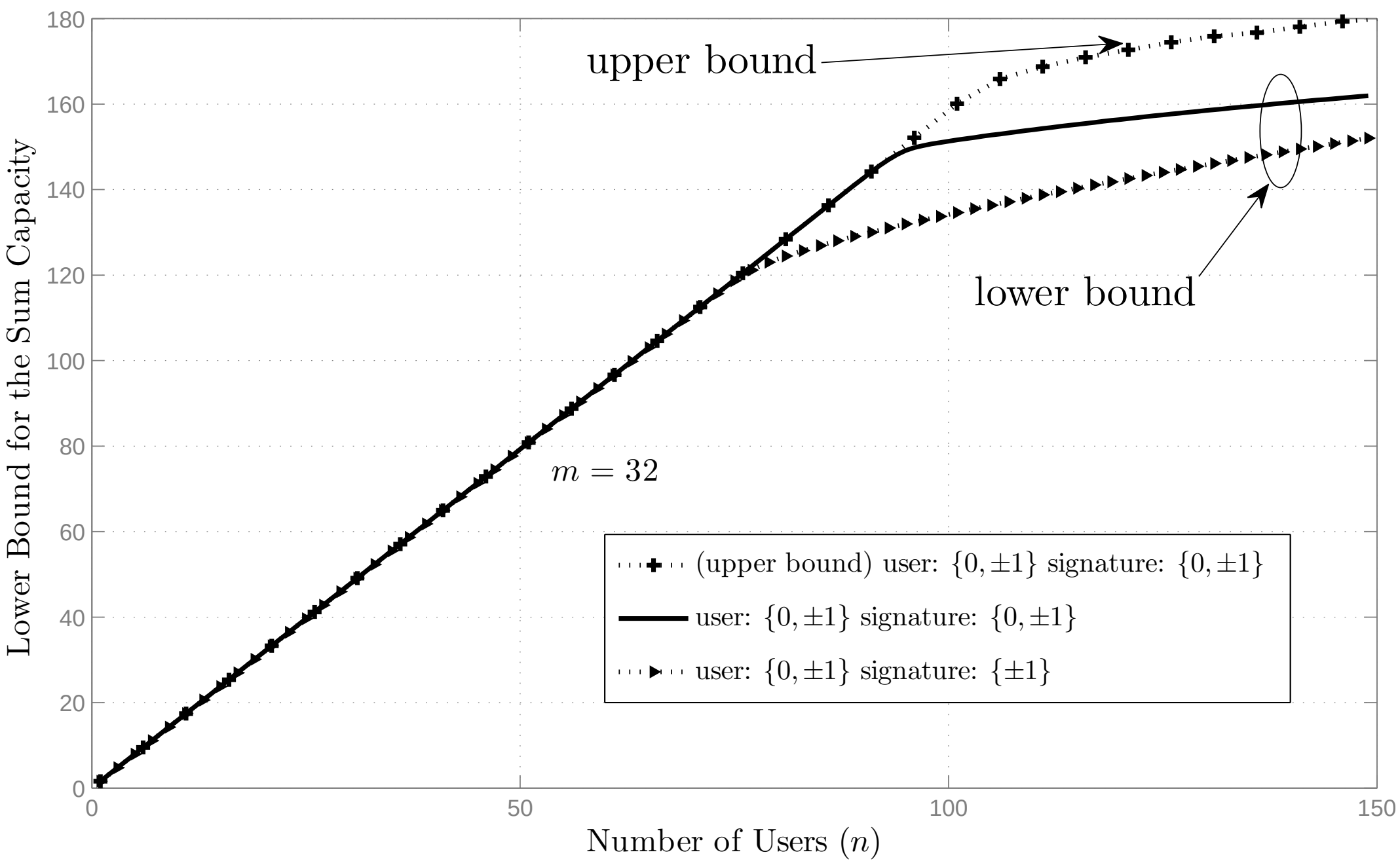

Fig. 10. The sum capacity lower and upper bounds versus the number of users for ternary input and binary/ternary signature matrices. The lower bound for the ternary signature matrix is better than that of the binary case.

Notice that the above formula is similar to (7). This implies that the lower bound for a binary/binary system with $2m$ chips is equivalent to the lower bound for binary/quaternary system with $m$ chips.
See Fig. 11 for the simulation of the above formula. The lower bound for the signature with AIN quaternary matrix is much better than the dependent case. This is again due to the absence of noise; the differences are less pronounced in the case of additive noise. On the same figure, the lower and upper bounds of binary/pentads are plotted. This simulation implies that our lower bounds are relatively tight.
In case of additive noise, the bounds are modified as shown in the following section:

## V. Sum Capacity lower Bound for General User Inputs and Matrices- the Noisy Case

Under additive noise scenario, the evaluation of $C\left(m, n, \mathcal{I}, \mathcal{S}, \eta\right)$ is an extremely difficult problem but we will derive a family of lower bounds for $C\left(m, n, \mathcal{I}, \mathcal{S}, \eta\right)$ for a class of random signature matrices. In the following general theorem, we let $\mathbf{A} = r\mathbf{B}$ as in (5) for a fixed value of $r$ and a randomly chosen $\mathbf{B}$ with distribution $\mathbb{P}^{\pi}$. In this case the power constraint (8) can be written in the following simpler form

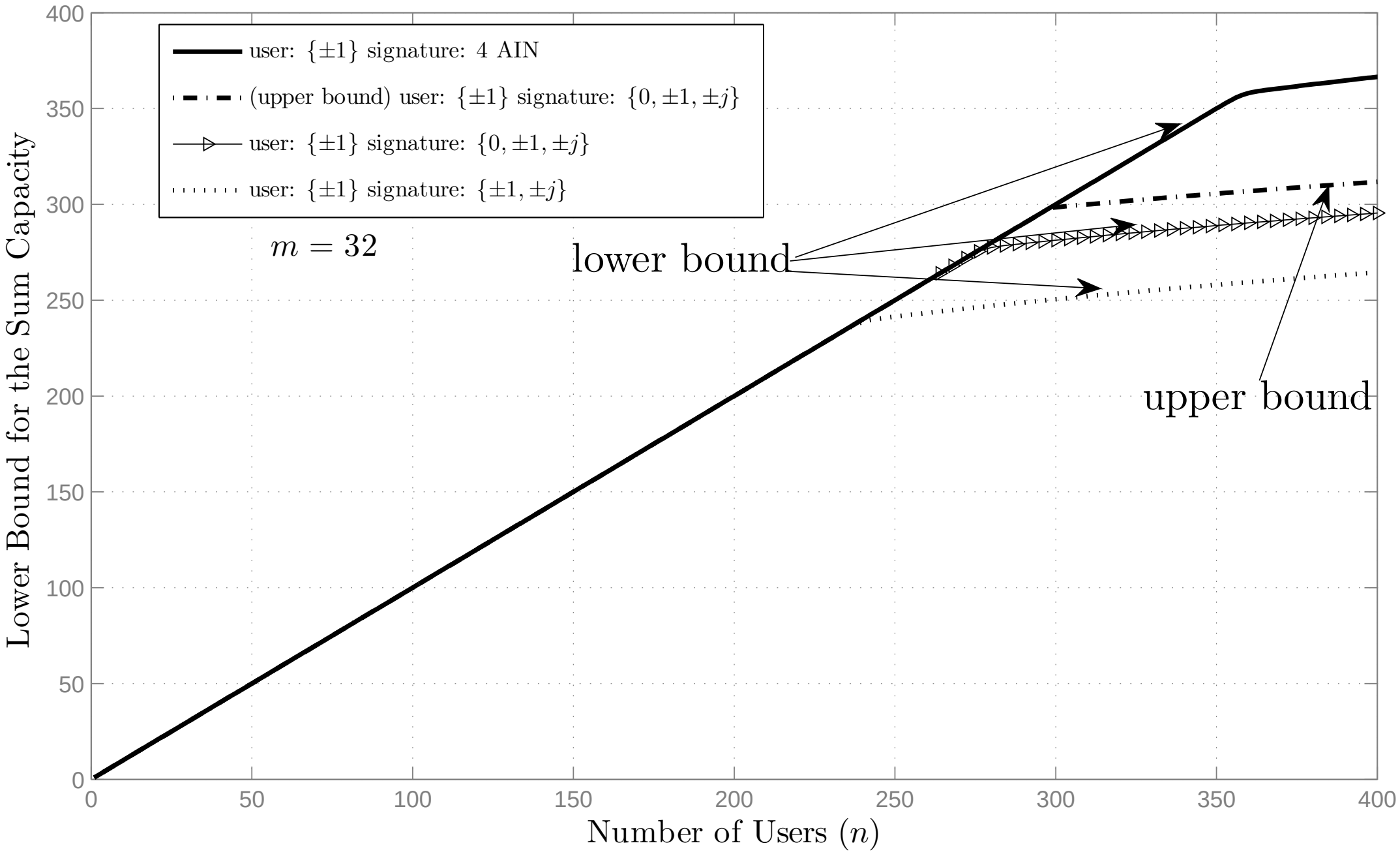

Fig. 11. The sum capacity lower and upper bounds versus the number of users for binary input and quaternary/pentad signature matrices.

after taking expectation over $\mathbb{P}^\pi$:

$$\frac{r^2}{m}\big(\sigma_p^2 nm\big(\sigma_\pi^2+\mu_\pi^2\big)+{\mu_p}^2 n^2 m\big(\sigma_\pi^2+\mu_\pi^2\big)\big)\leq n\eta\sigma_f^2 \tag{56}$$

or

$$r\leq\sqrt{\frac{\eta\sigma_f^2}{\big(\sigma_p^2+n\mu_p^2\big)\big(\sigma_\pi^2+\mu_\pi^2\big)}}. \tag{57}$$

### *A. Capacity Lower Bound*

The following theorem is the most general theorem for the sum capacity lower bound for any given input and signature matrix symbols with real/complex additive noise of arbitrary distribution:

*Theorem 21 (General Lower Bound for the Sum Capacity):*

$$C\left(m,n,\mathcal{I},\mathcal{S},\eta\right)\geq \sup_{\pi,p}\sup_{q}\left[-m\mathbb{E}\left(q\left(N_1\right)\right)-\log\mathbb{E}_{\tilde{X}}\left(\left(\mathbb{E}_{b,N_1}\left(2^{-q\left(N_1-r\frac{b^T\tilde{X}}{\sqrt{m}}\right)}\right)\right)^m\right)\right], \quad (58)$$

Where $p(\cdot)$ and $\pi(\cdot)$ are probability distributions on $\mathcal{I}$ and $\mathcal{S}$, respectively, and $q\left(\cdot\right)$ is any arbitrary function. Also $r$ satisfies the inequality given in (57), and $N_1$ is the first entry of the i.i.d. complex noise $N$ as defined in Subsection II-A. $b$ and $\tilde{X}$ are, respectively, vectors of length $n$ with i.i.d. entries of distribution $\pi(\cdot)$ and $\tilde{p}(\cdot)$ as explained in Definition 2 in Section IV.

For the proof, please refer to Appendix I.

In the important special case of Gaussian noise, Theorem 21 can be stated in a more explicit way by substituting $q\left(x\right)=\frac{\gamma}{2}|\frac{x}{\sigma}|^2\log e$ in (58); the result is given in the following corollary:

*Corollary 5 (General Lower Bound for the Sum Capacity for the Gaussian Noise):* For a given $\mathcal{I}$ and $\mathcal{S}$, when noise is complex Gaussian with independent real and imaginary parts of variance 1 (hence $\sigma_f^2=2$), (58) becomes

$$C\left(m,n,\mathcal{I},\mathcal{S},\eta\right)\geq \sup_{\pi,p}\sup_{\gamma}\{-m\left(\gamma\log e-\log\left(1+\gamma\right)\right)-\log\mathbb{E}_{\tilde{X}}\left(\left(\mathbb{E}_{b}\left(e^{\frac{-\gamma r^2}{2(1+\gamma)m}|b^T\tilde{X}|^2}\right)\right)^m\right)\}, \quad (59)$$

where $r=\sqrt{\frac{2\eta}{\left(\sigma_p^2+n\mu_p^2\right)\left(\sigma_\pi^2+\mu_\pi^2\right)}}$, $b$ and $\tilde{X}$ are, respectively, vectors of length $n$ with i.i.d. entries of distributions $\pi(\cdot)$ and $\tilde{p}(\cdot)$.

*Remark 3:* In the above corollary, the additive noise is taken as complex Gaussian with independent real and imaginary parts with zero means and equal variances. If the CDMA system was real, the imaginary part of the noise would not affect the capacity bounds. This implies that the bounds would have been the same if the noise was real with half the total complex variance. Having said this, we should note that since we have used the supremum over a family of Jensen's inequalities, the complex results are not analytically the same as the real one although the plots are very similar when we use numerical simulations.

*Remark 4:* The general noiseless Theorem 17 can be derived from the general noisy Theorem 21 or Corollary 5 when the additive noise power becomes negligible.

In the next 2 sections, we will derive and simulate special cases when the alphabet sizes are either finite ($|\mathcal{I}|$ and $|\mathcal{S}| < \infty$) or infinite (real/complex).

## VI. Results for Finite User Inputs and Arbitrary Signature Matrices- the Noisy Case

In this section, we will discuss lower bounds, upper bounds, asymptotics, and simulation results.

### A. Capacity Lower Bounds

In this subsection, we will develop bounds for $(q, l)$-ary and $(q, Real/Complex)$ CDMA systems.

*1) Lower Bound for Finite User Inputs and Signature Matrices:*
For the finite user inputs and signature matrices, the general Theorem 21 reduces to the following two corollaries:

*Corollary 6 (General Lower Bound for the Sum Capacity for $(q, l)$-ary CDMA Systems):*
Let $\mathcal{I} = \{\imath_1, \ldots, \imath_q\}$, $\tilde{\mathcal{I}} = \{\tilde{\imath}_0, \tilde{\imath}_1, \ldots, \tilde{\imath}_{\tilde{q}}\}$, $\tilde{\imath}_0 = 0$ with distribution $\tilde{p}$ and $\mathcal{S} = \{s_1, \ldots, s_l\}$. If $\tilde{p}(\tilde{\imath}_i) = \tilde{p}_i$ and $\pi(s_i) = \pi_i$, (59) becomes

$$C(m, n, \mathcal{I}, \mathcal{S}, \eta) \geq \sup_{\tilde{p}(\cdot), \pi(\cdot), \gamma} \Bigg\{ -m\left(\gamma \log e - \log\left(1+\gamma\right)\right) - \log \sum_{k_0 + \cdots + k_{\tilde{q}} = n} \binom{n}{k_0, \ldots, k_{\tilde{q}}} \prod_{j=0}^{\tilde{q}} \tilde{p}_j^{k_j} \times \left( \sum_{\substack{\sum_{i=1}^{l} u_{ij} = k_j \\ 1 \leq j \leq \tilde{q}}} \prod_{i=1}^{\tilde{q}} \binom{k_i}{u_{1i}, \ldots, u_{li}} \prod_{t=1}^{l} \pi_t^{\sum_{\alpha=1}^{\tilde{q}} u_{t\alpha}} e^{\frac{-\gamma r^2}{2(1+\gamma)m} \left| \sum_{i=1}^{l} s_i \left( \sum_{j=1}^{\tilde{q}} u_{ij} \tilde{\imath}_j \right) \right|^2} \right)^m \Bigg\}, \tag{60}$$

where $r = \sqrt{\frac{2\eta}{\left(\sigma_p^2 + n\mu_p^2\right)\left(\sigma_\pi^2 + \mu_\pi^2\right)}}$.

For the proof see Appendix J.

*Corollary 7 (Results for Finite User Inputs and Symmetric Symbols for Signature Matrices):*
Let $\mathcal{I} = \{\imath_1, \ldots, \imath_q\}$, $\tilde{\mathcal{I}} = \{\tilde{\imath}_0, \tilde{\imath}_1, \ldots, \tilde{\imath}_{\tilde{q}}\}$, $\tilde{\imath}_0 = 0$ with distribution $\tilde{p}(\cdot)$ and $\mathcal{S} = \{\pm s_1, \ldots, \pm s_{l'}\}$.

If $\tilde{p}(\tilde{\imath}_i)=\tilde{p}_i$ and $\pi(s_i)=\pi(-s_i)=\pi_i$, (59) becomes

$$C(m,n,\mathcal{I},\mathcal{S},\eta)\geq \sup_{\tilde{p}(\cdot),\pi(\cdot),\gamma}\{-m\left(\gamma\log e-\log\left(1+\gamma\right)\right)-\log\sum_{k_0+\cdots+k_{\tilde{q}}=n}\binom{n}{k_0,\ldots,k_{\tilde{q}}}2^{n-k_0}\prod_{j=0}^{\tilde{q}}\tilde{p}_j^{k_j}$$
$$\times\Bigg(\sum_{\substack{\sum_{i=1}^{l'}u_{ij}+v_{ij}=k_j\\ 1\leq j\leq\tilde{q}}}\prod_{i=1}^{\tilde{q}}\binom{k_i}{u_{1i},\ldots,u_{l'i},v_{1i},\ldots,v_{l'i}}$$
$$\prod_{t=1}^{l'}\pi_t{}^{\sum_{\alpha=1}^{\tilde{q}}u_{t\alpha}+v_{t\alpha}}e^{\frac{-\gamma r^2}{2(1+\gamma)m}|\sum_{i=1}^{l'}s_i(\sum_{j=1}^{\tilde{q}}(u_{ij}-v_{ij})\tilde{\imath}_j)|^2}\Bigg)^m\},\quad(61)$$

where $r=\sqrt{\frac{2\eta}{(\sigma_p^2+n\mu_p^2)\sigma_\pi^2}}$.

Also when $s_{l'}=0$, we have

$$C(m,n,\mathcal{I},\mathcal{S},\eta)\geq \sup_{\tilde{p}(\cdot),\pi(\cdot),\gamma}\{-m\left(\gamma\log e-\log\left(1+\gamma\right)\right)-\log\sum_{k_0+\cdots+k_{\tilde{q}}=n}\binom{n}{k_0,\ldots,k_{\tilde{q}}}2^{n-k_0}\prod_{j=0}^{\tilde{q}}\tilde{p}_j^{k_j}$$
$$\times\Bigg(\sum_{\substack{u_{l'j}+\sum_{i=1}^{l'-1}u_{ij}+v_{ij}=k_j\\ 1\leq j\leq\tilde{q}}}\prod_{i=1}^{\tilde{q}}\binom{k_i}{u_{1i},\ldots,u_{l'i},v_{1i},\ldots,v_{(l'-1)i}}$$
$$\times\prod_{t=1}^{l'-1}\pi_t{}^{\sum_{\alpha=1}^{\tilde{q}}u_{t\alpha}+v_{t\alpha}}\pi_{l'}{}^{\sum_{\alpha=1}^{\tilde{q}}u_{l'\alpha}}e^{\frac{-\gamma r^2}{2(1+\gamma)m}|\sum_{i=1}^{l'-1}s_i(\sum_{j=1}^{\tilde{q}}(u_{ij}-v_{ij})\tilde{\imath}_j)|^2}\Bigg)^m\}.\quad(62)$$

Refer to Appendix K for the proof.

*2) Lower Bound for Finite Input and Real/Complex CDMA Systems:*

In this subsection, we consider the case where the user data are finite and the signature matrix can have real or complex entries.

*Theorem 22 (Lower Bound for ($q$-ary, Real/Complex) CDMA Systems):*

For signature matrices with real or complex entries, let $\tilde{I}=\{\tilde{\imath}_0,\pm\tilde{\imath}_1,\ldots,\pm\tilde{\imath}_{\tilde{q}'}\}$, where $\tilde{\imath}_0=0$ , $\tilde{p}(0)=\tilde{p}_0$, and $\tilde{p}(+\tilde{\imath}_i) = \tilde{p}(-\tilde{\imath}_i) = \tilde{p}_i$.

$$C\left(m, n, \mathcal{I}, \mathbb{R}, \eta\right) \geq \sup_{\tilde{p}(\cdot),\gamma} \{-m\left(\gamma \log e - \log\left(1+\gamma\right)\right) - \log \sum_{\sum_{i=0}^{\tilde{q}'} k_i = n} \binom{n}{k_0, \ldots, k_{\tilde{q}'}} 2^{n-k_0} \prod_{j=0}^{\tilde{q}'} \tilde{p}_j^{k_j} \left(1 + \frac{2\gamma\eta \sum_{j=0}^{\tilde{q}'} k_j |\tilde{\imath}_j|^2}{m\left(1+\gamma\right)\left(\sigma_p^2 + n\mu_p^2\right)}\right)^{-\frac{m}{2}} \}. \tag{63}$$

For signature matrices with complex entries,

$$C\left(m, n, \mathcal{I}, \mathbb{C}, \eta\right) \geq \sup_{\tilde{p}(\cdot),\gamma} \{-m\left(\gamma \log e - \log\left(1+\gamma\right)\right) - \log \sum_{\sum_{i=0}^{\tilde{q}'} k_i = n} \binom{n}{k_0, \ldots, k_{\tilde{q}'}} 2^{n-k_0} \prod_{j=0}^{\tilde{q}'} \tilde{p}_j^{k_j} \left(1 + \frac{\gamma\eta \sum_{j=0}^{\tilde{q}'} k_j |\tilde{\imath}_j|^2}{m\left(1+\gamma\right)\left(\sigma_p^2 + n\mu_p^2\right)}\right)^{-m} \}. \tag{64}$$

See Appendix L for the proof.

Special cases of Corollaries 6, 7, and Theorem 22 when the user inputs are either binary or ternary are given below:

*Example 22 (Lower Bound for the Sum Capacity for Binary/ $l$-ary System):*
For $\mathcal{I} = \{\pm 1\}$ and $\mathcal{S} = \{s_1, \ldots, s_l\}$, if $p(1) = p_1$ and $\pi(s_i) = \pi_i$, (60) becomes

$$C(m, n, \mathcal{I}, \mathcal{S}, \eta) \geq \sup_{p_1, \pi(\cdot), \gamma} \{-m\gamma \log e - m \log\left(1+\gamma\right) - \log \sum_{k_1+k_2+k_3=n} \binom{n}{k_1, k_2, k_3} \left(p_1\left(1-p_1\right)\right)^{k_1+k_2} \left(p_1^2 + \left(1-p_1\right)^2\right)^{k_3} \times \left( \sum_{\substack{\sum_{i=1}^{l} u_i = k_1 \\ \sum_{i=1}^{l} v_i = k_2}} \binom{k_1}{u_1, \ldots, u_l} \binom{k_2}{v_1, \ldots, v_l} \prod_{i=1}^{l} {\pi_i}^{u_i+v_i} e^{\frac{-\gamma r^2}{2(1+\gamma)m} |2(\sum_{j=1}^{l}(u_j - v_j)s_j)|^2} \right)^m \}, \tag{65}$$

where $r = \sqrt{\frac{2\eta}{\left(\sigma_p^2 + n\mu_p^2\right)\left(\sigma_\pi^2 + \mu_\pi^2\right)}}$.

A special case of Corollary 7 when the user inputs are binary and the signature matrix is symmetric $2l'$-ary is given below:

*Example 23 (Lower Bound for the Sum Capacity for Binary/Symmetric $2l'$-ary System):* For $\mathcal{I} = \{\pm 1\}$ and $\mathcal{S} = \{\pm s_1, \ldots, \pm s_{l'}\}$ and if $\pi(+s_i) = \pi(-s_i) = \pi_i$ and $p(1) = \frac{1}{2}$, (61) becomes

$$C(m,n,\mathcal{I},\mathcal{S},\eta) \geq \sup_{\pi(\cdot),\gamma} \{-m\left(\gamma \log e - \log\left(1+\gamma\right)\right) - \log \sum_{k=0}^{n} \frac{\binom{n}{k}}{2^n}$$

$$\times \left( \sum_{\sum_{i=1}^{2l'} u_i = k} \binom{k}{u_1, \ldots, u_{2l'}} \prod_{i=1}^{l'} \pi_i^{u_i + u_{l'+i}} e^{\frac{-\gamma r^2}{2(1+\gamma)m} |2(\sum_{t=1}^{l'} (u_t - u_{t+l'}) s_t)|^2} \right)^m \}, \tag{66}$$

where $r = \sqrt{\frac{2\eta}{\sigma_\pi^2}}$. Also when $s_{l'} = 0$, the bound becomes

$$C(m,n,\mathcal{I},\mathcal{S},\eta) \geq \sup_{\pi(\cdot),\gamma} \{-m\left(\gamma \log e - \log\left(1+\gamma\right)\right) - \log \sum_{k=0}^{n} \frac{\binom{n}{k}}{2^n}$$

$$\times \left( \sum_{u_0 + \sum_{i=1}^{2(l'-1)} u_i = k} \binom{k}{u_0, \ldots, u_{2(l'-1)}} \prod_{i=1}^{l'-1} \pi_i^{u_i + u_{l'+i}} \pi_0^{u_0} e^{\frac{-\gamma r^2}{2(1+\gamma)m} |2(\sum_{t=1}^{l'-1} (u_t - u_{t+l'}) s_t)|^2} \right)^m \}. \tag{67}$$

A special case of Corollary 7 when the user inputs are ternary and the signature matrix has symmetric $2l'$-ary symbols is given below:

*Example 24 (Lower Bound for the Sum Capacity for Ternary/Symmetric $2l'$-ary Systems):* For $\mathcal{I} = \{0, \pm 1\}$ and $\mathcal{S} = \{\pm s_1, \ldots, \pm s_{l'}\}$ and if $\pi(+s_i) = \pi(-s_i) = \pi_i$ and $p(0) = p_0, p(+1) = p(-1) = p_1$ , (61) becomes

$$C(m,n,\mathcal{I},\mathcal{S},\eta) \geq \sup_{p_1,\pi(\cdot),\gamma} \{-m\left(\gamma \log e - \log\left(1+\gamma\right)\right)$$

$$- \log \sum_{k_1+k_2+k_3=n} \binom{n}{k_1, k_2, k_3} 2^{k_1+k_2} p_1^{2k_1} (2p_0 p_1)^{k_2} (2p_1^2 + p_0^2)^{k_3}$$

$$\left( \sum_{\substack{\sum_{i=1}^{2l'} u_i = k_1 \\ \sum_{i=1}^{2l'} v_i = k_2}} \binom{k_1}{u_1, \ldots, u_{2l'}} \binom{k_2}{v_1, \ldots, v_{2l'}} \right.$$

$$\left. \prod_{i=1}^{l'} \pi_i^{u_i + v_i + u_{l'+i} + v_{l'+i}} e^{\frac{-\gamma r^2}{2(1+\gamma)m} |2(\sum_{t=1}^{l'} (2(u_t - u_{l'+t}) + v_t - v_{l'+t}) s_t)|^2} \right)^m \}, \tag{68}$$

where $r=\sqrt{\frac{\eta}{p_1\sigma_\pi^2}}$. Also when $s_{l'}=0$, the bound becomes

$$C(m,n,\mathcal{I},\mathcal{S},\eta)\geq\sup_{p_1,\pi(\cdot),\gamma}\{-m\left(\gamma\log e-\log\left(1+\gamma\right)\right)$$
$$-\log\sum_{k_1+k_2+k_3=n}\binom{n}{k_1,k_2,k_3}2^{k_1+k_2}{p_1}^{2k_1}\left(2p_0p_1\right)^{k_2}\left(2p_1^2+p_0^2\right)^{k_3}$$
$$\times\left(\sum_{\substack{u_0+\sum_{i=1}^{2(l'-1)}u_i=k_1\\ v_0+\sum_{i=1}^{2(l'-1)}v_i=k_2}}\binom{k_1}{u_0,\ldots,u_{2(l'-1)}}\binom{k_2}{v_0,\ldots,v_{2(l'-1)}}\right.$$
$$\left.\times\prod_{i=1}^{l'-1}{\pi_i}^{u_i+v_i+u_{l'-1+i}+v_{l'-1+i}}{\pi_0}^{u_0+v_0}e^{\frac{-\gamma r^2}{2(1+\gamma)m}|2(\sum_{t=1}^{l'-1}(2(u_t-u_{l'-1+t})+v_t-v_{l'-1+t})s_t)|^2}\right)^m\}. \tag{69}$$

A special case of Theorem 22 when the user inputs are binary and the signature matrix is real is given below:

*Example 25 (Lower Bound for the Sum Capacity for Binary/Real Systems):*
For $\mathcal{I}=\{\pm1\}$ and $\mathcal{S}=\mathbb{R}$ and if $p(1)=\frac{1}{2}$, (63) becomes

$$C\left(m,n,\{\pm1\},\mathbb{R},\eta\right)\geq\sup_{\gamma}\left(-m\left(\gamma\log e-\log\left(1+\gamma\right)\right)-\log\left(\sum_{k=0}^{n}\frac{\binom{n}{k}}{2^n}\left((1+\frac{8k\gamma\eta}{m})\right)^{-\frac{m}{2}}\right)\right). \tag{70}$$

## B. Capacity Upper Bound

Below a general conjectured upper bound is given:

*Theorem 23 (A General Upper Bound for the Noisy Case):*
If $\mathcal{I}=\{\imath_1,\ldots,\imath_q\}$ with distribution $p(\imath_i)=p_i$ and $\mathcal{S}=\{s_1,\ldots,s_l\}\subset S^1$ (unit circle), we conjecture that the capacity is achieved when the user distributions are i.i.d. We thus have the following conjectured upper bound:

$$C\left(m,n,\mathcal{I},\mathcal{S}\right)\leq\max_{\substack{\sum_{i=1}^{l}u_i=n\\ p(\cdot)}}\left\{\min\left(n\mathbb{H}(\mathcal{I}),m\left(\mathbb{H}(\tilde{f})-\log(\frac{\pi e}{\eta})\right)\right)\right\}, \tag{71}$$

in which

$$\tilde{f}(z)=\sum_{\substack{\sum_{j=1}^{q}v_{ij}=u_i\\ 1\leq i\leq l}}\left(\prod_{k=1}^{l}\binom{u_k}{v_{k1},\ldots v_{kq}}\right)\left(\prod_{k=1}^{q}{p_k}^{\sum_{\alpha=1}^{l}v_{\alpha k}}\right)f(z-\frac{1}{\sqrt{m}}(\sum_{k=1}^{l}s_k\sum_{\alpha=1}^{q}v_{k\alpha}\imath_\alpha)), \tag{72}$$

and

$$f(x+\sqrt{-1}y)=\frac{\eta}{\pi}e^{-\eta(x^2+y^2)} \qquad (73)$$

Also, when $s_i = e^{\frac{2\pi i}{l}\sqrt{-1}}$ and $l$ divides $n$, we conjecture that $u_1 = u_2 = \cdots = u_l = \frac{n}{l}$. For the proof see Appendix M.

### *C. Simulation Results*

*1) Binary/Binary:* A special case of Example 23 is simulated in Fig.12. The simulation is for $(2,2)$-ary and when the spreading gain is $m = 16$. In the simulations, the input alphabets are fixed to $\pm 1$. Also simulation results show that when $\mathcal{S} = \{1, e^{j0.9\pi}\}$, the lower bound is maximum. This simulation shows that the choice of symbols for the input and the signature matrix affects the sum capacity. The best lower bound is the signature with $2$ AIN's.

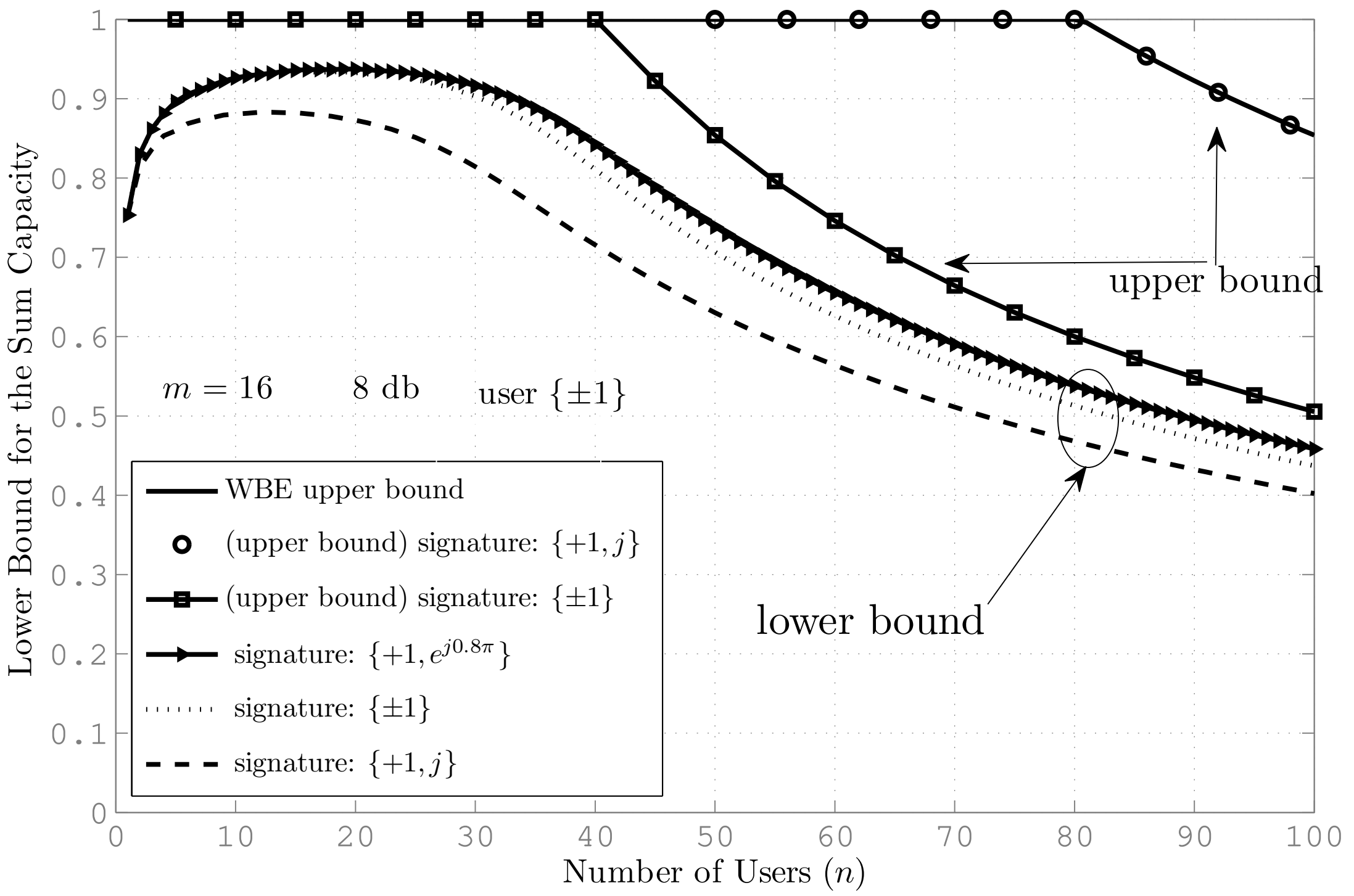

Fig. 12. The normalized sum capacity upper and lower bounds versus the number of users for binary input and binary signature matrix when all the probabilities are equal to $1/2$ for $m = 16$ and $\eta = 8$ dB.

*2) Binary/Ternary:* The simulation results of Examples 22 and 23 are shown in Fig.13 for the special case of $(2,3)$-ary when the number of chips is equal to $4$, $\eta = 8$ dB, and the binary input is equal to $\pm 1$. This figure shows a comparison of 3 different signature symbols; this simulation implies that the choice

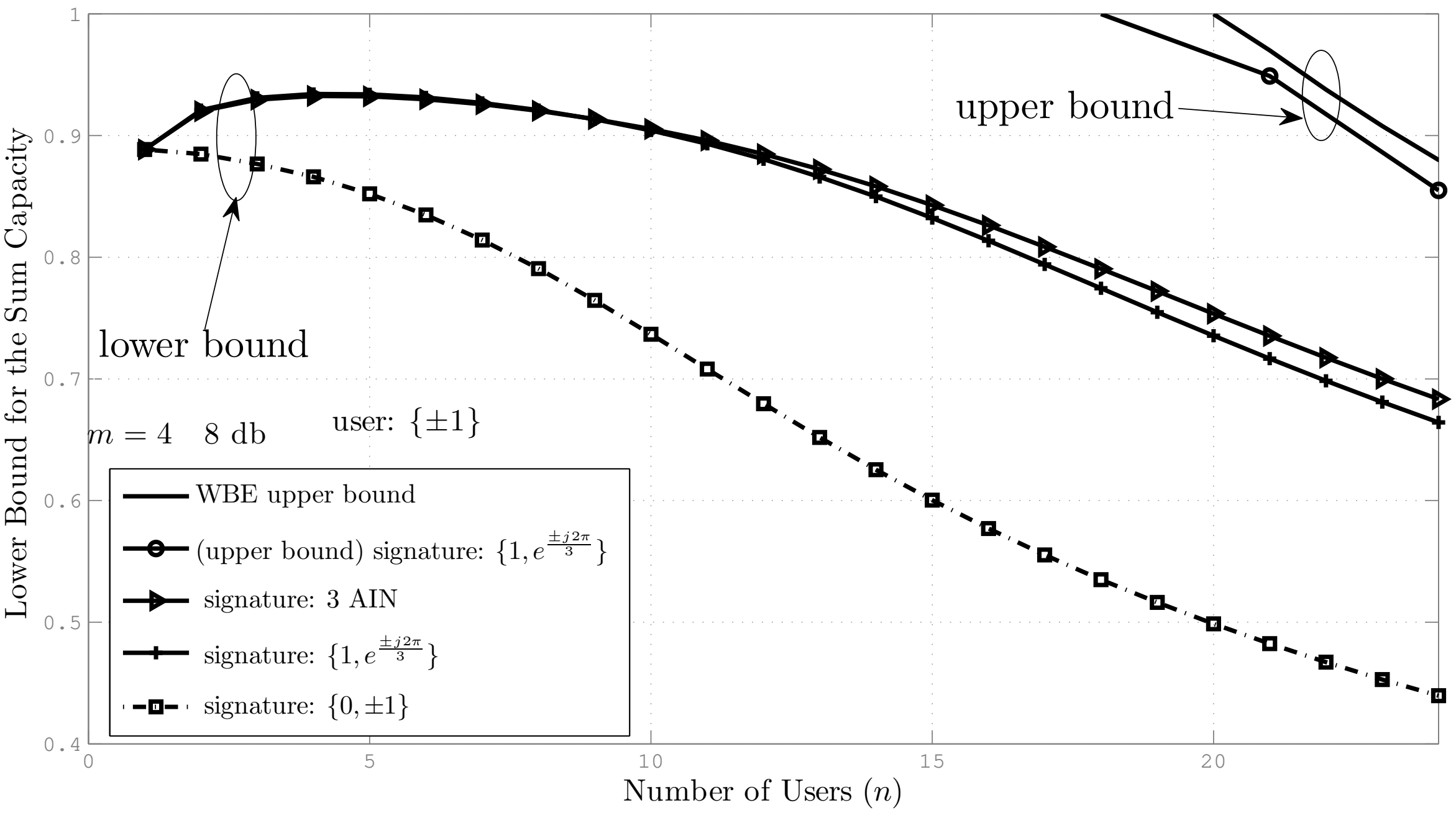

Fig. 13. The normalized sum capacity upper and lower bounds versus the number of users for $(2,3)$-ary when $m = 4$ and $\eta = 8$ dB.

of $\{0, \pm 1\}$ for the signature matrix is not as good as far as the capacity is concerned.

*3) Ternary/Ternary:* The simulation results of Examples 23 and 24 are shown in Fig.14 for the special cases of $(3,3)$-ary, $(2,3)$-ary and $(3,2)$-ary. The input alphabet symbols are $\{0, \pm 1\}$ for ternary and $\{\pm 1\}$ for binary users. The signature symbol is ternary $\{0, \pm 1\}$. As expected, the ternary/ternary system performs the best. For the quaternary matrix, the following example is instructive.

*4) Binary/Quaternary:* The simulation results of Examples 23 and 24 are shown in Fig.15 for two special cases of $(2,4)$-ary systems. The user input alphabet symbols are binary $\{\pm 1\}$. The signature symbols are either $\{\pm 1, \pm j\}$ or $\{1, e^{\frac{j\pi}{2.4}}, e^{\frac{2j\pi}{2.4}}, e^{\frac{3j\pi}{2.4}}\}$ that are AIN. This figure shows that the AIN case outperforms the "natural" quaternary signature matrix.

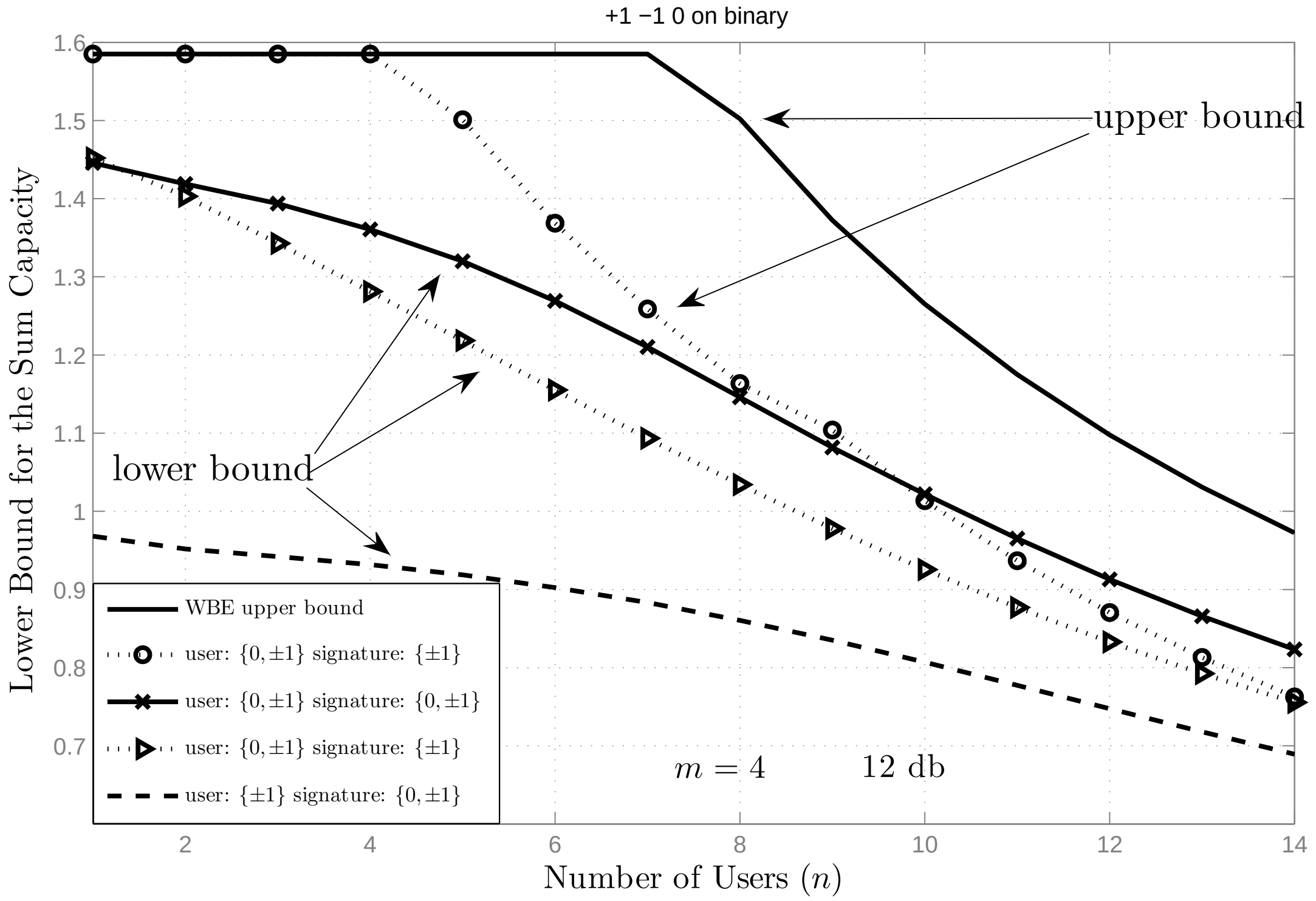

Fig. 14. The normalized sum capacity upper and lower bounds versus the number of users for ternary/ternary, binary/ternary and ternary/binary systems for $m = 4$, $\eta = 12$ dB.

*5) Binary/Arbitrary Matrix:* When the user inputs are binary, we have simulated the bounds given in Examples 22-25 for binary, ternary and complex quaternary signature matrices as depicted in Fig. 16; the plots are for $m = 16$ and $\eta = 20\ dB$. This figure shows that by increasing the cardinality of $\mathcal{S}$, the lower bound is improved. This is a significant result since by adding extra complexity at the transmitter and receiver sides, we can increase the capacity. On the other hand, our asymptotic results as well as those of [12], [40] imply that the sum capacity is independent of the signature alphabet.

### *D. Asymptotic Lower Bound for the Sum Capacity for Additive Gaussian Noise*

In this subsection, we consider the problem of estimating the capacity per user ($c = \frac{1}{n}C$) in the limit when the number of users $n$ and the chip rate $m$ go to infinity while the overloading factor $\beta = n/m$ and the normalized SNR $\eta$ are kept constant. We first prove an asymptotic formula for the expression appearing in the right-hand-side of Corollary 5.

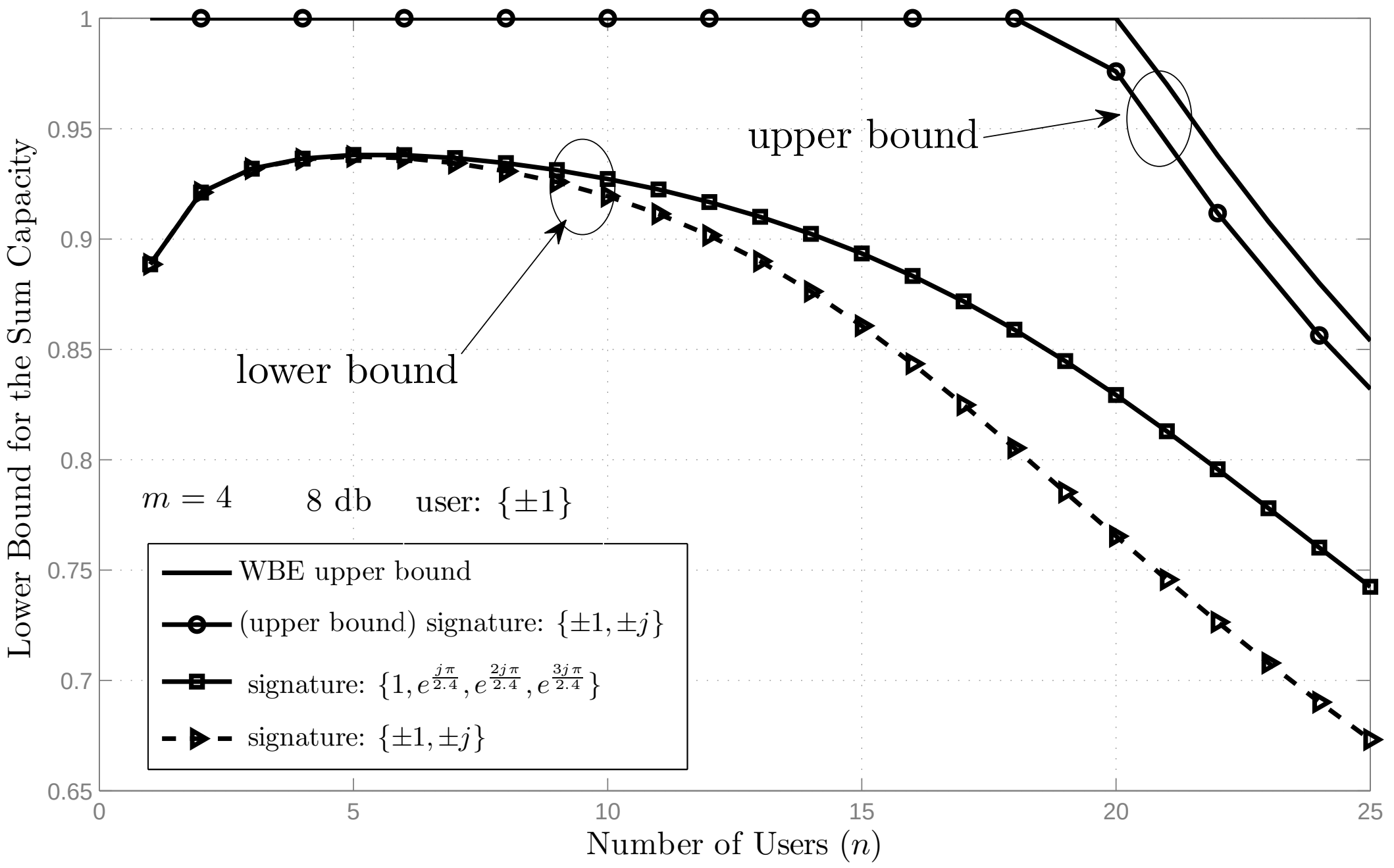

Fig. 15. The normalized sum capacity upper and lower bounds versus the number of users for binary/quaternary system for $m = 4$ and $\eta = 8$ dB.

*Theorem 24 (Asymptotic Lower Bound for Finite Input and Additive Gaussian Noise):* For given sets $\mathcal{S}$ and $\mathcal{I}$ (with $|\mathcal{I}| < \infty$), and the corresponding probability distributions $\pi(\cdot)$ and $p(\cdot)$ where $\mu_p = 0$, let $b$ and $\tilde{X}$ be vectors of length $n$ with i.i.d. entries of distributions $\pi(\cdot)$ and $\tilde{p}(\cdot)$, respectively. Then

$$\lim_{\substack{m,n\to\infty \\ n/m\to\beta}} \frac{1}{n}\left[-m\gamma\log e - \log \mathbb{E}_{\tilde{X}}\left(\left(\mathbb{E}_b\left(\frac{e^{\frac{-\gamma r^2}{2(1+\gamma)m}|b^T\tilde{X}|^2}}{1+\gamma}\right)\right)^m\right)\right]$$

$$= \sup_{\gamma}\left\{\inf_{\hat{p}(\cdot),\mu_\pi\mu_{\hat{P}}=0}\left\{\mathbb{D}\left(\hat{p}\|\tilde{p}\right) - \frac{1}{\beta}\left(\gamma\log e - \log\left(1+\gamma\right)\right)\right.\right.$$

$$\left.\left. +\frac{1}{2\beta}\left(\log\left(1+\frac{2\beta\eta\gamma\lambda_1}{(1+\gamma)\,\sigma_p^2(\sigma_\pi^2+\mu_\pi^2)}\right)+\log\left(1+\frac{2\beta\eta\gamma\lambda_2}{(1+\gamma)\,\sigma_p^2(\sigma_\pi^2+\mu_\pi^2)}\right)\right)\right\}\right\}, \tag{74}$$

where $\lambda_1$,$\lambda_2$ are eigenvalues of the covariance matrix of an r.v. which has the distribution of the product of two independent variables with distribution $\hat{p}$ and $\pi$, and $\mathbb{D}(\cdot\|\cdot)$ is the Kullback-Leibler distance. Notice that for a binary CDMA system, the above lower bound becomes identical to that of (25).
The proof is given in Appendix N.

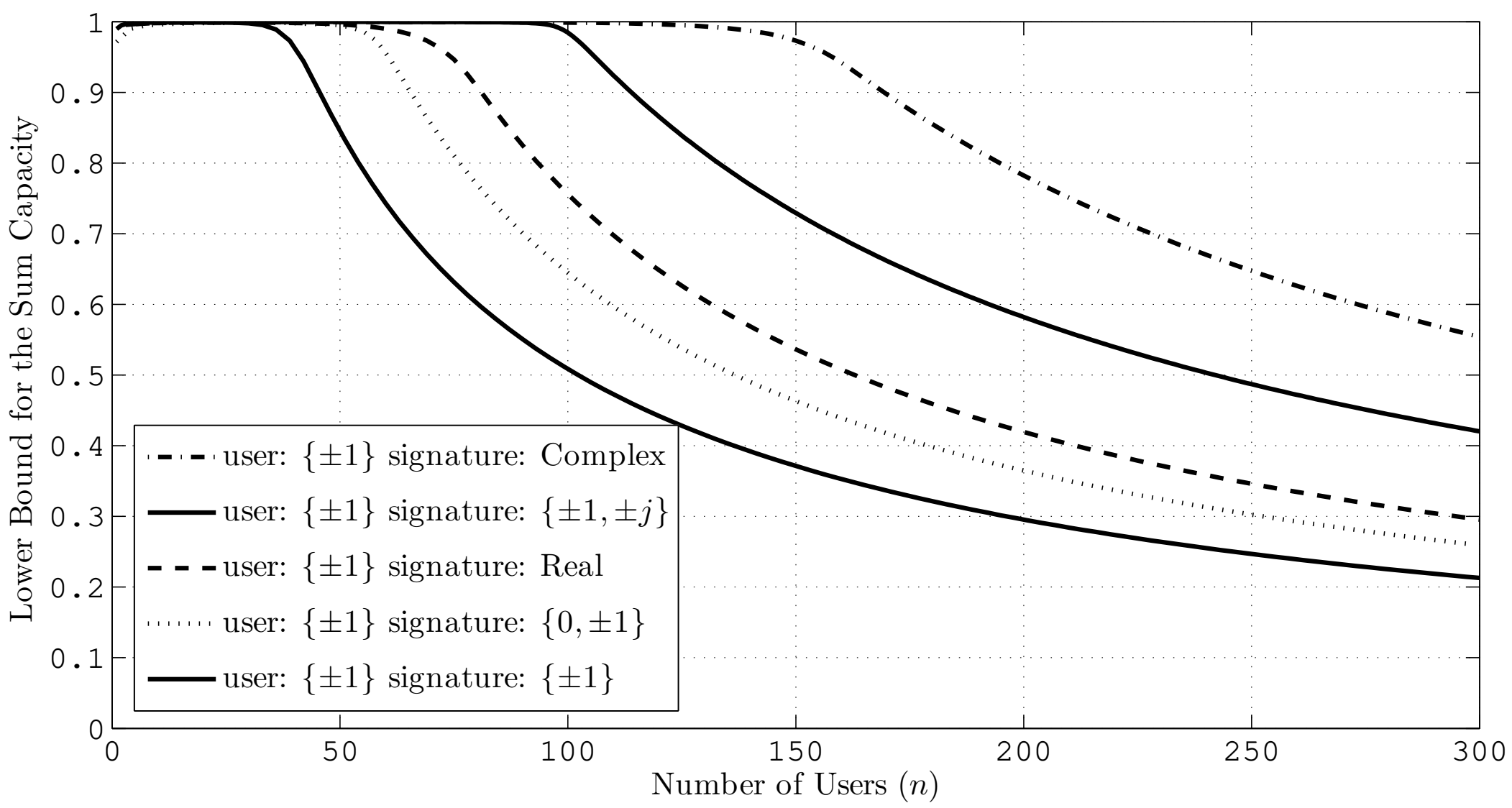

Fig. 16. The normalized lower bounds for the sum capacity versus the number of users for various signature alphabets. In this figure $\mathcal{I} = \{\pm 1\}$, $m = 16$ and $\eta = 20$ dB

This theorem implies that the asymptotic bound does not depend on the symbol sizes or types of the user inputs or the signature matrix but rather on the probabilities of $p(\cdot)$ and $\pi(\cdot)$. This result is plotted in Fig.17. This figure shows the normalized sum capacity bounds versus $\eta$ for binary, ternary, and quaternary inputs for $\beta = 1$ and $3$. Figure 18 is the simulation of several asymptotic results. This figure shows a comparison of the actual (soft decision) single user capacity for QPSK with the Guo-Verdu asymptotic result (9), and our asymptotic lower bound from Theorem 24 for $\mathcal{I} = \{\pm 1, \pm j\}$. For $\beta = 1$, we can use orthogonal Walsh codes and hence the sum capacity is equivalent to the single user QPSK. Clearly, the asymptotic average mutual information derived by Guo-Verdu [12] for joint decoding QPSK is a lower bound for the actual sum capacity; however, for the hard decision QPSK, the Guo-Verdu bound is slightly better. For $\beta = 3$, the Guo-Verdu bound is between our lower and WBE upper bounds. This phenomenon is similar to that of the Tanaka's and our bounds for the binary case [15].

In the next section, the previous theorems and corollaries are extended to the case when not all users are active.

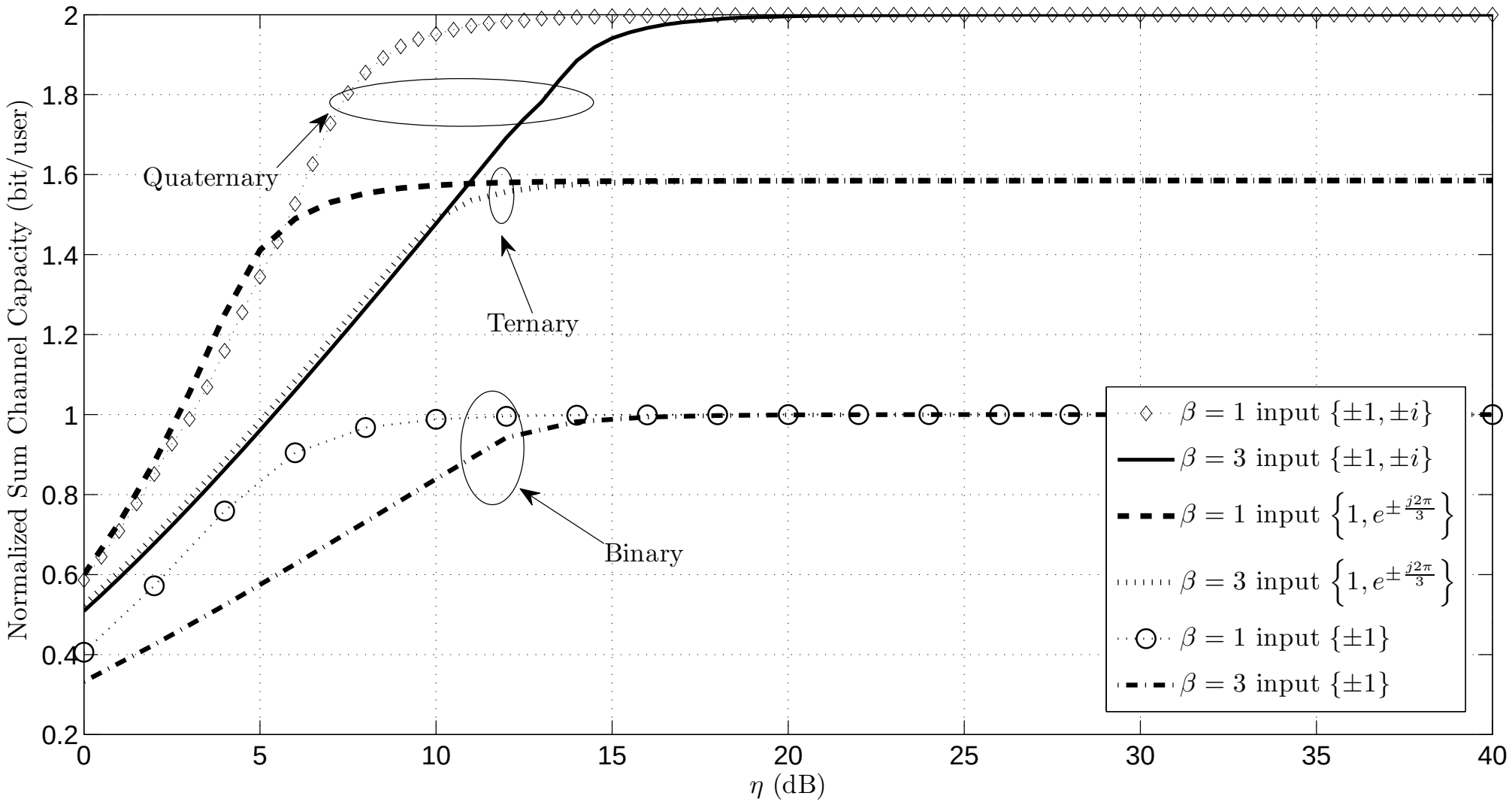

Fig. 17. The asymptotic lower bounds for the normalized sum capacity versus $\eta$ for binary, ternary, and quaternary inputs for $\beta = 1$ and 3. The matrix is irrelevant as far as the asymptotic results are concerned.

## VII. Capacity Bounds for Variable-Active Users

In Subsection III-E, we defined active users and overloaded codes for the binary case that were errorless (COWDA) with active user detection capability. In this section, we would like to find the sum capacity bounds for the variable-active users for AWGN channels. We consider two main scenarios: 1- $k$-active, and 2- $p_{act}$-active. The term "$k$-active users" implies that only $k$ out of $n$ users are active at any given time for a CDMA system (signature time length). The term $p_{act}$-active implies that each user is active with the probability $p_{act}$. This concept may also have applications to compressive sensing of sparse signals [46]; the signature $m \times n$ matrix becomes the sampling $m \times n$ matrix and the original signal is $k$-sparse. In the compressive sensing applications, typically, $2k < m < n$; this is is analogous to an overloaded CDMA system when only $k$-users are active. The following two subsections extend the result of Theorem 22 to the $k$- and $p_{act}$-active scenarios.

### *A. Capacity Lower Bounds for a $k$-Active CDMA System*

In this subsection, we will consider 3 scenarios: 1- Binary/Real, 2- Real/Real, and Binary/Binary CDMA systems

*1) Capacity Lower Bound for $k$-Active Binary/Real CDMA Systems:*

The following theorem extends the result of Theorem 22 to the $k$-active scenario:

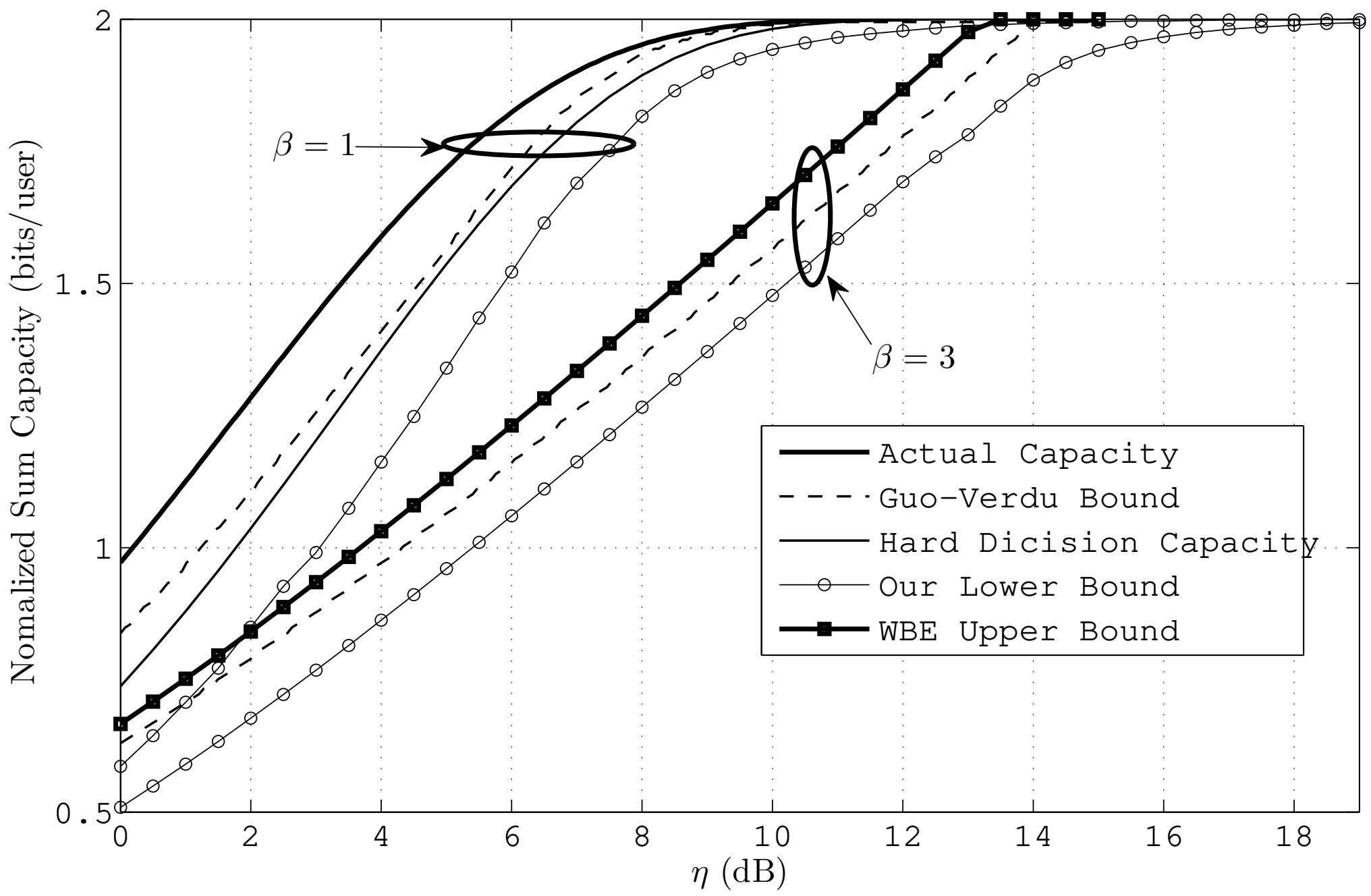

Fig. 18. The asymptotic actual, upper and lower bounds for the normalized sum capacity versus $\eta$ for QPSK inputs for $\beta = 1$ and 3. In this figure, our lower bound is compared to that of Guo-Verdu and the single user sum capacity.

*Theorem 25 (Capacity Lower Bound for $k$-Active CDMA Systems, $\mathcal{I} = \{\pm 1\}$, $\mathcal{S} = \mathbb{R}$):* If $X$ is a $k$-active binary input vector with i.i.d. entries and the distribution on $\mathcal{S}$ is $\mathcal{N}(0,1)$, then

$$C_{\text{active}}(m,n,k,\{\pm 1\},\mathbb{R},\eta) \geq \sup_{\gamma}\{-m(\gamma \log e - \log(1+\gamma)) - \log \frac{1}{2^k \binom{n}{k}^2} \sum_{i,j} 2^{i-2j} \binom{n}{j, k+j-i, i-2j, n-k} \left(1 + \frac{4in\gamma\eta}{km(\gamma+1)}\right)^{\frac{-m}{2}}\}. \tag{75}$$

The proof is given in Appendix O.1. The above lower bound is also simulated in Fig. 19 for $\eta = 20$dB, and will be discussed at the end of this subsection.

Below, we will discuss active users for real inputs and signature matrices:

*2) Capacity Lower Bound for $k$-Active Real/Real CDMA Systems:*

For the $k$-active user case when the probability distributions on $\mathcal{I} = \mathcal{S} = \mathbb{R}$ are $\mathcal{N}(0,1)$, we can find

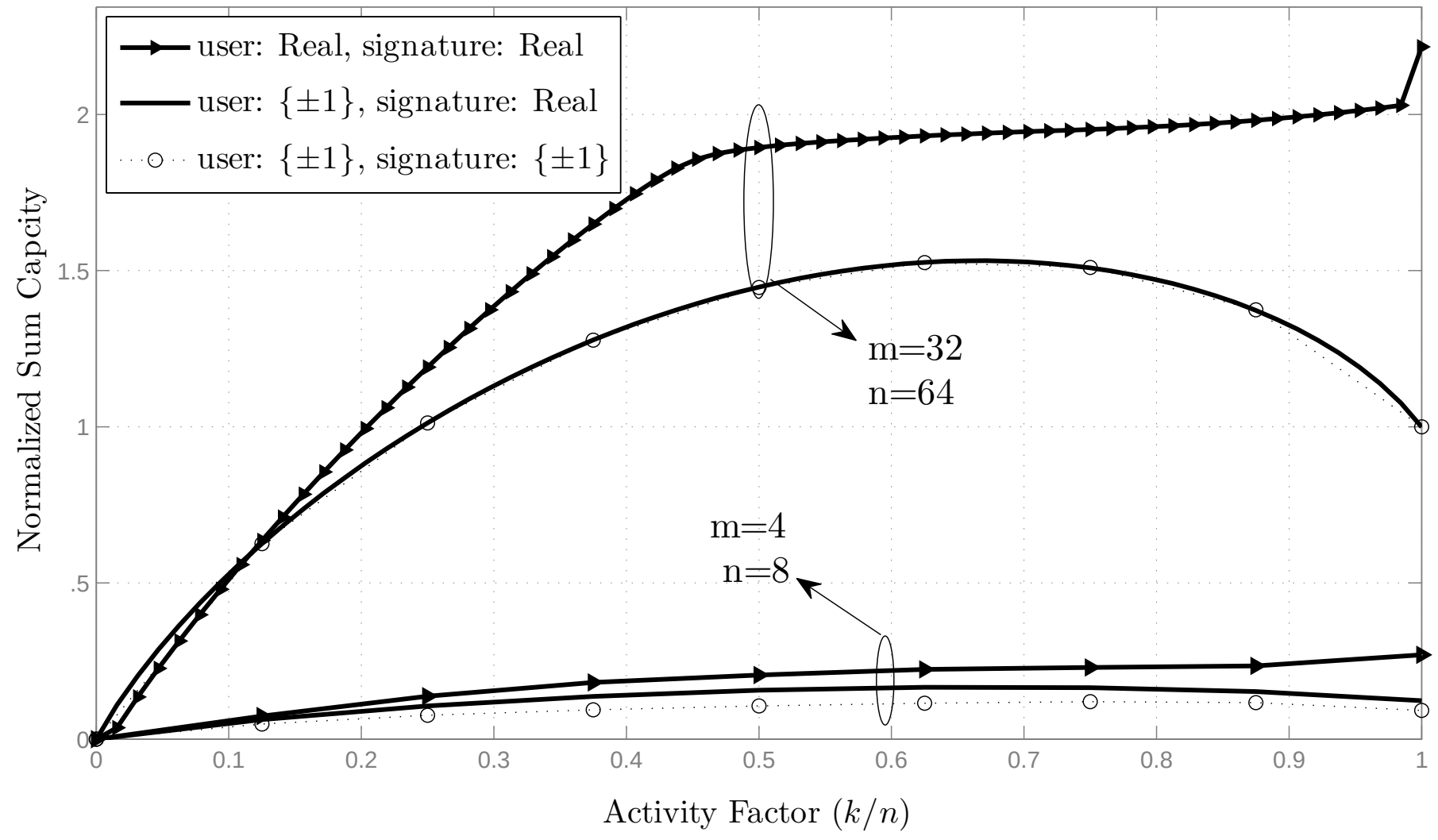

Fig. 19. The sum capacity lower bounds w.r.t. activity factor for various scenarios when $\eta = 20$dB.

the lower bound as shown below:

*Theorem 26 (Lower Bound for $k$-Active Real/Real CDMA Systems):*

$$C_{\text{active}}(m,n,k,\mathbb{R},\mathbb{R},\eta) \geq \sup_{\gamma}\{-m\left(\gamma \log e - \log\left(1+\gamma\right)\right) - \log \sum_{j=\max(0,2k-n)}^{k} \psi\left(j\right)\},$$

where

$$\psi\left(j\right) = \begin{cases} \frac{1}{2^k\Gamma(k)}\int_0^\infty \left(1+\frac{2\gamma x n\eta}{(1+\gamma)km}\right)^{\frac{-m}{2}} x^{k-1}e^{\frac{-x}{2}}dx & j=0 \\ \frac{\binom{k}{j}\binom{n-k}{k-j}}{\binom{n}{k}}\frac{1}{2^{k-j}\Gamma(k-j)2^{\frac{j}{2}}\Gamma\left(\frac{j}{2}\right)}\int_0^\infty\int_0^\infty \left(1+\frac{\gamma(x+2y)2n\eta}{(1+\gamma)km}\right)^{\frac{-m}{2}} x^{k-j-1}y^{\frac{j}{2}-1}e^{\frac{-x}{2}}e^{\frac{-y}{2}}dxdy & j>0 \\ \frac{1}{2^{\frac{k}{2}}\Gamma\left(\frac{k}{2}\right)}\int_0^\infty \left(1+\frac{4\gamma y n\eta}{(1+\gamma)km}\right)^{\frac{-m}{2}} y^{\frac{k}{2}-1}e^{\frac{-y}{2}}dy & j=k \end{cases}. \quad (76)$$

See Appendix O.2 for the proof.

The above lower bound is also simulated in Fig. 19 for $\eta = 20$dB; the explanation of this figure will be discussed in the next subsection. For the special case of real user inputs when all users are active ($k = n$), we have the actual sum capacity and therefore the lower sum capacity can be compared to the actual one. Simulation results depicted in Fig. 20 imply that the sum capacity lower bound is relatively tight and closely tracks the actual capacity. Also, on the same figure, the average capacity derived from (9) is plotted. This average is derived from randomly generating real signature matrices and applying (9) in the Preliminaries. Our lower bound is derived from this average value. The capacity lower bound for a binary/binary $k$-Active CDMA System is given below:

*3) Capacity Lower Bound for $k$-Active Binary/Binary CDMA Systems:*
When the binary users are inactive, the input is equivalent to a ternary $\{0, \pm 1\}$ system, where the probability of $0$ is equal to $k/n$. The lower bound is derived from the following theorem:

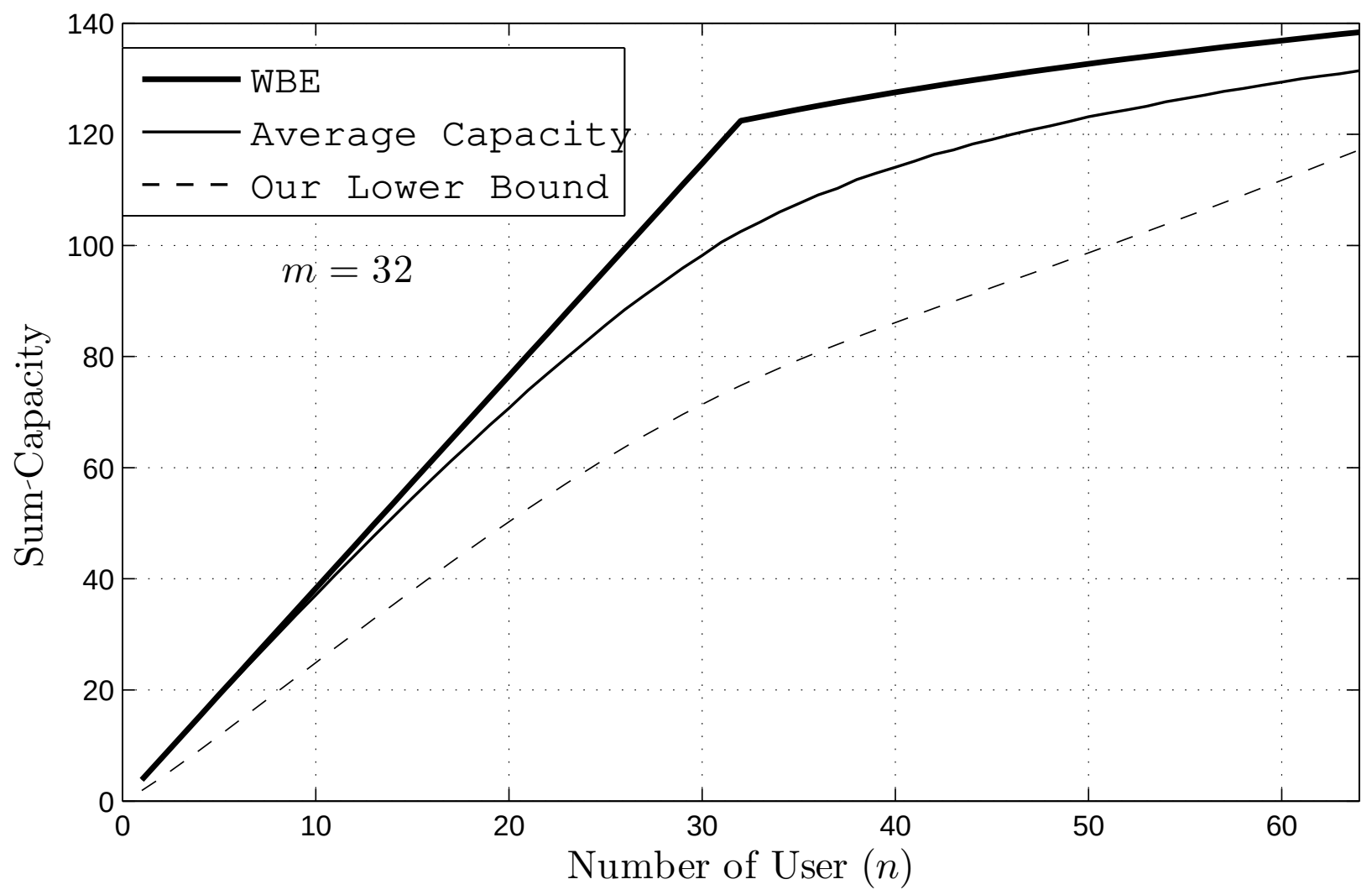

Fig. 20. The sum capacity lower bound is compared to the actual sum capacity for real valued inputs ($\eta = 20dB$). This plot highlights that our sum capacity lower bound tracks the average capacity.

*Theorem 27 (Capacity Lower Bound for $k$-Active CDMA Systems, $\mathcal{I} = \mathcal{S} = \{\pm 1\}$):*
If $X$ is a $k$-active vector with i.i.d. entries with binary distribution, then

$$C_{active}(m, n, k, \{\pm 1\}, \{\pm 1\}, \eta) \geq \sup_{\gamma}\{-m(\gamma \log e - \log(1+\gamma)) - \log(\mu(k, \eta, \gamma))\}, \tag{77}$$

where

$$\mu(k,\eta,\gamma) = \frac{1}{2^k\binom{n}{k}^2}\sum_{i,j}\binom{n}{i, j, \frac{j}{2}, k-i-\frac{j}{2}, n-k-j}2^{\frac{j}{2}}$$

$$\times\left(\frac{1}{2^{i+j}}\sum_{u,v}\binom{i}{u}\binom{j}{i+\frac{j+v}{2}-2u}e^{\frac{-2n\eta\gamma v^2}{mk(1+\gamma)}}\right)^m. \tag{78}$$

For the proof, refer to Appendix O.3. The above lower bounds are simulated in Fig. 19 with real values. This figure compares three scenarios, namely, real/real, binary/real and binary/binary for two different values of $m$ and $n$. For a large scaled system ($m = 32$ and $n = 64$), the binary/real and binary/binary CDMA systems coincide while for a small scaled system ($m = 4$, $n = 8$), the binary/binary system is somewhat less. These plots confirm that for large scaled systems, the lower bounds become independent of the signature alphabets.

### *B. Capacity Lower Bound When the Probability of Active Users is Known*

When the users are active with the probability $p_{act}$, the capacity lower bound can again be derived from Corollary 5. For large scale systems, the $k$-active CDMA case discussed in the previous subsections become equivalent to the case when $p_{act} = k/n$. In the following subsections, we will discuss various scenarios for the user inputs and signature matrices:

#### *1) Capacity Lower Bound for a Real/Real $p_{act}$-Active CDMA System:*

In this subsection, we shall derive the sum capacity lower bound for the $p_{act}$-active real inputs assuming the matrix is also real. We will also derive the asymptotic case for the large scale system.

*Theorem 28 (Active Users with Probability $p_{\text{act}}$, $\mathcal{I} = \mathcal{S} = \mathbb{R}$):*
For a CDMA system with real input, let $p_{\text{act}}$ be the probability that the users are active, then we have

$$C_{p_{act}}(m,n,\mathbb{R},\mathbb{R},\eta,p_{\text{act}}) \geq \sup_{\gamma}\{-m\left(\gamma\log e - \log\left(1+\gamma\right)\right)$$

$$-\log\sum_{k_1+k_2+k_3=n}\binom{n}{k_1,k_2,k_3}2^{k_2}p_{act}^{2k_1+k_2}\left(1-p_{act}\right)^{2(n-k_1)-k_2}\phi\left(k_1,k_2\right)\}, \tag{79}$$

where

$$\phi\left(k_1,k_2\right)=\begin{cases}1 & k_1=k_2=0\\ \frac{1}{2^{k_2/2}\Gamma(k_2/2)}\int_0^\infty\left(1+\frac{2\gamma\eta y}{(1+\gamma)mp_{act}}\right)^{-m/2}y^{k_2/2-1}e^{-y/2}\mathrm{d}y & k_1=0\\ \frac{1}{2^{k_1/2}\Gamma(k_1/2)}\int_0^\infty\left(1+\frac{4\gamma\eta x}{(1+\gamma)mp_{act}}\right)^{-m/2}x^{k_1/2-1}e^{-x/2}\mathrm{d}x & k_2=0\\ \frac{1}{2^{k_1/2}2^{k_2/2}\Gamma(k_1/2)\Gamma(k_2/2)} & \\ \times\int_0^\infty\int_0^\infty\left(1+\frac{2\gamma\eta(2x+y)}{(1+\gamma)mp_{act}}\right)^{-m/2}x^{k_1/2-1}y^{k_2/2-1}e^{-x/2}e^{-y/2}\mathrm{d}x\mathrm{d}y & k_1k_2\neq 0\end{cases}. \tag{80}$$

See Appendix O.4 for the proof. The above bound is simulated in Fig. 21 and will be discussed at the end of this section.

*2) Asymptotic Lower Bound for a Real/Real $p_{act}$-Active CDMA System:*

The asymptotic sum capacity for the above equations is given below:

*Theorem 29:* Asymptotic Lower Bound for a Real/Real $p_{act}$-Active CDMA System

$$\lim_{\substack{n,m\to\infty\\ \frac{n}{m}\to\beta}}\frac{1}{n}C_{p_{act}}\left(m,n,\mathbb{R},\mathbb{R},\eta,p_{\text{act}}\right)\geq-\frac{1}{\beta}\left(\gamma\log e-\log\left(1+\gamma\right)\right)-\sup_{x\geq 0}\{(F\left(x\right)-I\left(x\right)\}, \tag{81}$$

where

$$F\left(x\right)=\frac{-1}{2\beta}\log\left(1+\frac{2\gamma\eta\beta}{\left(1+\gamma\right)p_{\text{act}}}x\right) \tag{82}$$

and

$$I\left(x\right)=\sup_t\{\log e\times xt-\log\left(\left(1-p_{act}\right)^2+\frac{2p_{act}\left(1-p_{act}\right)}{\sqrt{1-2t}}+\frac{p_{act}^2}{\sqrt{1-4t}}\right)\} \tag{83}$$

For the proof see Appendix O.5.

*3) Capacity Lower Bound for a Binary/Real $p_{act}$-Active CDMA System:*

When the inputs are binary (bipolar) and the signature matrix is real, we have the following theorem for the $p_{act}$-active CDMA System:

*Theorem 30 (Binary Input with Real Signature):*

$$C_{p_{act}}(m,n,\{\pm1\},\mathbb{R},\eta,p_{\text{act}})\geq\sup_\gamma\{-m(\gamma\log e-\log\left(1+\gamma\right))-\log\sum_{k_1+k_2+k_3=n}2^{k_2-k_1}\binom{n}{k_1,k_2,k_3}$$
$$\times {p_{\text{act}}}^{2k_1+k_2}(1-p_{\text{act}})^{k_2}\left(1+\frac{3}{2}p_{\text{act}}^2-2p_{\text{act}}\right)^{k_3}\left(1+\frac{2\gamma\eta(4k_1+k_2)}{mp_{\text{act}}(\gamma+1)}\right)^{-\frac{m}{2}}\}. \tag{84}$$

This theorem can be derived from Theorem 22 when $\mathcal{I} = \{0, \pm 1\}$. The simulation of the above bound is shown in Fig. 21 and will be discussed in the next subsection. The asymptotic formula for the $p_{act}$ binary input case is the same as the general asymptotic derivation given in (74) when the input is ternary and the probability of 0 is $1 - p_{act}$. The simulations of the above formula along with the asymptotic bound derived from (74) are depicted in Fig. 22. This figure shows that for even small values of $m = 16$, the asymptotic bounds and the finite bounds derived from the above evaluation for $\beta = 1, 2, 3$ coincide.

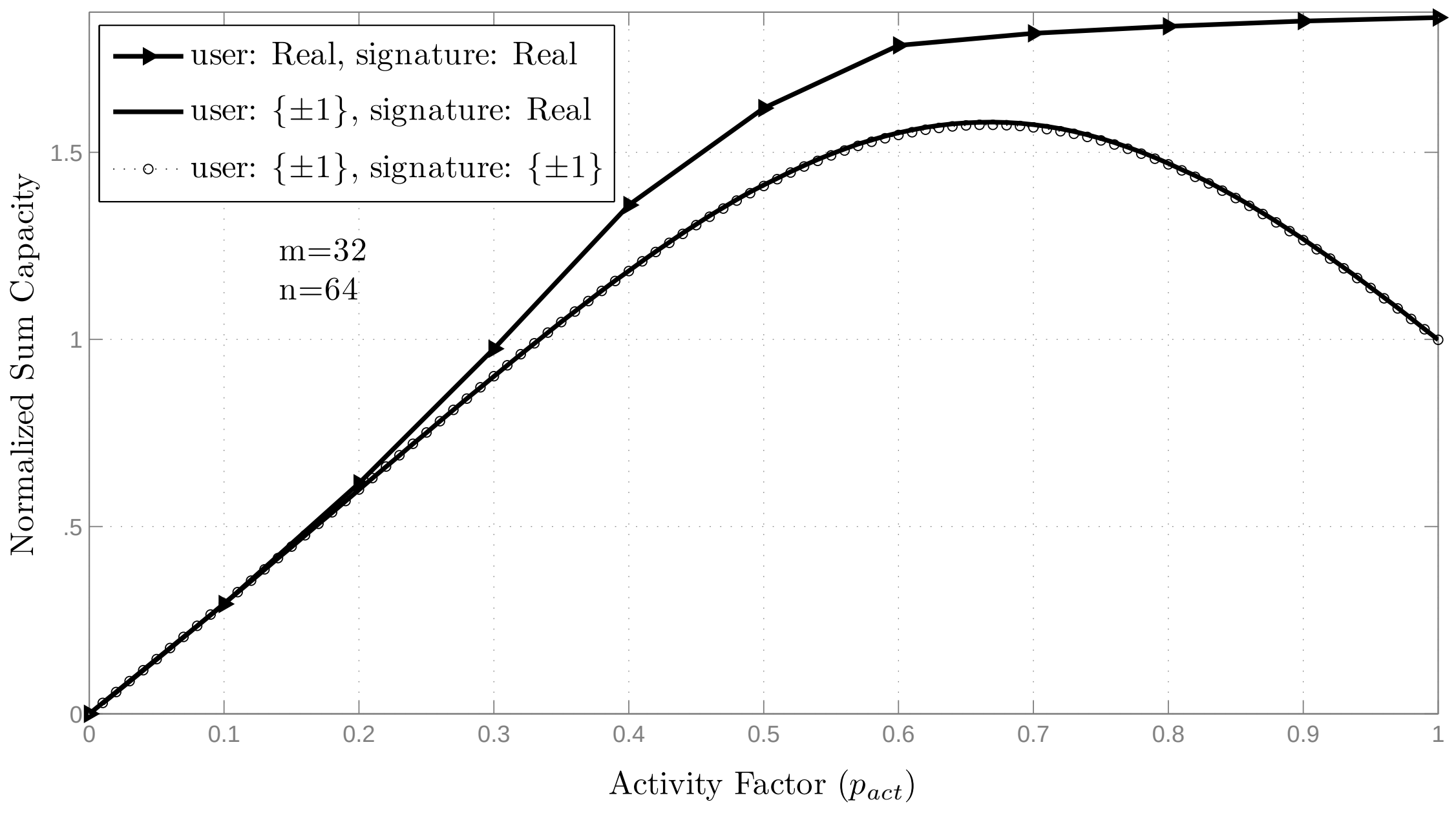

Fig. 21. The normalized sum capacity lower bound for $p_{act}$-active CDMA when $m = 32, n = 64$ and $\eta = 20dB$. This figure shows that for large scale systems, the bound is independent of the signature matrix.

*4) Capacity Lower Bound for a Binary/Binary $p_{act}$-Active CDMA System:*

When the inputs and the signature matrix are binary (bipolar), we have the following theorem for the $p_{act}$-active CDMA system:

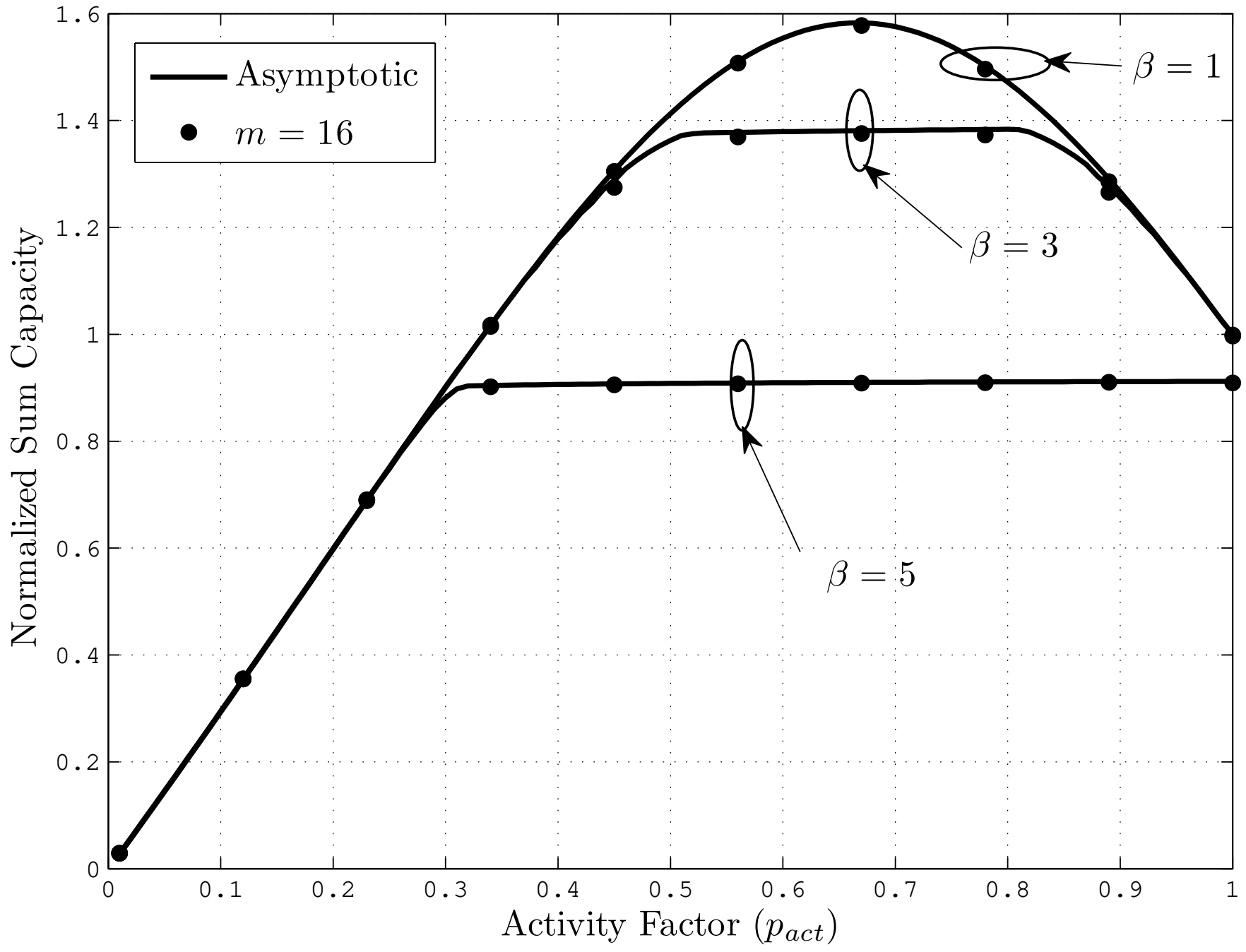

Fig. 22. The normalized asymptotic sum capacity lower bounds w.r.t. $p_{act}$ as compared to finite cases for binary input. Here $\eta = 20dB$, $m = 16$, $\beta = 1, 2, 3$, and the signature matrix is arbitrary.

*Theorem 31 (Binary Input and Binary Signature):*

$$C_{p_{act}}(m, n, \{\pm1\}, \{\pm1\}, \eta, p_{\text{act}}) \geq \sup_{\gamma}\{-m(\gamma \log e - \log(1+\gamma))$$
$$-\log \sum_{k_1+k_2+k_3=n} 2^{k_2-k_1} \binom{n}{k_1, k_2, k_3} {p_{\text{act}}}^{2k_1+k_2} (1-p_{\text{act}})^{k_2} \left(1 + \frac{3}{2}p_{\text{act}}^2 - 2p_{\text{act}}\right)^{k_3}$$
$$\times \left(\sum_{i,j} 2^{-k_1-k_2} \binom{k_1}{i} \binom{k_2}{j} e^{\frac{-2\eta\gamma(4i+2j-2k_1-k_2)^2}{m p_{\text{act}}(1+\gamma)}}\right)^m \}. \qquad (85)$$

For the proof please refer to Appendix O.6. Fig. 21 is the simulation of the sum capacity for the $p_{act}$-active CDMA case for the above examples. The simulations are for $m = 32$, $n = 64$ and $\eta = 20dB$. This figure shows that for large scale systems, the bound is independent of the signature matrix. The lower bounds have a maximum for $p_{act} \approx 0.65$ because of the extra information needed for the determination of active users which is approximately equal to $\log \binom{n}{np_{act}}$.

## VIII. Summary of the Main Results, Conclusion, and Future Work

In this paper, we have attempted to reveal some of the unknowns regarding CDMA. Our first concern was related to the developments of overloaded errorless codes (referred to here as GCO code) for general, finite/infinite, and real/complex users and signature matrices. A general theorem (Theorem 14) was developed to construct larger GCO matrices from the smaller ones. Various examples were given for special binary/ternary user inputs and signature matrices. The same procedure was extended to complex signatures. A general upper bound for the overloading factor ($n/m$) was developed and simulated for a special case. From these theorems, practical ML algorithms were suggested to decode large size GCO codes. Simulation results for special cases showed that GCO matrices outperformed the WBE matrices. The fairness of COW/GCO codes were simulated and discussed; GCO codes were not as fair as the WBE codes but even the worst GCO code was better than the best WBE code for moderate SNR values. If some of the users are not active, the special class of GCO codes (named COWDA) were simulated and compared to WBE and COW codes. COWDA matrices lack the same performance as the COW codes where there is no need for active user detection but they outperform the WBE codes.

Our second concern was on the evaluation of the bounds for the sum capacity for arbitrary input symbols and signature matrices. The next 4 sections were devoted to this problem. Section IV is related to the noiseless case. In this section, a general theorem (Theorem 17) was developed to find a lower bound for the sum capacity. From this theorem, a corollary for finite input/matrix and examples for the special cases were derived and simulated. The simulations show that there is a linear region, where the lower bound is very tight up to a point at which the linearity suddenly breaks down in a severely loaded region ($n_{th}$). We conjecture that the maximum number of GCO users is $n_{th}(m)$ for any given $m$. Also, the simulations showed that by choosing proper symbols for the signature matrix (such as algebraically independent numbers) and a probability law for the symbols, the lower bound could be drastically improved. The simulations also showed that by choosing matrix entries from larger sets ($|\mathcal{I}|$), we could increase the linear region of the lower bound and hence $n_{th}$.

In Section V, we developed a general theorem (Theorem 21) for the sum capacity lower bound for an arbitrary CDMA system for any types of symbols, size, probability distributions, and additive noise. The general noiseless Theorem 17 can be derived from the general noisy Theorem 21 when the additive noise power becomes negligible. The important special case of additive Gaussian noise is given in Corollary 5.

Section VI is the special case of finite symbols for the user inputs and signature matrices. Corollaries 22-24 were special cases for binary/ternary user inputs and $l$-ary signature matrices. Many examples and simulations were given in this section. Again, the dependence of the lower bound on the symbols, sizes, and the noise level is apparent. In this section, an asymptotic lower bound was derived, which implies that for large scale systems the bound does not depend on the signature alphabets but rather on their inputs. Subsection VI-A.2 is related to the development of the sum capacity lower bound for binary users but real or complex signature matrices. The simulations show that by changing the signature matrix from a finite set to an infinite real/complex number, the sum capacity can be improved significantly. This is a noteworthy result since by adding extra complexity at the transmitter and receiver sides, we can increase the capacity. However, our asymptotic results, as well as those of [12], [40], imply that the sum capacity is independent of the signature alphabets. Nevertheless, our simulations for finite $n$ and $m$ show significant differences for the sum capacity lower bounds for different signature symbols, probabilities, and cardinalities.

The sum capacity lower bound was extended to the $k$-active case in Section VII, which may have applications to compressed sensing. In this section, we developed the sum capacity lower bound for the special cases of binary/real, real/real, and binary/binary $k$-active users; the lower bounds were also simulated and compared with respect to the $k$-activity. For the special case of real user inputs when all the users are active, we have the actual sum capacity and therefore the sum capacity lower bound was compared to the actual one. Simulation results showed that the sum capacity lower bound was relatively tight and closely tracked the actual capacity. We also developed similar results for $p_{\text{act}}$-activity with simulations; in addition, an asymptotic lower bound for the real/real case was derived.

As for future work, we suggest to study the effects of fading due to multipath on injectivity of GCO matrices and the evaluation of the sum capacity bounds. The problem of asynchronous CDMA would be a productive area of investigation as indicated by several attempts already made [30], [47]–[49]. The extension of our results to near-far effects is another challenging task [50]. Also, the consideration of generalized users [51]–[54] is yet another interesting topic to look into[7].

[7] A generalized user chooses its data from a set of variable vectors with different symbols.

## Appendix

### A. Proof of Theorem 14, A General Method for Constructing GCO Matrices

The necessary and sufficient condition for $\mathbf{B}$ to be one-to-one is Ker$\{\mathbf{B}\}\bigcap\left\{\frac{\mathcal{I}-\mathcal{I}}{d}\right\}^n=\{0\}^n$. Now if $Z\in$ Ker$\{B\}\bigcap\left\{\frac{\mathcal{I}-\mathcal{I}}{d}\right\}^n$, we claim that $Z=0$. We know that

$$\mathbf{B}Z=0,\ \text{where}\ \ Z=\left[Z_1^T,\dots,Z_{w+1}^T\right]^T, \tag{86}$$

and each $Z_i$ is an $n\times 1$ vector for $i=1,2,,\dots,w$, $Z_{w+1}$ is an $m\times 1$ vector and the entries of $Z_i$ are from $\mathcal{I}-\mathcal{I}$ for $i=1,\dots,w+1$. Thus (86) implies that:

$$k\mathbf{H}_w\left[\mathbf{C}Z_1,\dots,\mathbf{C}Z_w\right]^T=-\left[\mathbf{M}_1Z_{w+1},\dots,\mathbf{M}_wZ_{w+1}\right]^T. \tag{87}$$

Multiplying by $\mathbf{H}_w^T$, we get

$$kw\mathbf{I}_w\left[\mathbf{C}Z_1,\dots,\mathbf{C}Z_w\right]^T=-\mathbf{H}_w^T[\mathbf{M}_1Z_{w+1},\dots,\mathbf{M}_wZ_{w+1}]^T. \tag{88}$$

We assume that the first column of $\mathbf{H}_w$ does not have any $-1$. Thus, we obtain

$$kw(\mathbf{C}Z_1)=-\sum_{i=1}^{w}\mathbf{M}_iZ_{w+1}. \tag{89}$$

Hence

$$\mathbf{C}Z_1=\frac{-1}{kw}\left(\sum_{i=1}^{w}\mathbf{M}_i\right)Z_{w+1}=\begin{pmatrix}\frac{-t}{kw} & \dots & 0\\ \vdots & \ddots & \vdots\\ 0 & \dots & \frac{-t}{kw}\end{pmatrix}Z_{w+1}. \tag{90}$$

The left hand side of (90) is an integer vector and because

$$\left\{\frac{uwk}{t}:u\in\mathbb{Z}\right\}\cap\frac{\mathcal{I}-\mathcal{I}}{d}=\{0\}, \tag{91}$$

all entries of $Z_{w+1}$ are $0$, therefore

$$kw\mathbf{I}_w\left[\mathbf{C}Z_1,\dots,\mathbf{C}Z_w\right]^T=0. \tag{92}$$

Hence

$$(\mathbf{C}Z_i)=0\qquad i=1,2,\dots,w. \tag{93}$$

Since $\mathbf{C}$ is a GCO$\,(m,n,\mathcal{I},\mathcal{S})$ matrix, $Z_i=0$ for $i=1,2,\dots,w$, which implies $Z=0$. This completes the first part. Now let $\mathcal{S}$ consists only of odd numbers, then $Z_{w+1}$ is a $(m-1)\times 1$ vector and (90)

becomes:

$$\mathbf{C}Z_1 = \frac{-1}{kw}\left(\sum_{i=1}^{w}\mathbf{M}_i\right)Z_{w+1} = \begin{pmatrix} \frac{-t}{kw} & \dots & 0 \\ \vdots & \ddots & \vdots \\ 0 & \dots & \frac{-t}{kw} \\ 0 & \dots & 0 \end{pmatrix} Z_{w+1}. \tag{94}$$

The left hand side of (94) is an integer vector with the same parity as its entries and because the last entry of the right hand side vector in (94) is $0$, all the entries should be even and again because

$$\left\{\frac{2uwk}{t} : u \in \mathbb{Z}\right\} \cap \frac{\mathcal{I}-\mathcal{I}}{d} = \{0\}, \tag{95}$$

all the entries of $Z_{w+1}$ are $0$, and the rest of the proof is similar to (92) and (93). □

### B. *Proof of Corollary 1, Constructing Binary GCO Matrices*

$\mathcal{S}$ consists only of odd numbers $\pm 1$. We set $k = 1$ and $w = 4r$ in Theorem 14; note that $\mathbf{M}_i$'s are binary matrices such that $\sum_{i=1}^{4r} \mathbf{M}_i = 2\tilde{\mathbf{I}}_m$. Thus, $t = 2$ and it suffices to show that

$$\{4ru : u \in \mathbb{Z}\} \cap (\mathcal{I} - \mathcal{I}) = \{0\}. \tag{96}$$

The above equation is true because the minimum nonzero positive entry of $\{4ru : u \in \mathbb{Z}\}$ is $4r$ and the maximum positive entry of $\mathcal{I} - \mathcal{I}$ is $2q$, and by our hypothesis, $4r \geq 2q + 1$. □

### C. *Proof of Theorem 15, Generation of Overloaded Complex GCO Matrices for Arbitrary Input Integers*

The necessary and sufficient condition for $\mathbf{B}$ to be one-to-one is Ker$\{\mathbf{B}\} \bigcap\ (\mathcal{I} - \mathcal{I})^n = \{0\}^n$. Now if $Z \in$ Ker$\{\mathbf{B}\} \bigcap\ (\mathcal{I} - \mathcal{I})^n$, we claim that $Z = 0$. Suppose that

$$\mathbf{B}\Big(Z_0 \quad Z_1 \quad \dots Z_k \quad Y_1 \quad Y_2 \quad \dots \quad Y_k\Big)^T = \mathbf{0}, \tag{97}$$

where $Z_i$ and $Y_i$ are, respectively, $1 \times n$ and $1 \times (m-1)$ vectors. Since $\{1, \xi_1, \dots, \xi_k\}$ are independent numbers, the coefficient vector that is multiplied by $\xi_i$ is the zero vector for $i = 1, 2, \dots, k$. Thus, we have

$$\xi_i\big(\mathbf{D}_i Z_i^T\big) + \xi_i \mathbf{M}^i Y_i^T = 0 \quad \text{for} \quad i = 1, 2, \dots, k. \tag{98}$$

Hence

$$\big(\mathbf{D}_i \ \mathbf{M}^i\big)(Z_i Y_i)^T = 0 \quad \text{for} \quad i = 1, 2, \dots, k. \tag{99}$$

because $\big[\mathbf{D}_i \ \mathbf{M}^i\big]$ is a GCO $(m, n, \mathcal{I}, \{0, \pm 1\})$, We must have $(Z_i Y_i)^T = \mathbf{0}$ for $i = 1, 2, \dots, k$; because $\mathbf{D}_0$ is injective on $\mathcal{I}$, $Z_0 = \mathbf{0}$, and thus $Z = \mathbf{0}$.

*D. Proof of Theorem 16, A General Upper Bound for GCO* $(m, n, \mathcal{I}, \mathcal{S})$

Let the input multiuser data be defined by the random vector $X = [x_1, \ldots, x_n]^T$, where $x_i$'s are i.i.d. random variables from the set $\mathcal{I}$ with uniform distribution. Since $x_i$'s are independent, we have $\mathbb{H}(X) = n \log_2 |\mathcal{I}|$. Now, let the transmitted CDMA random vector be defined by $Y = \mathbf{A}X$, where $\mathbf{A}$ is a GCO $(m, n, \mathcal{I}, \mathcal{S})$. Since $\mathbf{A}$ is one-to-one over $\mathcal{I}^n$, we get $\mathbb{H}(Y) = \mathbb{H}(X) = n \log_2 |\mathcal{I}|$. But we have $\mathbb{H}(Y) \leq \sum_{i=1}^m \mathbb{H}(y_i) \leq m \max_{(a_1,\ldots,a_n)\in\mathcal{S}^n} \mathbb{H}(y_1)$, where $y_1 = \sum_{j=1}^n a_j x_j$. □

*E. Proof of Theorem 17, A General Lower Bound for the Noiseless Case*

For a given $A$,

$$\begin{aligned}
\mathbb{I}(X;Y) = \mathbb{H}(Y) &= -\sum_y \mathbb{P}(Y=y)\log\mathbb{P}(Y=y) \\
&= -\sum_x \mathbb{P}(X=x)\log\mathbb{P}(Y=\mathbf{A}x) \\
&= -\sum_x \mathbb{P}_X(x)\log\left[\sum_{x'}\mathbb{P}_X(x')1_{\{\mathbf{A}x=\mathbf{A}x'\}}\right] \\
\geq -\log\left[\sum_{x,x'}\mathbb{P}_X(x)\mathbb{P}_X(x')1_{\{\mathbf{A}(\mathbf{x}-\mathbf{x}')=\mathbf{0}\}}\right] &= -\log\left[\sum_u \mathbb{P}_{\tilde{X}}(u)1_{\{\mathbf{A}u=0\}}\right],
\end{aligned} \tag{100}$$

where the inequality is due to Jensen. Now taking expectation with respect to $\mathbf{A}$ and using Jensen inequality once more, we have

$$\begin{aligned}
\mathbb{E}_A\left(\mathbb{I}(X;Y)\right) &\geq \mathbb{E}_A\left(-\log\sum_u \mathbb{P}_{\tilde{X}}(u)1_{\{\mathbf{A}u=0\}}\right) \\
&\geq -\log\left(\mathbb{E}_A\left(\sum_u \mathbb{P}_{\tilde{X}}(u)1_{\{\mathbf{A}u=0\}}\right)\right) \\
&= -\log\left(\sum_u \mathbb{P}_{\tilde{X}}(u)\mathbb{P}\left(\mathbf{A}u=0\right)\right).
\end{aligned} \tag{101}$$

But since the rows of $\mathbf{A}$ are i.i.d., $(\mathbf{A}u)_1, \ldots, (\mathbf{A}u)_m$ are likewise i.i.d., and thus we get

$$\mathbb{P}\left(\mathbf{A}u=0\right) = \mathbb{P}\left((\mathbf{A}u)_1=0,\ldots,(\mathbf{A}u)_m=0\right) = \mathbb{P}\left((\mathbf{A}u)_1=0\right)^m = \mathbb{P}\left(a.u=0\right)^m. \tag{102}$$

Therefore, $\mathbb{E}_A\left(\mathbb{I}(X;Y)\right) \geq -\log\left(\sum_u \mathbb{P}_{\tilde{X}}(u)\mathbb{P}(a.u=0)^m\right) = -\log\mathbb{E}_{\tilde{X}}\left(\left(\mathbb{P}(a.\tilde{X}=0)\right)^m\right)$. □

### F. Proof of Theorem 19, Noiseless Upper Bound

Let $\mathcal{I} = \{\imath_1, \ldots, \imath_q\}$ and $\mathcal{S} = \{s_1, \ldots, s_l\}$. Clearly, the capacity is maximum when the numbers $\imath_i s_j$ are AIN's; for this case we have:

$$\mathbb{I}(X;Y) = \mathbb{H}(Y) \leq \sum_{i=1}^{m} \mathbb{H}(Y_i) \leq m \cdot \max_{1 \leq k \leq m} \mathbb{H}(Y_k). \tag{103}$$

But $Y_k = \sum_{j=1}^{n} a_{kj} X_j$ and if the number of $s_i$'s in the vector $[a_{k1}, \ldots, a_{kn}]$ is $u_i$, then $Y_k = s_1 z_1 + \cdots + s_l z_l$ where $z_i$ has the distribution: $\text{Mult}\left(u_i, p(\imath_1), p(\imath_2), \ldots, p(\imath_q)\right)$ and $s_i$'s are independent. The algebraically independence condition implies that the map $(z_1, \ldots, z_l) \to Y_k$ is $1-1$ and therefore $\mathbb{H}(Y_k) = \sum_{i=1}^{l} \mathbb{H}(z_i)$. On the other hand, according to a theorem from Shepp and Olkin [55], the entropy of a multinomial distribution $\text{Mult}(t, \theta_1, \ldots, \theta_t)$ is concave and symmetric with respect to $\theta_1, \ldots, \theta_t$ and hence attains the maximum at $\theta_1 = \cdots = \theta_t = \frac{1}{t}$. □

### G. Proof of Theorem 20, Asymptotic Noiseless lower Bound

To prove, we need the following lemma:

*Lemma 3:* Let $\Sigma = \{\xi_1, \ldots, \xi_d\}$ be a finite set of numbers and $\sigma$ a non-degenerate probability distribution on $\Sigma$ with $\mu_\sigma = 0$.

If $X_1, X_2, \ldots$ are i.i.d. with distribution $\sigma$ then

$$\mathbb{P}(X_1 + X_2 + \cdots + X_n = 0) = O\left(n^{-\frac{\text{rank}(\Sigma)}{2}}\right). \tag{104}$$

*Proof of Lemma 3:*

Let $k_i = \#\{j \in \{1, \ldots, n\} | X_j = \xi_i\}$ and $\Gamma = \left\{(c_1, \ldots, c_d) \in \mathbb{Z}^d | c_1\xi_1 + \cdots + c_d\xi_d = 0\right\}$. Clearly $\Gamma$ is a $(d - \text{rank}(\Sigma))$-dimensional lattice in $\mathbb{Z}^d$ and $X_1 + \cdots + X_n = 0$ iff $k = (k_1, \ldots, k_d) \in \Gamma$. But $k_j \sim B(n, \sigma_j)$ and hence $k'_j = \frac{k_j - n\sigma_j}{\sqrt{n}} \approx \mathcal{N}(0, \sigma_j(1-\sigma_j))$. But since $\mu_\sigma = 0$ thus if $k \in \Gamma$, $k'_1\xi_1 + \cdots + k'_d\xi_d = 0$ and so we have $k \in \Gamma$ iff $k' = (k'_1, \ldots, k'_d) \in \frac{1}{\sqrt{n}}(\Gamma - \sqrt{n}\sigma)$.

Now $k'$ has an asymptotically fixed non-degenerate d-dimensional Gaussian distribution and $\Gamma$ is a lattice of co-dimension $\text{rank}(\Sigma)$ and hence $\frac{1}{\sqrt{n}}(\Gamma - \sqrt{n}\sigma)$ is of $O\left(\frac{1}{\sqrt{n}^{\text{rank}(\Sigma)}}\right)$ measure. □

*Proof of Theorem:*

We know that

$$\frac{1}{n}C\left(m,n,\mathcal{I},\mathcal{S}\right)\geq-\frac{1}{n}\log\mathbb{E}_{\tilde{X}}\left(\mathbb{P}\left(a.\tilde{X}=0\right)^{m}\right). \tag{105}$$

Now let $\hat{p}$ be a probability distribution on $\tilde{\mathcal{I}}$ and let $\tilde{X}=\hat{x}$ where the empirical distribution of $\hat{x}$ is $\hat{p}$. Moreover, assume that $\hat{p}$ is supported on $J\subseteq\tilde{\mathcal{I}}$. For large values of $n$, $a.\tilde{x}$ may be considered as sum of n i.i.d. r.v.'s with the non-degenerate distribution $\hat{p}\otimes\pi$ from $J\cdot\mathcal{S}$ and hence according to Lemma 3,

$$\mathbb{P}\left(a.\hat{x}=0\right)=O\left(n^{-\frac{\text{rank}(J\cdot\mathcal{S})}{2}}\right). \tag{106}$$

Note that this probability might dependent on $\hat{p}$ and $\pi$ so we rewrite (106) as:

$$\mathbb{P}\left(a.\hat{x}=0\right)\leq f\left(\hat{p},\pi\right)n^{-\frac{\text{rank}(J\cdot\mathcal{S})}{2}} \tag{107}$$

and therefore:

$$\mathbb{P}\left(a.\hat{x}=0\right)^{m}\leq 2^{m\log f(\hat{p},\pi)-\frac{m\log n\,\text{rank}(J\cdot\mathcal{S})}{2}}=2^{-n\left(\frac{\text{rank}(J\cdot\mathcal{S})}{2\xi}-\frac{m}{n}\log f(\hat{p},\pi)\right)}=2^{-n\left(\frac{\text{rank}(J\cdot\mathcal{S})}{2\xi}+o(1)\right)}. \tag{108}$$

But recall that the probability of appearance of $\hat{p}$ on the empirical distribution of $\tilde{X}$ is approximately $2^{-nD(\hat{p}\|\tilde{p})}$ and therefore:

$$\lim_{\substack{n,m\to\infty\\ \frac{n}{m\log n}\to\xi}}\frac{1}{n}C\left(m,n,\mathcal{I},\mathcal{S}\right)\geq\max_{\substack{J\subseteq\tilde{\mathcal{I}}\\ \hat{p}\text{ on}J}}\lim_{n\to\infty}-\frac{1}{n}\log\left(2^{-nD(\hat{p}\|\tilde{p})}.2^{-n\left(\frac{\text{rank}(J\cdot\mathcal{S})}{2}+O(1)\right)}\right)=$$

$$\min_{\substack{J\subseteq\tilde{\mathcal{I}}\\ \hat{p}\text{on}J}}\left\{D\left(\hat{p}\|\tilde{p}\right)+\frac{\text{rank}\left(J\cdot\mathcal{S}\right)}{2\xi}\right\}. \tag{109}$$

But it is easy to see that $\min_{p\text{ supported on J}}D\left(\hat{p}\|\tilde{p}\right)=-\log\tilde{p}\left(J\right)$ which is attained where $\hat{p}$ is proportional to $\tilde{p}$ on $J$ and this finishes the proof. □

*H. Proof of Examples of Corollary 4*

*1) Proof of Example 12, Binary Wireless CDMA:*

Note that $\tilde{\mathcal{I}}=\{+2,-2,0\}$ with probabilities $\tilde{p}(+2)=\tilde{p}(-2)=p_1(1-p_1)$, and $\tilde{p}(0)={p_1}^2+(1-p_1)^2$. Let $\tilde{x}_1\in\mathcal{B}(k_1,k_2,k_3)$; suppose that $a^T.\tilde{x}_1=0$ and $a^T$ has $\alpha$ "+1" entries and $\beta$ "+1" entries in places in where $\tilde{x}_1$ is equal to $+2$ and $-2$, respectively. We conclude that $2(\alpha-(k_1-\alpha)-\beta+(k_2-\beta))=0$ or $\beta=\alpha+(k_2-k_1)/2$. The number of such $a^T$ is equal to $\binom{k_1}{\alpha}\binom{k_2}{\beta}$ with the probability ${\pi_1}^{\alpha+\beta}(1-\pi_1)^{k_1+k_2-\alpha-\beta}$; thus $\theta(k_1,k_2,k_3,\mathcal{S})=\sum^{k_1}_{\substack{\alpha=0\\ \beta=\alpha+\frac{k_2-k_1}{2}}}\binom{k_1}{\alpha}\binom{k_2}{\beta}{\pi_1}^{\alpha+\beta}(1-\pi_1)^{k_1+k_2-\alpha-\beta}$ and this completes the proof.

If $\pi_1 = p_1 = 1/2$, then due to symmetry it is easy to show that if the number of nonzero entries in $\tilde{x}$ is a fix number like $k$ then the value of $p(a^T.\tilde{x} = 0)$ remains constant, i.e. $\pm 2$ behave like a single symbol and $\tilde{\mathcal{I}} = \{\pm 2, 0\}$. Thus if $\tilde{x}_1 \in \mathcal{B}\{k, k_1\}$ and $a^T.\tilde{x} = 0$, then $k$ must be even and $p(a^T.\tilde{x} = 0)$ is equal to $\binom{k}{\frac{k}{2}}/2^k$. Hence if $q = p = 1/2$, the bound reduces to the bound in [14]. $\square$

*2) Proof of Example 13, Binary Optical CDMA:*

The proof is similar to the proof of example 12. $\square$

*3) Proof of Example 14, Binary Complex Signature (Uniform Distribution) :*

Note that $\tilde{\mathcal{I}} = \{+2, -2, 0\}$ with probabilities $\tilde{p}_{\imath_1} = \tilde{p}_{\imath_2} = \frac{1}{4}, \tilde{p}_{\imath_3} = \frac{1}{2}$. Let $\tilde{x}_1 \in \mathcal{B}(k_1, k_2, k_3)$; suppose that $a^T.\tilde{x}_1 = 0$ and $a^T$ has $\alpha$ "+1" and $\beta$ "+1" entries in positions where $\tilde{x}_1$ is equal to $+2$ and $-2$, respectively. We conclude that $2(k_1 - \alpha - k_2 + \beta)j + 2(\alpha - \beta) = 0$; thus $\alpha = \beta, k_1 = k_2 - k$ . The number of such $a^T$'s is $\sum_{\alpha=0}^{k} \binom{k}{\alpha}\binom{k}{\alpha}$ because we should choose $\alpha$ entries from those $+2$ entries in $\tilde{x}_1$ and $\alpha$ entries from those $-2$ entries and let the corresponding entries in $a^T$ be $1$ and let the remaining $2k - \alpha$ entries be $j$. Each $a^T$ has probability $\left(\frac{1}{2}\right)^{2k}$ and since $\sum_{\alpha=0}^{k} \binom{k}{\alpha}\binom{k}{\alpha} = \binom{2k}{k}$, we get $\theta(k, k, n-2k, \mathcal{S}) = \left(\frac{1}{2}\right)^{2k} \sum_{\alpha=0}^{k} \binom{k}{\alpha}\binom{k}{\alpha} = \left(\frac{1}{2}\right)^{2k}\binom{2k}{k}$. $\square$

*4) Proof of Example 15, Binary/Ternary, $\mathcal{I} = \{\pm 1\}$, $\mathcal{S} = \{0, \pm 1\}$, $\pi(0) = \pi_0$ and $\pi(+1) = \pi(-1)$:*

If $p(+1) = p(-1) = \frac{1}{2}$, it is easy to show that if the number of nonzero entries in $\tilde{x}$ is equal to $k$, then the value of $p(a^T.\tilde{x} = 0)$ remains constant. In other words $\pm 2$ behave like a single symbol and $\tilde{\mathcal{I}} = \{\pm 2, 0\}$. Now, suppose that $\tilde{x}_1 \in \mathcal{B}\{k, k_1\}$ and the nonzero entries of $\tilde{x}_1$ are all $+2$ and $a^T.\tilde{x}_1 = 0$ . If $a^T$ has $\alpha$ "+1" and $\beta$ "$-1$" entries, in positions where $\tilde{x}_1$ is equal to $2$, we conclude that $\alpha = \beta$ . The number of such $a^T$ is $\binom{k}{2\alpha}\binom{2\alpha}{\alpha}$ each with probability $\pi_0^{k-2\alpha}\pi_1^{2\alpha}$. Thus $\theta(k, k_1, \mathcal{S}) = \left(\sum_{\alpha=0}^{k} \binom{k}{2\alpha}\binom{2\alpha}{\alpha}\pi_0^{k-2\alpha}\pi_1^{2\alpha}\right)$. $\square$

*5) Proof of Example 16 Binary/Ternary Uniform distribution:*

The proof is similar to proof of example 17.

*6) Proof of Example 17, Binary/Ternary Uniform Distribution:*

Note that $\tilde{\mathcal{I}} = \{+2, -2, 0\}$ with probabilities $\tilde{p}_{\imath_1} = \tilde{p}_{\imath_2} = \frac{1}{4}, \tilde{p}_{\imath_3} = \frac{1}{2}$; Let $\tilde{x}_1 \in \mathcal{B}(k_1, k_2, k_3)$; suppose that $a^T.\tilde{x}_1 = c_1 w^2 + c_2 w + c_3 = 0$. Because $w$ is a the third root of unity, one can deduce that $c_1 = c_2 = c_3 = c$. If we denote $s$ as the sum of entries of $\tilde{x}_1$, we have $c_1 + c_2 + c_3 = s = 2(k_1 - k_2)$, which yields $c = \frac{2(k_1-k_2)}{3}$. In order to satisfy $a^T.\tilde{x}_1 = 0$, one should choose $\alpha_1$ entries such that $\tilde{x}_1$ is equal to $+2$, and $\alpha_1 - \frac{c}{2}$ entries such that $\tilde{x}_1$ is equal to $-2$. Let the corresponding entries in $a^T$ be $w^2$ and the same argument for $w$ and $1$ entries of $a^T$. Thus the number of $a^{T\prime}s$ which satisfy $a^T.\tilde{x}_1 = 0$ is equal to $\left(\sum \binom{k_1}{\alpha_1}\binom{k_1-\alpha_1}{\alpha_2}\binom{k_2}{\alpha_1 - \frac{k_1-k_2}{3}}\binom{k_2-\alpha_1+\frac{k_1-k_2}{3}}{\alpha_2-\frac{k_1-k_2}{3}}\right)$ with probability $\frac{1}{3^{k_1+k_2}}$. $\square$

*7) Proof of Example 18, Ternary/Binary System,* $\mathcal{I} = \{0, \pm 1\}, \mathcal{S} = \{\pm 1\}$, $p\,(0) = p_0$, $p(+1) = p(-1) = p_1$ *and* $\pi(+1) = \frac{1}{2}$*:*

Because $\pi(+1) = \pi(-1) = \frac{1}{2}$ and $p(+1) = p(-1)$, it is easy to show that if the number of $\pm 2$ and $\pm 1$ entries in $\tilde{x}$ are fixed, then the value of $p(a^T.\tilde{x} = 0)$ remains constant. In other words, $\pm 2$ and $\pm 1$ behave like $+2$ and $+1$ symbols. Let $\tilde{\mathcal{I}} = \{\imath_1, \imath_2, \imath_3\} = \{\pm 2, \pm 1, 0\}$ with the probabilities: $\tilde{p}_{\imath_1} = 2{p_1}^2, \tilde{p}_{\imath_2} = 4p_1p_0, \tilde{p}_{\imath_3} = {p_0}^2 + 2{p_1}^2$. Let $\tilde{x}_1 \in \mathcal{B}(k_1, k_2, k_3)$ and $a^T.\tilde{x}_1 = 0$; from the above argument we can assume that the nonzero entries of $\tilde{x}_1$ are all $+2$ or $+1$. If $a^T$ has $\alpha$ "$+1$" entries where $\tilde{x}_1 = +2$ and $\beta$ "$+1$" entries where $\tilde{x}_1 = +1$, we get $a^T.\tilde{x} = 2(2\alpha - k_1) + 2\beta - k_2 = 0$, which yields $\beta = \frac{k_2}{2} + k_1 - 2\alpha$. The number of such $a^T$, is $\binom{k_1}{\alpha}\binom{k_2}{\beta}$ with probability $(\frac{1}{2})^{k_1+k_2}$. By applying Corollary 4, we get (52).

*8) Proof of Example 19, Ternary Wireless System,* $\mathcal{I} = \mathcal{S} = \{0, \pm 1\}$, $p\,(0) = p_0$, $p(+1) = p(-1) = p_1$, $\pi(0) = \pi_0$ *and* $\pi(+1) = \pi(-1) = \pi_1$*:*

By the same argument in the proof of Example 18, let $\tilde{\mathcal{I}} = \{\imath_1, \imath_2, \imath_3\} = \{\pm 2, \pm 1, 0\}$ with the probabilities $\tilde{p}_{\imath_1} = 2{p_1}^2, \tilde{p}_{\imath_2} = 4p_1p_0$ and $\tilde{p}_{\imath_3} = {p_0}^2 + 2{p_1}^2$. Let $\tilde{x}_1 \in \mathcal{B}(k_1, k_2, k_3)$ and $a^T.\tilde{x}_1 = 0$; we can assume that the nonzero entries of $\tilde{x}_1$ are all $+2$ or $+1$. If $a^T$ has $\alpha$ "$+1$" and $\beta$ "$-1$" entries when $\tilde{x}_1 = +2$, respectively, and $\alpha'$ "$+1$" and $\beta'$ "$-1$" entries when $\tilde{x}_1 = +1$, we get $a^T.\tilde{x} = 2(\alpha - \beta) + \alpha' - \beta' = 0$, hence $\beta' = 2(\alpha - \beta) + \alpha'$. The number of such $a^T$'s is $\binom{k_1}{\alpha}\binom{k_1-\alpha}{\beta}\binom{k_2}{\alpha'}\binom{k_2-\alpha'}{\beta'}$ with the probability ${\pi_1}^{\alpha+\beta+\alpha'+\beta'}{\pi_0}^{k_1+k_2-\alpha-\beta-\alpha'-\beta'}$. By applying Corollary 4 we get (53).

*9) Proof of Example 20, Binary/Quaternary System for* $\mathcal{I} = \{\pm 1\}$, $\mathcal{S}$ *is a set of four AIN's:*

Let $\mathcal{S} = \{s_1, s_2, s_3, s_4\}$ and note that $\tilde{\mathcal{I}} = \{+2, -2, 0\}$ with the probabilities $\tilde{p}_{\imath_1} = \tilde{p}_{\imath_2} = \frac{1}{4}$ and $\tilde{p}_{\imath_3} = \frac{1}{2}$. Let $\tilde{x}_1 \in \mathcal{B}(k_1, k_2, k_3)$ and assume that $a^T.\tilde{x}_1 = c_1 s_1 + c_2 s_2 + c_3 s_3 + c_4 s_4 = 0$. Because $s_i$'s are independent numbers, one has $c_1 = c_2 = c_3 = c_4 = 0$; thus $0 = c_1 + c_2 + c_3 + c_4 = 2(k_1 - k_2)$ which yields $k_1 = k_2 = k$. Note that to satisfy $a^T.\tilde{x}_1 = 0$, one should choose $\alpha_1$ entries when $\tilde{x}_1 = +2$, and $\alpha_1$ entries when $\tilde{x}_1 = -2$, and let the corresponding entries in $a^T$ be $s_1$ and the same argument for $s_2, s_3$ and $s_4$. Thus the number of $a^T$'s which satisfy $a^T.\tilde{x}_1 = 0$ is equal to $\sum_{\substack{\alpha_1,\alpha_2,\alpha_3,\alpha_4\\ \sum\alpha_i = k}} \left(\frac{k!}{\alpha_1!\alpha_2!\alpha_3!\alpha_4!}\right)^2$ with probability $(\frac{1}{4})^{2k}$. The following identities yield (54):

$$\sum_{\substack{\alpha_1,\alpha_2,\alpha_3,\alpha_4\\ \sum\alpha_i = k}} \left(\frac{k!}{\alpha_1!\alpha_2!\alpha_3!\alpha_4!}\right)^2$$

$$= \sum_{\alpha_1,\alpha_2} \left(\frac{k!}{\alpha_1!\alpha_2!}\right)^2 \frac{1}{(\alpha_3+\alpha_4)!^2} \sum_{\alpha_3,\alpha_4} \left(\frac{(\alpha_3+\alpha_4)!}{\alpha_3!\alpha_4!}\right)^2 = \sum_{\alpha_1,\alpha_2} \left(\frac{k!}{\alpha_1!\alpha_2!}\right)^2 \frac{1}{(\alpha_3+\alpha_4)!^2} \binom{2(\alpha_3+\alpha_4)}{\alpha_3+\alpha_4}$$

$$= \sum_{\substack{\alpha_1,\alpha_2,m\\ \alpha_1+\alpha_2+m=k}} \left(\frac{k!}{\alpha_1!\alpha_2!}\right)^2 \frac{1}{m!^2}\binom{2m}{m} = \sum_m \frac{1}{m!^2}\binom{2m}{m}\frac{k!^2}{(k-m)!^2}\sum_{\alpha_1,\alpha_2}\left(\frac{(k-m)!}{\alpha_1!\alpha_2!}\right)^2$$

$$= \sum_{m=0}^{k} \frac{1}{m!^2}\binom{2m}{m}\frac{k!^2}{(k-m)!^2}\binom{2k-2m}{k-m} = \sum_{m=0}^{k}\binom{k}{m}^2\binom{2m}{m}\binom{2k-2m}{k-m} \qquad (110)$$

□

*10) Proof of Example 21, Binary/Quaternary System for $\mathcal{I} = \{\pm 1\}$, $\mathcal{S} = \{\pm 1, \pm i\}$ :*

If $p(+1) = p(-1) = \frac{1}{2}$ and $\pi$ is a uniform distribution, it is easy to show that if the number of nonzero entries in $\tilde{x}$ is equal to $k$, the value of $p(a^T.\tilde{x} = 0)$ remains constant; i.e $\pm 2$ behave like the single symbol $+2$ and $\tilde{\mathcal{I}} = \{\pm 2, 0\}$. Now assume that $\tilde{x}_1 \in \mathcal{B}\{k', k_1\}$ and the nonzero entries of $\tilde{x}_1$ are all $+2$ and $a^T.\tilde{x}_1 = 0$ . If $a^T$ has $\alpha$ "$+1$", $\beta$ "$-1$", $\alpha'$ "$j$", and $\beta'$ "$-j$" entries in positions where $\tilde{x}_1$ is equal to 2, we get $2(\alpha-\beta)+2(\alpha'-\beta')j = 0$; thus, $\alpha = \beta, \alpha' = \beta'$, so $k' = 2k$ is an even number. The number of such $a^T$'s is $\sum_{i=0}^{k}\binom{2k}{2i}\binom{2i}{i}\binom{2k-2i}{k-i}$ with probability $\left(\frac{1}{4}\right)^{2k}$. Hence, $\theta(k', k_1, \mathcal{S}) = \left(\sum_{i=0}^{k}\binom{2k}{2i}\binom{2i}{i}\binom{2k-2i}{k-i}\left(\frac{1}{4}\right)^{2k}\right)$. The combinatorial identities yield $\left(\sum_{i=0}^{k}\binom{2k}{2i}\binom{2i}{i}\binom{2k-2i}{k-i}\left(\frac{1}{4}\right)^{2k}\right) = \left(\frac{\binom{2k}{k}}{2^{2k}}\right)^2$. □

### I. Proof of Theorem 21, General Lower Bound for the Sum Capacity

For a fixed $\mathbf{A}$, we have

$$\begin{aligned}\mathbb{I}(X;Y) = h(Y) - h(N) &= \mathbb{E}_Y(-\log f_Y(Y)) - \mathbb{E}_N(-\log f_N(N)) \\ &= \mathbb{E}_{X,N}\left(-\log f_Y\left(\frac{1}{\sqrt{m}}\mathbf{A}X + N\right)\right) - \mathbb{E}_N(-\log f_N(N)) \\ &= \mathbb{E}_{X,N}\left(-\log \frac{f_Y\left(\frac{1}{\sqrt{m}}\mathbf{A}X + N\right)}{f_N(N)}\right).\end{aligned} \tag{111}$$

But clearly $f_Y(y) = \sum_x \mathbb{P}_X(x) f_N\left(y - \frac{1}{\sqrt{m}}\mathbf{A}X\right)$, and thus

$$\begin{aligned}\mathbb{I}(X;Y) &= \mathbb{E}_{X,N}\left(-\log \frac{\sum_x \mathbb{P}_X(x) f_N\left(\frac{1}{\sqrt{m}}\mathbf{A}(X-x) + N\right)}{f_N(N)}\right) \\ &\geq \mathbb{E}_N\left(-\log \mathbb{E}_X\left(\frac{\sum_x \mathbb{P}_X(x) f_N\left(\frac{1}{\sqrt{m}}\mathbf{A}(X-x) + N\right)}{f_N(N)}\right)\right) \\ &= \mathbb{E}_N\left(-\log\left(\sum_{x,x'} \mathbb{P}_X(x)\mathbb{P}_X(x') \frac{f_N\left(\frac{1}{\sqrt{m}}\mathbf{A}(x'-x) + N\right)}{f_N(N)}\right)\right).\end{aligned} \tag{112}$$

Now let $A$ be a random matrix with i.i.d. entries. By taking expectation with respect to $A$, we get

$$\begin{aligned}\mathbb{E}_A(\mathbb{I}(X;Y)) &\geq \mathbb{E}_{A,N}\left(-\log\left(2^{\sum q(N_i)} \cdot \sum_{x,x'} \mathbb{P}_X(x)\mathbb{P}_X(x') \frac{f_N\left(\frac{1}{\sqrt{m}}A(x'-x) + N\right)}{2^{\sum q(N_i)} f_N(N)}\right)\right) \\ &= -\mathbb{E}_N\left(\sum q(N_i)\right) - \mathbb{E}_{A,N}\left(\sum_{x,x'} \mathbb{P}_X(x)\mathbb{P}_X(x') \frac{f_N\left(\frac{1}{\sqrt{m}}A(x'-x) + N\right)}{2^{\sum q(N_i)} f_N(N)}\right).\end{aligned} \tag{113}$$

Using the Jensen's inequality again, we will have

$$\mathbb{E}_A(\mathbb{I}(X;Y)) \geq -m\mathbb{E}(q(N_1)) - \log\left(\sum_{x,x'} \mathbb{P}_X(x)\mathbb{P}_X(x') \mathbb{E}_{N,A}\left(\frac{f_N\left(\frac{1}{\sqrt{m}}A(x'-x) + N\right)}{2^{\sum q(N_i)} f_N(N)}\right)\right). \tag{114}$$

Now, note that $f_N(x) = \prod_{j=1}^m f_{N_j}(x_j) = \prod_{j=1}^m f(x_j)$ and hence

$$\frac{f_N\left(\frac{1}{\sqrt{m}}A(x'-x) + N\right)}{2^{\sum q(N_i)} f_N(N)} = \prod_{j=1}^m \frac{f\left(\frac{1}{\sqrt{m}}A(x'-x)_j + N_j\right)}{2^{q(N_j)} f(N_j)}. \tag{115}$$

For fixed $x$ and $x'$, $A(x'-x)_j$ and $N_j$ are independent for $j=1,\ldots,m$.

Thus, the expectation and product operators commute; we then get

$$\mathbb{E}_{N,A}\left(\frac{f_N\left(\frac{1}{\sqrt{m}}A\left(x'-x\right)+N\right)}{2^{\sum q(N_i)}f_N\left(N\right)}\right)=\prod_{j=1}^{m}\mathbb{E}_{N_j,A}\left(\frac{f\left(\frac{1}{\sqrt{m}}A\left(x'-x\right)_j+N_j\right)}{2^{q(N_j)}f\left(N_j\right)}\right)$$
$$=\left(\mathbb{E}_{N_1,A}\left(\frac{f\left(\frac{1}{\sqrt{m}}A\left(x'-x\right)_1+N_1\right)}{2^{q(N_1)}f\left(N_1\right)}\right)\right)^m. \quad (116)$$

But

$$\mathbb{E}_{N_1}\left(\frac{f\left(\frac{1}{\sqrt{m}}A\left(x'-x\right)_1+N_1\right)}{2^{q(N_1)}f\left(N_1\right)}\right)=\int\frac{f\left(\frac{1}{\sqrt{m}}A\left(x'-x\right)_1+u\right)}{2^{q(u)}f\left(u\right)}f\left(u\right)du$$
$$=\int f\left(\frac{1}{\sqrt{m}}A\left(x'-x\right)_1+u\right)2^{-q(u)}du=\int f\left(u\right)2^{-q\left(u-\frac{1}{\sqrt{m}}A(x'-x)_1\right)}du$$
$$=\mathbb{E}_{N_1}\left(2^{-q\left(N_1-\frac{1}{\sqrt{m}}A(x'-x)_1\right)}\right). \quad (117)$$

After substituting in the previous relations, we get

$$\mathbb{E}_A(\mathbb{I}(X;Y))\geq -m\mathbb{E}(q(N_1))-\log\left(\sum_{x,x'}\mathbb{P}_X(x)\mathbb{P}_X(x')\left(\mathbb{E}_{N_1,A}\left(2^{-q\left(N_1-\frac{1}{\sqrt{m}}(A(x'-x))_1\right)}\right)\right)^m\right)$$
$$=-m\mathbb{E}(q(N_1))-\log\left(\mathbb{E}_{\tilde{X}}\left(\mathbb{E}_{N_1,a}\left(2^{-q\left(N_1-\frac{a^T.\tilde{X}}{\sqrt{m}}\right)}\right)\right)^m\right). \quad (118)$$

Now since $A=rB$, we get

$$\mathbb{E}_A(\mathbb{I}(X;Y))\geq -m\mathbb{E}\left(q\left(N_1\right)\right)-\log\left(\mathbb{E}_{\tilde{X}}\left(\mathbb{E}_{N_1,b}\left(2^{-q\left(N_1-\frac{rb^T\tilde{X}}{\sqrt{m}}\right)}\right)\right)^m\right). \quad (119)$$

The right-hand side of the above inequality is a lower bound for the average mutual information where $A$ and $X$ are chosen at random with appropriate probability distributions. The choice of $r$ guarantees the SNR constraint on the average for this class of random matrices. Since the capacity is the maximization over $\pi(\cdot)$ and $p(\cdot)$, the inequality holds. □

Note that although in the proof of the above theorem, the SNR condition is only satisfied on the average, because of the measure concentration phenomenon [13], almost all matrices (w.r.t $\mathbb{P}^\pi$) result in almost the same power as well as almost the same mutual information. Thus, a typical matrix simultaneously satisfies both the SNR condition and the bound for mutual information with high probability.

To obtain a true bound for very small values of $m$ and $n$, one can start from the equation

$$\mathbb{E}_A\left(\mathbb{I}\left(X;Y\right)\right)=\mathbb{P}\left(\mathcal{Q}\right)\times\mathbb{E}_A\left(\mathbb{I}\left(X;Y\right)|\mathcal{Q}\right)+\left(1-\mathbb{P}\left(\mathcal{Q}\right)\right)\times\mathbb{E}_A\left(\mathbb{I}\left(X;Y\right)|\mathcal{Q}^c\right), \quad (120)$$

where $\mathcal{Q}$ denotes the event that the (random) matrix $A$ satisfies the SNR condition. If $|\mathcal{I}| = q$, then

$$\mathbb{E}_A\left(\mathbb{I}\left(X;Y\right)|\mathcal{Q}\right) \geq \frac{1}{\mathbb{P}\left(\mathcal{Q}\right)}\left(\mathbb{E}_A\left(\mathbb{I}\left(X;Y\right)\right) - \left(1-\mathbb{P}\left(\mathcal{Q}\right)\right)n\log q\right). \quad (121)$$

Now, it remains to evaluate a lower bound for $\mathbb{P}\left(\mathcal{Q}\right)$. But

$$\mathbb{P}\left(\mathcal{Q}\right) = 1 - \mathbb{P}\left(\sigma_p^2 \mathrm{tr}\left(\mathbf{BB}^*\right) + \mu_p^2 \mathrm{tr}\left(\mathbf{BJB}^*\right) \geq \frac{n\eta\sigma_f^2 m}{r^2}\right), \quad (122)$$

which can be controlled using the Markov inequality in the general case or using Chernoff like inequalities if $\mu_p = 0$ since $\mathrm{tr}\left(\mathbf{BB}^*\right)$ is the summation of i.i.d. random variables.

### *J. Proof of Corollary 6, General Lower Bound for the Sum Capacity for $(q,l)$-ary CDMA Systems*

Suppose $\tilde{x}$ has $k_0$ number of $0$'s and $k_i$ number of $\imath_i$'s. The number of such $\tilde{x}$'s is $\frac{n!}{k_0!k_1!\dots k_{\tilde{q}}!}$ with probability $\prod_{j=0}^{\tilde{q}} \tilde{p}_j^{k_j}$. For such $\tilde{x}$'s, suppose $u_{ij}$ is the number of $s_i\tilde{\imath}_j$ terms in the product $b^T\tilde{x}$; thus $u_{1j} + u_{2j} + \cdots + u_{lj} = k_j$ and if $u_{ij}$'s are fixed numbers, the number of corresponding $b^T$ vectors is $\prod_{i=1}^{\tilde{q}} \frac{k_i!}{u_{1i}!u_{2i}!\dots u_{li}!}$ with probability $\prod_{t=1}^{l} {\pi_t}^{\sum_{\alpha=1}^{\tilde{q}} u_{t\alpha}}$.
Therefore, we have

$$\left(\mathbb{E}_b\left(\frac{e^{\frac{-\gamma r^2}{2(1+\gamma)m}|b^T\tilde{X}|^2}}{1+\gamma}\right)\right)^m =$$

$$\left(\sum_{\substack{\sum_{i=1}^{l} u_{ij} = k_j \\ 1\leq j\leq \tilde{q}}} \prod_{i=1}^{\tilde{q}} \binom{k_i}{u_{1i},\dots,u_{li}} \prod_{t=1}^{l} {\pi_t}^{\sum_{\alpha=1}^{\tilde{q}} u_{t\alpha}} e^{\frac{-\gamma r^2}{2(1+\gamma)m}|\sum_{i=1}^{l} s_i(\sum_{j=1}^{\tilde{q}} u_{ij}\tilde{\imath}_j)|^2}\right)^m. \quad (123)$$

□

### *K. Proof of Corollary 7, Results for Finite User Inputs and Symmetric Signature Matrices- Noisy Case*

Suppose $\tilde{x}$ has $k_0$ number of $0$'s and $k_i$ number of $\pm\imath_i$'s. One can easily see that from the symmetric distribution on the signature alphabets, the expected value on the vector $b$ remains constant; i.e., $\pm\tilde{\imath}_i$ behaves similar to the single symbol $+\tilde{\imath}_i$. The rest of the proof is the same as Corollary 6. The second part is exactly the same as the first part and is straightforward. □

*L. Proof of Theorem 22, Lower Bound for Binary Input and Real/Complex Valued Signature Matrix*

We use Corollary 5 where $\pi(\cdot)$ is the distribution of a standard Gaussian real r.v. for the real case, or the distribution of a Gaussian complex r.v. with independent standard Gaussian coordinates for the complex case. In this case, $\tilde{X}$ is a vector of i.i.d. entries with values $\tilde{I} = \{\tilde{\imath}_0, \pm\tilde{\imath}_1, \pm\tilde{\imath}_2, \ldots, \pm\tilde{\imath}_{\tilde{q}'}\}$. By noting that the distribution is symmetric and Gaussian, it is easy to see that if the number of $\pm\tilde{\imath}_i$ entries of $\tilde{X}$ is fixed, then the expected value on the vector $b$ remains constant. Let $k_i$ be the number of $\pm\tilde{\imath}_i$ entries of $\tilde{x}$; the number of such $\tilde{x}$'s is $\frac{n!}{k_0!k_1!\ldots k_{\tilde{q}'}!}2^{k_1+k_2+\cdots+k_{\tilde{q}'}}$ with probability $\prod_{j=0}^{\tilde{q}'}\tilde{p}_j^{k_j}$. $b^T\tilde{x}$ is a linear combination of independent Gaussian random variables, which is again Gaussian. By using the fact that $r = \sqrt{\frac{2\eta}{{\sigma_p}^2+n{\mu_p}^2}}$ for the real case and $r = \sqrt{\frac{\eta}{{\sigma_p}^2+n{\mu_p}^2}}$ for the complex case, (63) and (64) are easily derived.

*M. Proof of Theorem 23, A General Upper Bound for the Noisy Case*

Our conjecture is based on i.i.d. user input distribution. Note that
$\mathbb{I}(X;Y) = \mathbb{H}(Y) - \mathbb{H}(Y|X) = \mathbb{H}(Y) - \mathbb{H}(N) = \mathbb{H}(Y_1, Y_2, \ldots, Y_m) - m\mathbb{H}(N_1) \leq m(\mathbb{H}(Y_1) - \mathbb{H}(N_1))$. When $\mathcal{S} = \{s_1, s_2, \ldots, s_q\}$, $\mathbb{H}(Y_1)$ is maximum when we have $u_i$ number of $s_i$ in vector $Y_1$. Let $p$ be the i.i.d. product distribution which maximizes $\mathbb{I}(X;Y)$; thus, $C(m,n,\mathcal{I},\mathcal{S},\eta) \leq n\mathbb{H}(\mathcal{I})$. Suppose $v_{ij}$ is the number of $s_i\imath_j$ terms in $Y_1 = \frac{1}{\sqrt{m}}\sum_{i=1}^{n} a_{1i}x_i + N$; the number of corresponding $x$'s is $\prod_{k=1}^{l}\frac{u_k!}{v_{k1}!v_{k2}!\ldots v_{kq}!}$ with probability $\prod_{k=1}^{q}{p_k}^{\sum_{\alpha=1}^{l} v_{\alpha k}}$. Therefore, $Y_1$ has the following distribution: in which

$$\tilde{f}(z) = \sum_{\substack{\sum_{j=1}^{q} v_{ij}=u_i \\ 1\leq i\leq l}} \left(\prod_{k=1}^{l}\binom{u_k}{v_{k1},\ldots v_{kq}}\right)\left(\prod_{k=1}^{q}{p_k}^{\sum_{\alpha=1}^{l} v_{\alpha k}}\right) f(z - \frac{1}{\sqrt{m}}(\sum_{k=1}^{l} s_k\sum_{\alpha=1}^{q} v_{k\alpha}\imath_\alpha)), \tag{124}$$

where

$$f(x+\sqrt{-1}y) = \frac{\eta}{\pi}e^{-\eta(x^2+y^2)} \tag{125}$$

is the distribution of a complex Gaussian.
Also, when $s_i = e^{\frac{2\pi i}{l}\sqrt{-1}}$ and $l$ divides $n$, we conjecture that $u_1 = u_2 = \cdots = u_l = \frac{n}{l}$. . □

*N. Proof of Theorem 24, Asymptotic Lower Bound for Finite Input and Additive Gaussian Noise*

$$\begin{aligned}
&\lim_{\substack{m,n\to\infty \\ n/m\to\beta}}\frac{1}{n}\left[-m\gamma\log e - \log\mathbb{E}_{\tilde{X}}\left(\left(\mathbb{E}_b\left(\frac{e^{\frac{-\gamma r^2}{2(1+\gamma)m}|b^T\tilde{X}|^2}}{1+\gamma}\right)\right)^m\right)\right] \\
&= -\frac{1}{\beta}\gamma\log e - \lim_{n\to\infty}\frac{1}{n}\log\mathbb{E}_{\tilde{X}}\left(\left(\mathbb{E}_b\left(\frac{e^{-\frac{\beta\gamma r^2}{2(1+\gamma)}|\frac{b^T\tilde{X}}{\sqrt{n}}|^2}}{1+\gamma}\right)\right)^{\frac{n}{\beta}}\right).
\end{aligned} \tag{126}$$

Now, let $n$ be a large number and $\tilde{X} = \hat{x}$; name the empirical probability measure induced by $\hat{x}$ on $\tilde{\mathcal{I}}$ as $\hat{p}$ (i.e., a proportion of $\hat{p}(\imath)$ of entries of $\hat{x}$ are $\imath$). In this case, $\mathbb{E}\left(\frac{b^T\tilde{X}}{\sqrt{n}}\right) = \sqrt{n}\mu_\pi\mu_{\hat{p}}$. Therefore, when $\mu_\pi, \mu_{\hat{p}} \neq 0$, $\mathbb{E}_b\left(\frac{e^{\frac{-\beta\gamma r^2}{2(1+\gamma)}|\frac{b^T\tilde{X}}{\sqrt{n}}|^2}}{1+\gamma}\right)$ is exponentially small and negligible. When $\mu_\pi\mu_{\hat{p}} = 0$, according to central limit theorem $\frac{b^T\tilde{X}}{\sqrt{n}}$ is a complex Gaussian random variable $\mathcal{Z}$. Thus, we have:

$$\mathbb{E}_b\left(\frac{e^{\frac{-\beta\gamma r^2}{2(1+\gamma)}|\frac{b^T\tilde{X}}{\sqrt{n}}|^2}}{1+\gamma}\right) \approx \mathbb{E}\left(\frac{e^{\frac{-\beta\eta}{({\sigma_p}^2+n{\mu_p}^2)({\sigma_\pi}^2+{\mu_\pi}^2)}\frac{\gamma}{1+\gamma}\mathcal{Z}^2}}{1+\gamma}\right) = TERM. \tag{127}$$

If $\mu_p \neq 0$, the above expression approaches $\frac{1}{1+\gamma}$ which yields a trivial bound. Hence, we assume that $\mu_p = 0$; thus we have

$$TERM = \sqrt{\frac{1}{1+\gamma+\frac{2\beta\eta\gamma\lambda_1}{\sigma_p^2(\sigma_\pi^2+\mu_\pi^2)}}\frac{1}{1+\gamma+\frac{2\beta\eta\gamma\lambda_2}{\sigma_p^2(\sigma_\pi^2+\mu_\pi^2)}}}, \tag{128}$$

where $\lambda_1$,$\lambda_2$ are eigenvalues of the covariance matrix of an r.v. which has the distribution of the product of two independent variables with distribution $\hat{p}$ and $\pi$, and hence

$$\left(\mathbb{E}_b\left(\frac{e^{\frac{-\beta\gamma r^2}{2(1+\gamma)}|\frac{b^T\tilde{X}}{\sqrt{n}}|^2}}{1+\gamma}\right)\right)^{\frac{n}{\beta}} \approx 2^{\frac{-n}{2\beta}[\log(1+\gamma+2\beta\eta\gamma\frac{\lambda_1}{({\sigma_p}^2)({\sigma_\pi}^2+{\mu_\pi}^2)})+\log(1+\gamma+2\beta\eta\gamma\frac{\lambda_2}{({\sigma_p}^2)({\sigma_\pi}^2+{\mu_\pi}^2)})]}. \tag{129}$$

On the other hand, by Sanov theorem [56] about the large deviation probability of empirical measures, the probability of appearance of $\hat{p}$ is asymptotically equal to $2^{-n\mathbb{D}(\hat{p}\|\tilde{p})}$, where $\mathbb{D}(\hat{p}\|\tilde{p})$ is the Kullback-Leibler distance of $\hat{p}$ and $\tilde{p}$. Thus, we have

$$-\frac{1}{\beta}\gamma\log(e) - \frac{1}{n}\log\mathbb{E}_{\tilde{X}}\left(\left(\mathbb{E}_b\left(\frac{e^{\frac{\beta\gamma r^2}{2(1+\gamma)}|\frac{b^T\tilde{X}}{\sqrt{n}}|}}{1+\gamma}\right)\right)^{\frac{n}{\beta}}\right) = -\frac{1}{\beta}\gamma\log e -$$

$$\sup_{\hat{p}}\left[-\mathbb{D}(\hat{p}\|\tilde{p}) - \frac{\log e}{2\beta}\left(\log\left(1+\gamma+\frac{2\beta\eta\gamma\lambda_1}{\sigma_p^2(\sigma_\pi^2+\mu_\pi^2)}\right)+\log\left(1+\gamma+\frac{2\beta\eta\gamma\lambda_2}{\sigma_p^2(\sigma_\pi^2+\mu_\pi^2)}\right)\right]\right). \tag{130}$$

From the above equation and (126) , we get the desired result. □

*O. Proof of Active User Theorems*

*1) Proof of Theorem 25, Capacity Lower Bound for Binary/Real Active System:*

We use Corollary 5 where $\pi(\cdot)$ is an $\mathcal{N}(0,1)$ random variable and $p(\cdot)$ is uniform distribution on $\{\pm 1\}$. We have $\tilde{X} = X - X'$ where $X$ and $X'$ are two $k$-active vectors. We evaluate the expectation conditioned on $\|\tilde{X}\|^2 = 2i$ for some $i$ in the interval $[0, 2k]$. Assume that the corresponding entries in $X$ and $\dot{X}$ are of the form $(+1,-1), (-1,+1), (-1,-1), (+1,+1), (0,-1), (0,+1), (+1,+1), (-1,0)$ $, (0,0)$ $k_1, k_2, \ldots, k_9$ times, respectively. Now in order to satisfy the condition $\|\tilde{X}\|^2 = 2i$, we require

the following equations:

$$\begin{cases} \sum_i k_i - (k_7 + k_8 + k_9) = k \\ \sum k_i = n \\ \sum k_i - (k_5 + k_6 + k_9) = k \\ 4k_1 + 4k_2 + k_5 + k_6 + k_7 + k_8 = m = 2i \in [0, 4k] \end{cases}$$

The simplification of the above equations lead to:

$$\begin{cases} k_1 + k_2 = j \\ k_3 + k_4 = k - i + j \\ k_5 + k_6 = i - 2j \\ k_7 + k_8 = i - 2j \\ k_9 = n + 2j - k - i \end{cases}$$

The probability that the above equation are satisfied simultaneously is equal to $\frac{\sum_j 2^{i-2j} \binom{n}{j,k+j-i,i-2j,n-k}}{2^k \binom{n}{k}^2}$ and the expectation term in (59) is $\left(1 + \frac{4in\gamma\eta}{mk(1+\gamma)}\right)^{-\frac{m}{2}}$. By varying the value of $i$ in $[0, 2k]$, we get the desired result. □

*2) Proof of Theorem 26, Lower Bound for $k$-Active Real/Real CDMA Systems:*

Let $\pi(\cdot)$ be the standard Gaussian distribution on $\mathbb{R}$. According to Corollary 5, we have

$$\mathbb{E}_B\left(\mathbb{I}\left(X;Y\right)\right) \geq -m\gamma \log e - \log \mathbb{E}_{\hat{X}}\left(\left(\mathbb{E}_b\left(\frac{e^{\frac{-\gamma r^2}{2(1+\gamma)m}|b^T\tilde{X}|^2}}{1+\gamma}\right)\right)^m\right), \tag{131}$$

$b^T\tilde{X} \sim |\tilde{X}|Z$ and hence

$$\mathbb{E}_b\left(e^{\frac{-\gamma r^2}{2(1+\gamma)m}|b^T\tilde{X}|^2}\right) = \mathbb{E}\left(e^{\left(\frac{-\gamma r^2|\tilde{X}|^2}{2(1+\gamma)m}\right)Z^2}\right) = \frac{1}{\sqrt{1 + \frac{r^2\gamma|\hat{X}|^2}{(1+\gamma)m}}}. \tag{132}$$

Thus,

$$\mathbb{E}_B\left(\mathbb{I}\left(X;Y\right)\right) \geq -m\left(\gamma \log e - \log\left(1+\gamma\right)\right) - \log \mathbb{E}_{\tilde{X}}\left(\left(1 + \frac{r^2\gamma|\tilde{X}|^2}{\left(1+\gamma\right)m}\right)^{\frac{-m}{2}}\right), \tag{133}$$

where $r = \sqrt{\frac{2\eta}{\frac{k}{n}}}$ and $\tilde{X} = X - X'$ is the difference of two independent $k$-active vectors. Conditioning on the number of common non-zero elements of $X$ and $X'$ to be $j$, one has $\|\tilde{X}\|^2 \sim \chi^2_{2(k-j)} + 2\chi^2_j$ and therefore, we derive (76). □

*3) Proof of Theorem 27, Capacity Lower Bound for Binary/Binary Active System:*
We use Corollary 5 where $\pi(\cdot)$ and $p(\cdot)$ are uniform distributions on $\{\pm 1\}$. We evaluate the expectation conditioned on the vector $\tilde{X}$ to have $i$ elements in $\{\pm 2\}$ and $j$ elements in $\{\pm 1\}$. Obviously $0 \le i+\frac{j}{2} \le k$ and $j \in 2\mathbb{Z}$, the probability of this condition is equal to $\frac{2^{\frac{j}{2}}}{2^k\binom{n}{k}^2}\binom{n}{i,j,\frac{j}{2},k-i-\frac{j}{2},n-k-j}$ and the expectation term in (59) becomes

$$\left(\frac{1}{2^{i+j}}\sum_{u,v}\binom{i}{u}\binom{j}{i+\frac{j+v}{2}-2u}e^{\frac{-2n\eta\gamma v^2}{mk(1+\gamma)}}\right)^m. \tag{134}$$

Now by averaging over all possible values of $i$ and $j$ we get (77). □

*4) Proof of Theorem 28, Active Users with Probability $p_{\text{act}}$, $\mathcal{I}=\mathcal{S}=\mathbb{R}$:*
From Corollary 5 for any realization of $\tilde{X}$, we have $b^T\tilde{X} \sim \mathcal{N}\left(0, |\tilde{X}|^2\right)$ and hence

$$\mathbb{E}_b\left(e^{\frac{-\gamma r^2}{2(1+\gamma)m}|b^T\tilde{X}|^2}\right) = \frac{1}{\sqrt{1+\frac{\gamma r^2|\tilde{X}|^2}{(1+\gamma)m}}}. \tag{135}$$

In the $p_{\text{act}}$-active case,

$$r = \sqrt{\frac{2\eta}{\sigma_\pi^2\sigma_p^2}} = \sqrt{\frac{2\eta}{p_{\text{act}}}}, \tag{136}$$

therefore:

$$C_{\text{active}}(m,n,\mathbb{R},\mathbb{R},\eta,p_{\text{act}}) \ge \sup_\gamma\{-m\left(\gamma\log e - \log\left(1+\gamma\right)\right) - \log\mathbb{E}_{\tilde{X}}\left(\left(1+\frac{2\gamma\eta|\tilde{X}|^2}{\left(1+\gamma\right)mp_{\text{act}}}\right)^{-\frac{m}{2}}\right)\} \tag{137}$$

Now $\tilde{X} = X - X'$, where $X$ and $X'$ are independent with i.i.d. entries which are $\mathcal{N}(0,1)$ with probability $p_{\text{act}}$ and $0$ with probability $1-p_{\text{act}}$. Hence the entries of $\tilde{X}$ are

$$\begin{cases} 0 & \text{with probability } \left(1-p_{act}\right)^2 \\ \mathcal{N}(0,1) & \text{with probability } 2p_{act}\left(1-p_{act}\right) \\ \mathcal{N}(0,2) & \text{with probability } p_{act}^2 \end{cases} \tag{138}$$

Now conditioning on the number of $\mathcal{N}(0,2)$ and $\mathcal{N}(0,1)$ elements of $\tilde{X}$ to be $k_1$ and $k_2$, respectively with the probability

$$\binom{n}{k_1,k_2,n-k_1-k_2}p_{act}^{2k_1}\left(2p_{act}\left(1-p_{act}\right)\right)^{k_2}\left(1-p_{act}\right)^{2(n-k_1-k_2)}. \tag{139}$$

We have $\|\tilde{X}\|^2 \sim 2{\chi_{k_1}}^2+{\chi_{k_2}}^2$ and thus we arrive at (79). □

*5) Proof of Theorem 29, Asymptotic Lower Bound for a Real/Real $p_{act}$-Active CDMA System:* From Corollary 5 and (136), we have

$$I(X;Y) \geq -m[\gamma \log e - \log(1+\gamma)] - \log \mathbb{E}_{\tilde{X}} \left( \left( 1 + \frac{2\gamma\eta|\tilde{X}|^2}{(1+\gamma)\, m p_{\text{act}}} \right)^{-m/2} \right), \tag{140}$$

where $\tilde{X} = X - X'$. We assume that elements of $X$ or $X'$ are zero with probability $1 - p_{act}$ and $\mathcal{N}(0,1)$ with probability $p_{act}$. Now we have,

$$\begin{aligned}\lim_{\substack{m,n\to\infty\\ n/m\to\beta}} \frac{1}{n}\mathbb{I}(X;Y) &\geq \frac{-1}{\beta}(\gamma \log e - \log(1+\gamma)) - \lim_{\substack{m,n\to\infty\\ n/m\to\beta}} \frac{1}{n}\log \mathbb{E}\left( \left(1 + \frac{2\gamma\eta|\tilde{X}|^2}{(1+\gamma)\, m p_{act}}\right)^{-\frac{m}{2}} \right)\\ &= \frac{-1}{\beta}(\gamma \log e - \log(1+\gamma)) - \lim_{\substack{m,n\to\infty\\ n/m\to\beta}} \frac{1}{n}\log \mathbb{E}\left( 2^{nF\left(\frac{|\tilde{X}|^2}{n}\right)} \right),\end{aligned} \tag{141}$$

where $F(x) = \frac{-1}{2\beta}\log\left(1 + \frac{2\beta\gamma\eta}{(1+\gamma)p_{act}}x\right)$. Now by Varadhan's lemma [57], the last expression is equal to $\sup_x \{F(x) - I(x)\}$, where $I(x)$ is the rate function for the sequence of random variables $\frac{\|\tilde{X}\|^2}{n}$. But $\|\tilde{X}\|^2$ is the sum of $n$ i.i.d. random variables $Y_i (1 \leq i \leq n)$ of the form

$$Y_i \sim \begin{cases} 0 & (1-p_{act})^2, \\ Z^2 & 2p_{act}(1-p_{act}), \\ 2Z^2 & p_{act}^2 \end{cases} \qquad Z \sim \mathcal{N}(0,1). \tag{142}$$

The Cramer's theorem [57] gives the rate function as:

$$I(x) = \sup_t \{\log e \times xt - \log M(t)\}, \tag{143}$$

where

$$\begin{aligned}M(t) = \mathbb{E}\left(e^{tY_i}\right) = (1-p_{\text{act}})^2\mathbb{E}\left(e^0\right) + 2p_{\text{act}}(1-p_{\text{act}})\mathbb{E}\left(e^{tZ^2}\right) + p_{\text{act}}^2\mathbb{E}\left(e^{2tZ^2}\right) = \\ (1-p_{\text{act}})^2 + \frac{2p_{\text{act}}(1-p_{\text{act}})}{\sqrt{1-2t}} + \frac{p_{\text{act}}^2}{\sqrt{1-4t}}.\end{aligned} \tag{144}$$

□

*6) Proof of Theorem 31, Binary Input and Binary Signature (p-Active):* We use Corollary 5 where $\pi(\cdot)$ is the uniformly distributed on $\{\pm 1\}$ and $p(\cdot)$ is uniform distribution on $\{\pm 1\}$. In this case $\tilde{X}$ is a vector of i.i.d. entries with values in $\{0, \pm 1, \pm 2\}$, where $\tilde{p}(0) = (1-p_{\text{act}})^2 + \frac{p_{\text{act}}^2}{2}$ and $\tilde{p}(\pm 1) = 2p_{\text{act}}(1-p_{\text{act}})$ and $\tilde{p}(\pm 2) = \frac{{p_{\text{act}}}^2}{2}$. Now, applying the condition on vector $\tilde{X}$ to have $k_1$ elements in $\{\pm 2\}$ and $k_2$ elements in $\{\pm 1\}$, we have $0 \leq k_1 + k_2 \leq n$. However the probability of this condition is equal to $2^{k_2-k_1}\binom{n}{k_1,k_2,n-k_1-k_2}{p_{\text{act}}}^{2k_1+k_2}\left(1+\frac{3}{2}{p_{\text{act}}}^2-2p_{\text{act}}\right)^{n-k_1-k_2}$. The distribution of $b^T\tilde{X}$ is equal to the sum of $k_1$ uniform distribution on $\{\pm 2\}$ and $k_2$ uniform distribution on $\{\pm 1\}$. we can assume that $i$ element of these $k_1$ elements are $\{+2\}$ and $j$ element of these $k_2$ elements are $\{+1\}$.Therefore, the conditional expectation is equal to

$$\left(\sum_{k_1,k_2} 2^{-k_1-k_2}\binom{k_1}{i}\binom{k_2}{j}e^{\frac{-2\eta\gamma(4i+2j-2k_1-k_2)^2}{mp_{\text{act}}(1+\gamma)}}\right)^m$$

Now by varying for all possible values of $k_1$ and $k_2$, the theorem is derived.

## Acknowledgment

This work would not have been possible without the great efforts of our brilliant and humble undergraduate students. Specifically, A. A. Makhdoumi for editing the GCO section and providing Figs.1-6, and P. Kabir for providing the simulation results for the active users. We would like to sincerely thank the academic staff and the students of Advanced Communications Research Institute (ACRI) and Center of Excellence on Multi-access Communications Systems of Sharif University of Technology, specially, Profs. J.A. Salehi, M.R. Aref, Nasiri, Drs. H. Saidi, B. Seif and G. Abed-Hotari. We are also very indebted to our students specially V. Montazer-Hodjat. A. Amini, M. Ferdosizadeh, A. Haghi and R. Khosravi-Farsani for their helpful comments. We would also like to thank Profs Arfaei and Alishahian from Institute of Physics and Mathamatics (IPM) for providing the facility to the authors for several days and nights to jump start this project. Finally, one of the authors, F. Marvasti, would like to dedicate this paper to his late mother who passed away unexpectedly; she was a great advocate of academic excellence.

## References

[1] T. Ojanpera and R. Prasad, *Wideband CDMA For Third Generation Mobile Communications: Universal Personal Communications*, 1st ed. Norwood, MA, USA: Addison-Wesley, 1998.

[2] F. Chung, J. Salehi, and V. Wei, “Optical orthogonal codes: design, analysis and applications,” *IEEE Trans. Inf. Theory*, vol. 35, no. 3, pp. 595–604, May 1989.

[3] S. Mashhadi and J. Salehi, “Code-division multiple-access techniques in optical fiber networks part III: Optical AND gate receiver structure with generalized optical orthogonal codes,” *IEEE Trans. Commun.*, vol. 54, no. 6, pp. 1349–1349, July 2006.

[4] J. Salehi, “Emerging OCDMA communication systems and data networks,” *Journal of Optical Networking*, vol. 6, no. 9, pp. 1138–1178, September 2007.

[5] S. Verdu, *Multiuser Detection*. New York, USA: Cambridge University Press, 1998.

[6] A. J. Viterbi, *CDMA : Principles of Spread Spectrum Communication*, 1st ed. New York, USA: Addison-Wesley, 1995.

[7] J. L. Massey and T. Mittelholzer, "Welch's bound and sequence sets for code-division multiple-access systems," *in Sequences II, Methods in Communication, Security, and Computer Sciences, R. Capocelli, A. De Santis, and U. Vaccaro, Eds. New York: Springer-Verlag*, 1993.

[8] M. Rupf and J. L. Massey, "Optimum sequences multisets for synchronous code-division multiple-access channels," *IEEE Trans. Inform. Theory*, vol. 40, pp. 1261–1266, July 1994.

[9] L. Welch, "Lower bound on the maximum cross correlation of signals," *IEEE Trans. Inf. Theory*, vol. 20, no. 3, pp. 397–399, May 1974.

[10] T. Tanaka, "A statistical-mechanics approach to large- system analysis of CDMA multiuser detectors," *IEEE Trans. Inform. Theory*, vol. 48, no. 11, pp. 2888–2910, November 2002.

[11] A. Montanari and D. N. C. Tse, "Analysis of belief propagation for non- linear problems: The example of CDMA (or: How to prove tanaka's formula)," *in Proc. IEEE Information Theory Workshop, Punta del Este, Uruguay*, March 2006.

[12] D. Guo and S. Verdu, "Randomly spread CDMA: Asymptotics via statistical physics," *IEEE Trans. Inf. Theory*, vol. 51, no. 6, pp. 1983–2010, March 2005.

[13] S. B. Koroda and N. Makris, "On the concentration of the capacity for a code division multiple access system," *in Proc. IEEE int. Symp. Information Theory (ISIT)*, June 2007.

[14] P. Pad, F. Marvasti, K. Alishahi, and S. Akbari, "A class of errorless codes for over-loaded synchronous wireless and optical CDMA systems," *IEEE Trans. Inform. Theory*, vol. 55, no. 6, pp. 2705–2715, June 2009.

[15] K. Alishahi, F. Marvasti, V. Aref, and P. Pad, "Bounds on the sum capacity of synchronous binary CDMA channels," *IEEE Trans. Inform. Theory*, vol. 55, no. 8, pp. 3577–3593, August 2009.

[16] S. P. Ponnaluri and T. Guess, "Effects of spreading and training on capacity in overloaded CDMA," *IEEE Trans. Commun.*, vol. 54, pp. 523–526, April 2008.

[17] R. van der Hofstad and M. J. Klok, "Performance of DS-CDMA systems with optimal hard-decision parallel interference cancellation," *IEEE Trans. Inf. Theory*, vol. 49, no. 11, pp. 2918–2940, November 2003.

[18] G. Xue, J. Weng, T. Le-Ngoc, and S. Tahar, "Adaptive multistage parallel interference cancellation for CDMA," *IEEE Journal on Selected Areas in Communications*, vol. 17, no. 10, pp. 1815–1827, October 1999.

[19] D. Guo, L. Rasmussen, S. Sun, and T. Lim, "A matrix-algebraic approach to linear parallel interference cancellation in CDMA," *IEEE Trans. Commun.*, vol. 48, no. 1, pp. 152–161, January 2000.

[20] M. Kobayashi, J. Boutros, and G. Caire, "Successive interference cancellation with SISO decoding and EM channel estimation," *IEEE Journal on Selected Areas in Communications*, vol. 19, no. 8, pp. 1450–1460, August 2001.

[21] F. Vanhaverbeke, M. Moeneclaey, and H. Sari, "DS-CDMA with two sets of orthogonal spreading sequences and iterative detection," *IEEE Commun. Letters*, vol. 4, no. 9, pp. 289–291, September 2000.

[22] A. Kapur and M. Varanasi, "Multiuser detection for over-loaded CDMA systems," *IEEE Trans. Inf. Theory*, vol. 49, no. 7, pp. 1728–1742, July 2003.

[23] C. Rose, "CDMA codeword optimization: interference avoidance and convergence via class warfare," *IEEE Trans. Inform. Theory*, vol. 47, pp. 2368–2382, September 2001.

[24] G. N. Karystinos and D. A. Pados, "The maximum squared correlation, total asymptotic efficiency, and sum capacity of minimum total-squared-correlation binary signature sets," *IEEE Trans. Inform. Theory*, vol. 51, pp. 348–355, January 2005.

[25] D. Djoinin and V. Bhagrava, "New results on low complexity detectors for over-saturated CDMA systems," *in Proc. IEEE GLOBECOM*, November 2001.

[26] F. Marvasti, *Nonuniform Sampling: Theory and Practice*. Springer, 2001.

[27] P. Anigstein and V. Anantharam, “Ensuring convergence of the MMSE iteration for interference avoidance to the global optimum,” *IEEE Trans. Inform. Theory*, vol. 49, pp. 873–885, April 2003.

[28] M. Faraji, P. Pad, and F. Marvasti, “A new method for constructing large size wbe codes with low complexity ml decoder,” *IEEE Commun. Letters*, October 2008.

[29] P. Pad, M. Soltanolkotabi, S. Hadikhanlou, A. Enayati, and F.Marvasti, “Errorless codes for over-loaded CDMA with active user detection,” *ICC’09, Dresden, Germany*, June 2009.

[30] F. Marvasti, M. Ferdowsizadeh, and P. Pad, “Iterative synchronous and asynchronous multi-user detection with optimum soft limiter.” *Patent applicaion 12/122668*, May 2008.

[31] P. Pad, F. Marvasti, K. Alishahi, and S. Akbari, “Errorless codes for over-loaded synchronous CDMA systems and evaluation of channel capacity bounds,” *in Proc. IEEE Int. Symp. Inf. Theory (ISIT), Toronto, Canada*, June 2008.

[32] W. H. Mow, “Recursive constructions of detecting matrices for multiuser coding: A unifying approach,” *IEEE Trans. Inf. Theory*, vol. 55, no. 1, pp. 93–98, January 2009.

[33] S. Söderberg and H. S. Shapiro, “A combinatory detection problem,” *American Mathematical Monthly*, vol. 70, no. 10, pp. 1066–1070, December 1963.

[34] R. Erdös and A. Rényi, “On two problems of information theory,” *Magyar Tud. Akad. Mat. Kutato  Int. Ko zl.*, pp. 229–243, 1963.

[35] B. Lindström, “On a combinatory detection problem,” *Publ. Hung. Acad. Sci.*, vol. 9, pp. 195–207, 1964.

[36] D. G. Cantor and W. H. Mills, “Determining a subset from certain combinatorial properties,” *Can. J. Math.*, vol. 18, pp. 42–48, 1966.

[37] B. Lindström and J. N. Srivastava, “Determining subsets by unramified experiments,” *In A Survey of Statistical Design and Linear Models. New York: North-Holland*, 1975.

[38] S. S. Martirossian and G. H. Khachatrian, “Construction of signature codes and the coin weighing problem,” *Probl. Inf. Trans.*, vol. 25, pp. 334–335, October-December 1989.

[39] B. Lindström, “On möbius functions and a problem in combinatorial number theory,” *Can. Math. Bull.*, vol. 14, pp. 513–516, 1971.

[40] S. B. Korada and A. Montanari, “Applications of Lindeberg principle in communications and statistical learning,” *ArXiv:1004.0557*, April 2010.

[41] S. B. Koroda and N. Macris, “Tight bounds on the capacity of binary input random CDMA systems,” *accepted in IEEE Trans. Inform. Theory, ArXiv: 0803.1454*, March 2008.

[42] S. Dashmiz, P. Pad, and F. Marvasti, “New bounds for binary and ternary overloaded CDMA,” *ArXiv:0901.1683v2*, January 2009.

[43] M. Akhavan-Bahabadi and M. Shiva, “Double orthogonal codes for increasing capacity in MC-CDMA systems,” *Wireless Optical Communication Network, WCON*, pp. 468–471, March 2005.

[44] G. N. Karystinos and D. A. Pados, “New bounds on the total squared correlation and optimum design of DS-CDMA binary signature sets,” *IEEE Trans. Commun.*, vol. 51, no. 1, pp. 48–51, January 2003.

[45] S. Dashmiz, M. R. Takapoui, P. Pad, and F. Marvasti, “New bounds for the sum capacity of binary and nonbinary synchronous CDMA systems,” *in Proc. IEEE Int. Symp. Inf. Theory (ISIT)*, June 2010.

[46] D. Donoho, “Compressed sensing,” *IEEE Trans. Inf. Theory*, vol. 52, no. 4, pp. 1289 – 1306, April 2006.

[47] S. Verdu, “The capacity region of the symbol-asynchronous gaussian multiple-access channel,” *IEEE Trans. Inf. Theory*, vol. 35, pp. 733–751, July 1989.

[48] J. Luo, S. Ulukus, and A. Ephremides, “Optimal sequences and sum capacity of symbol asynchronous CDMA systems,” *IEEE Trans. Inf. Theory*, vol. 51, no. 8, pp. 2760–2769, August 2005.

[49] S. Ulukus and R. D. Yates, “User capacity of asynchronous CDMA systems with matched filter receivers and optimum signature sequences,” *IEEE Trans. Inf. Theory*, vol. 50, no. 5, pp. 903–909, May 2004.

[50] M. H. Shafinia, P. Kabir, P. Pad, S. M. Mansouri, and F. Marvasti, “Errorless codes for CDMA system with near-far effect,” *ICC’10, Capetown, South Africa*, May 2010.

[51] S. C. Chang and E. J. W. Jr, “Coding for T-user multiple-access channels,” *IEEE Trans. Inf. Theory*, vol. 25, no. 6, November 1979.

[52] S. C. Chang, “Further results on coding for T-user multiple-access channels,” *IEEE Trans. Inf. Theory*, vol. 30, no. 2, pp. 411–415, March 1984.

[53] T. J. Ferguson, “Generalized T-user codes for multiple-access channels,” *IEEE Trans. Inf. Theory*, vol. 28, no. 5, pp. 775–778, September 1982.

[54] G. H. Khachatrian and S. S. Martirossian, “Codes for T-user noiseless adder channel,” *Probl. Control Inf. Theory*, vol. 16, no. 3, pp. 187–192, 1987.

[55] L. A. Shepp and I. Olkin, “Entropy of the sum of independent Bernoulli random variables and of the multinoial distribution,” *Technical Report 131, Department of Statistics, Stanford University, Stanford, California*, 1978.

[56] A. Dembo and O. Zeitouni, *Large Deviation Techniques and Application*, 2nd ed. Springer, 2009.

[57] S. Varadhan, “Asymptotic probabilities and differential equations,” *Comm. Pure Appl. Math*, vol. 19, pp. 261–286, 1966.

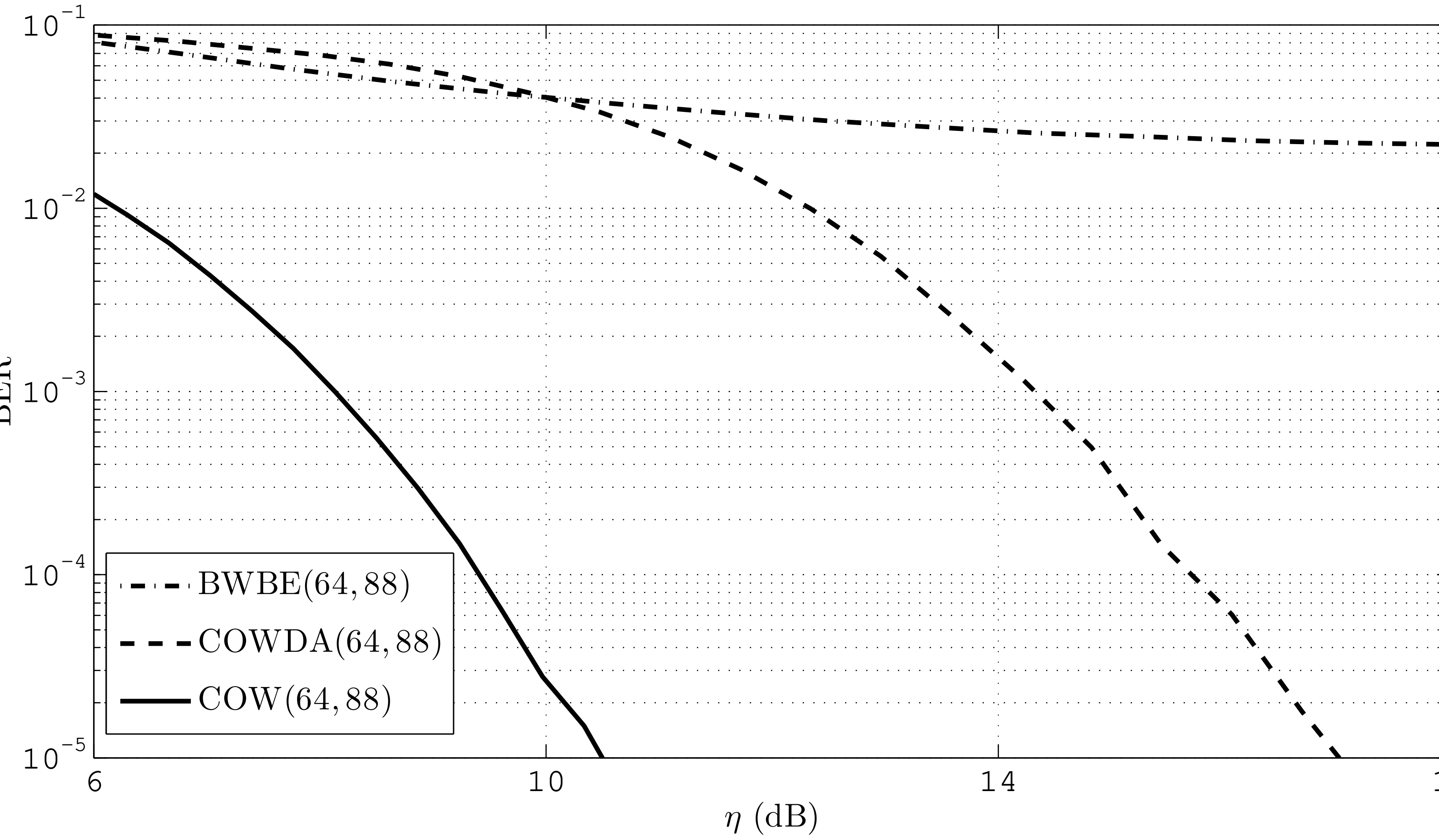

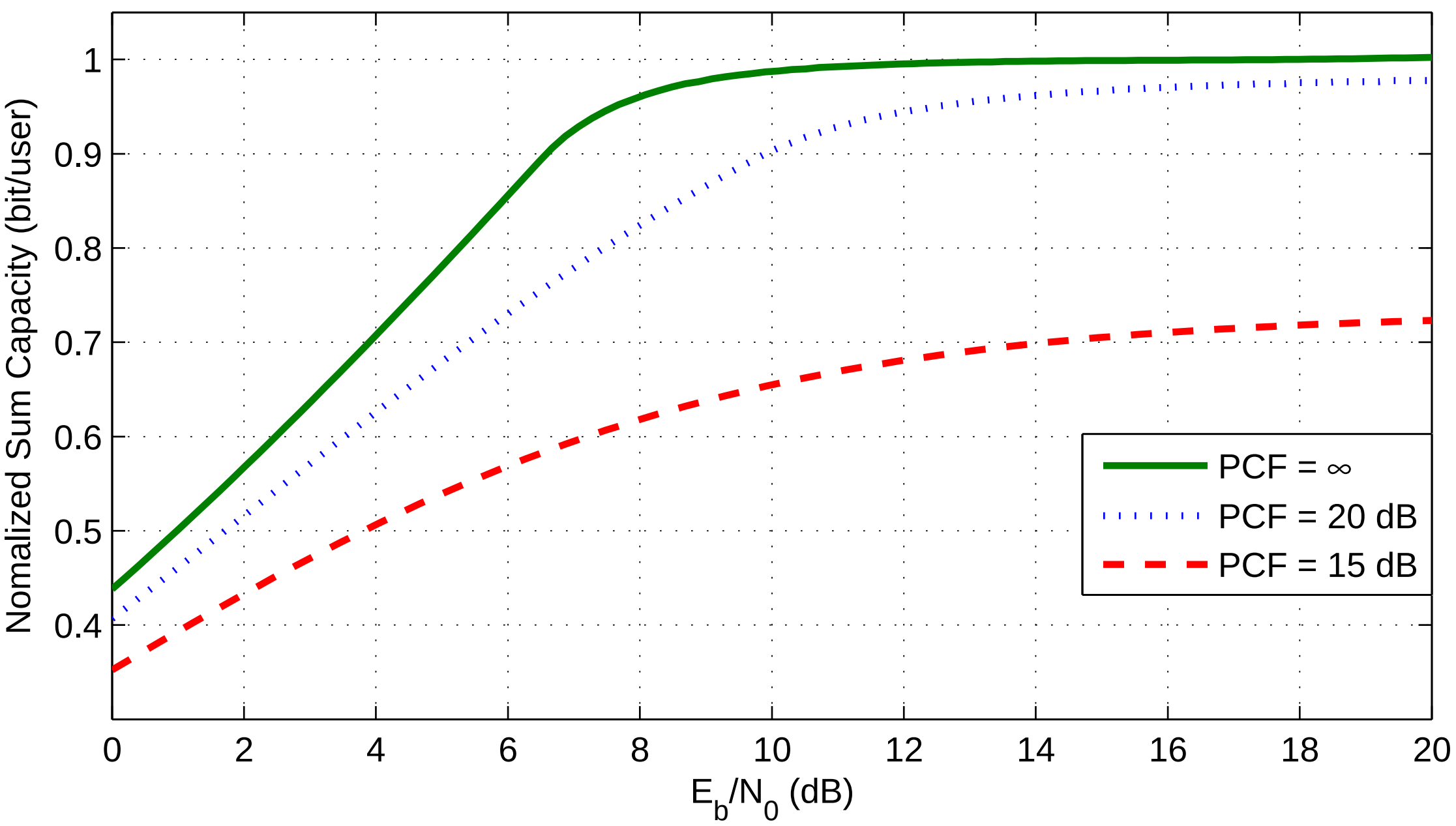

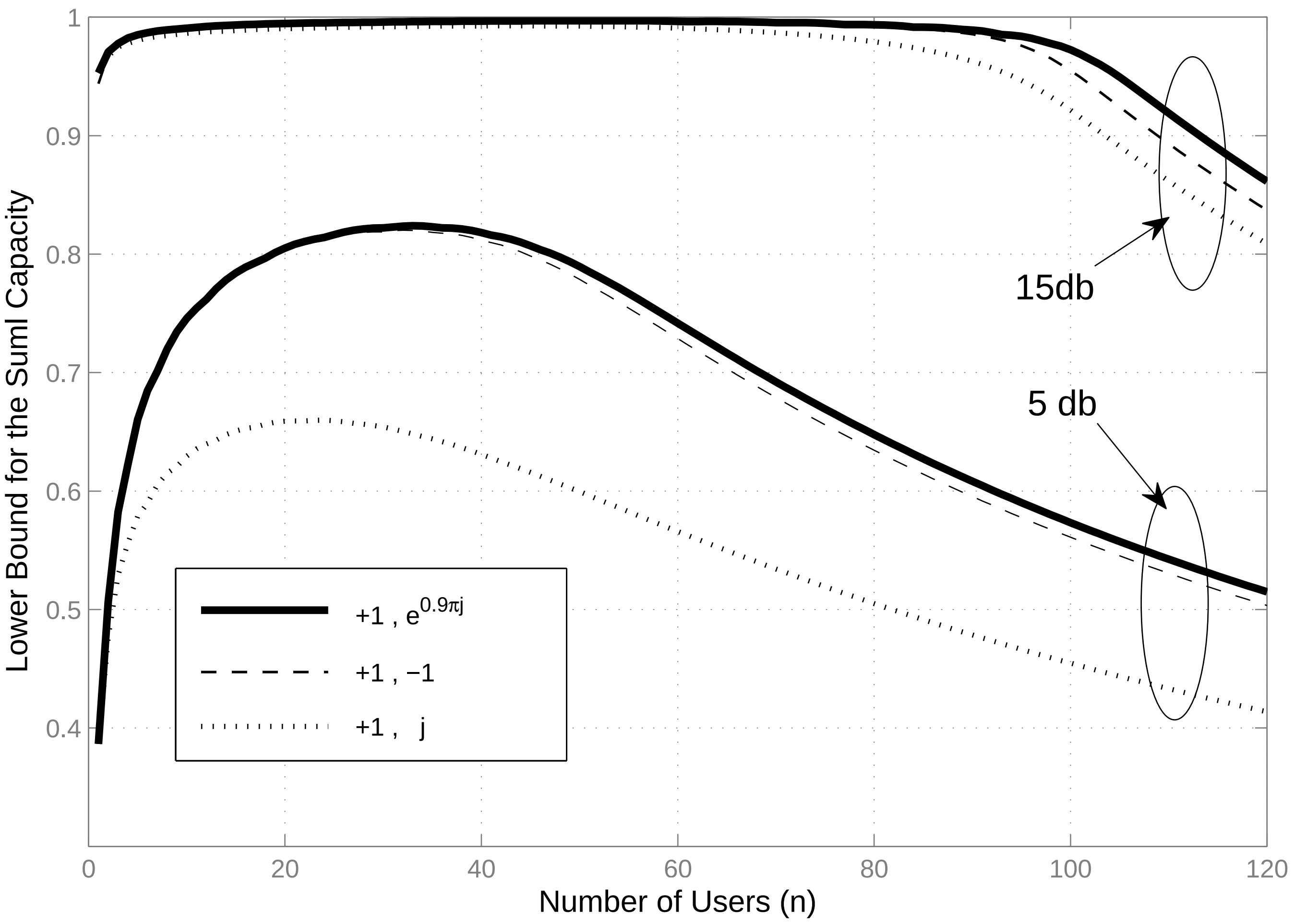

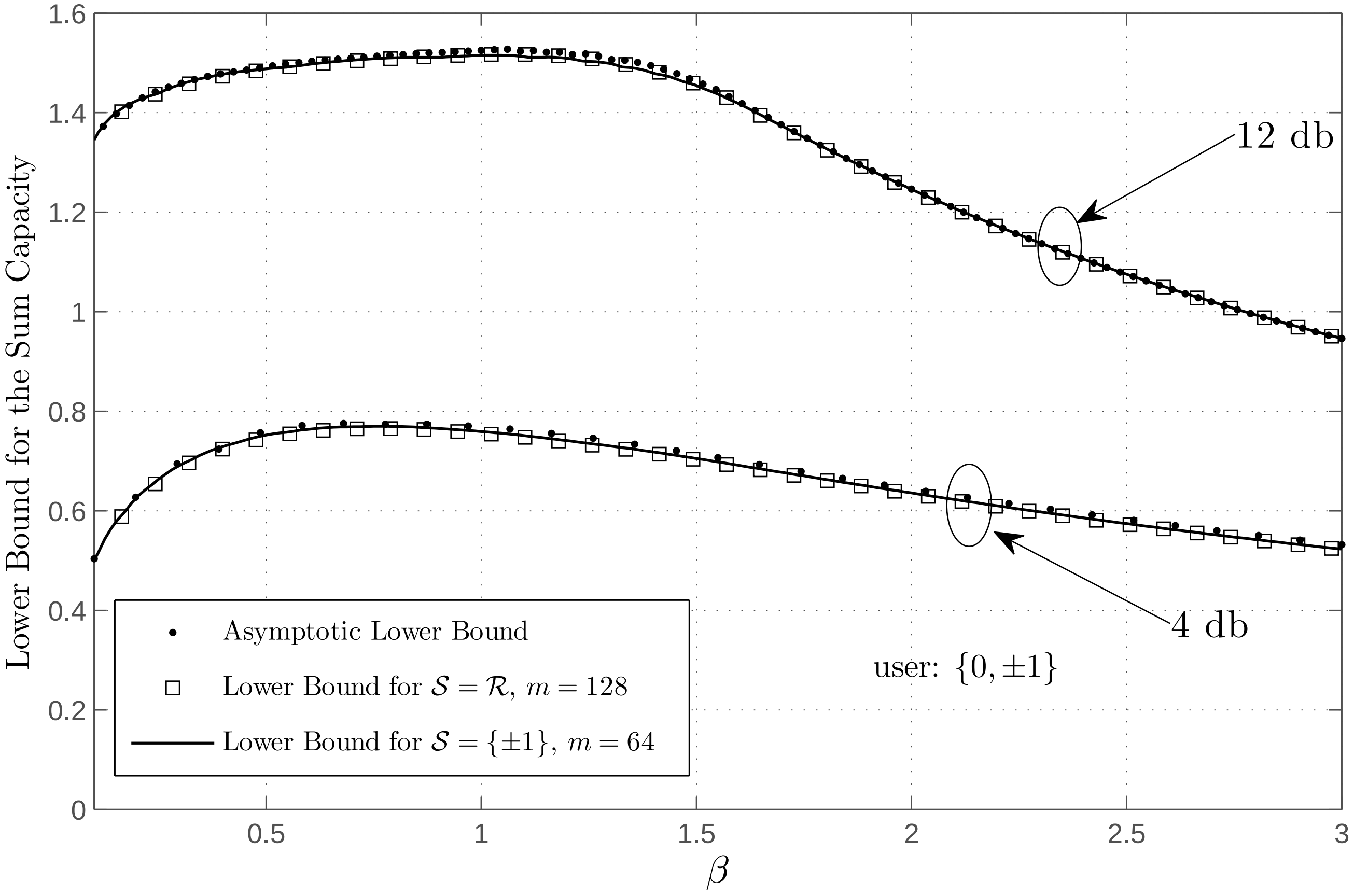

Lower Bound for the Sum Capacity
β
12 db
4 db
user: {0, ±1}
Asymptotic Lower Bound
Lower Bound for $\mathcal{S} = \mathcal{R}$, $m = 128$
Lower Bound for $\mathcal{S} = \{\pm 1\}$, $m = 64$

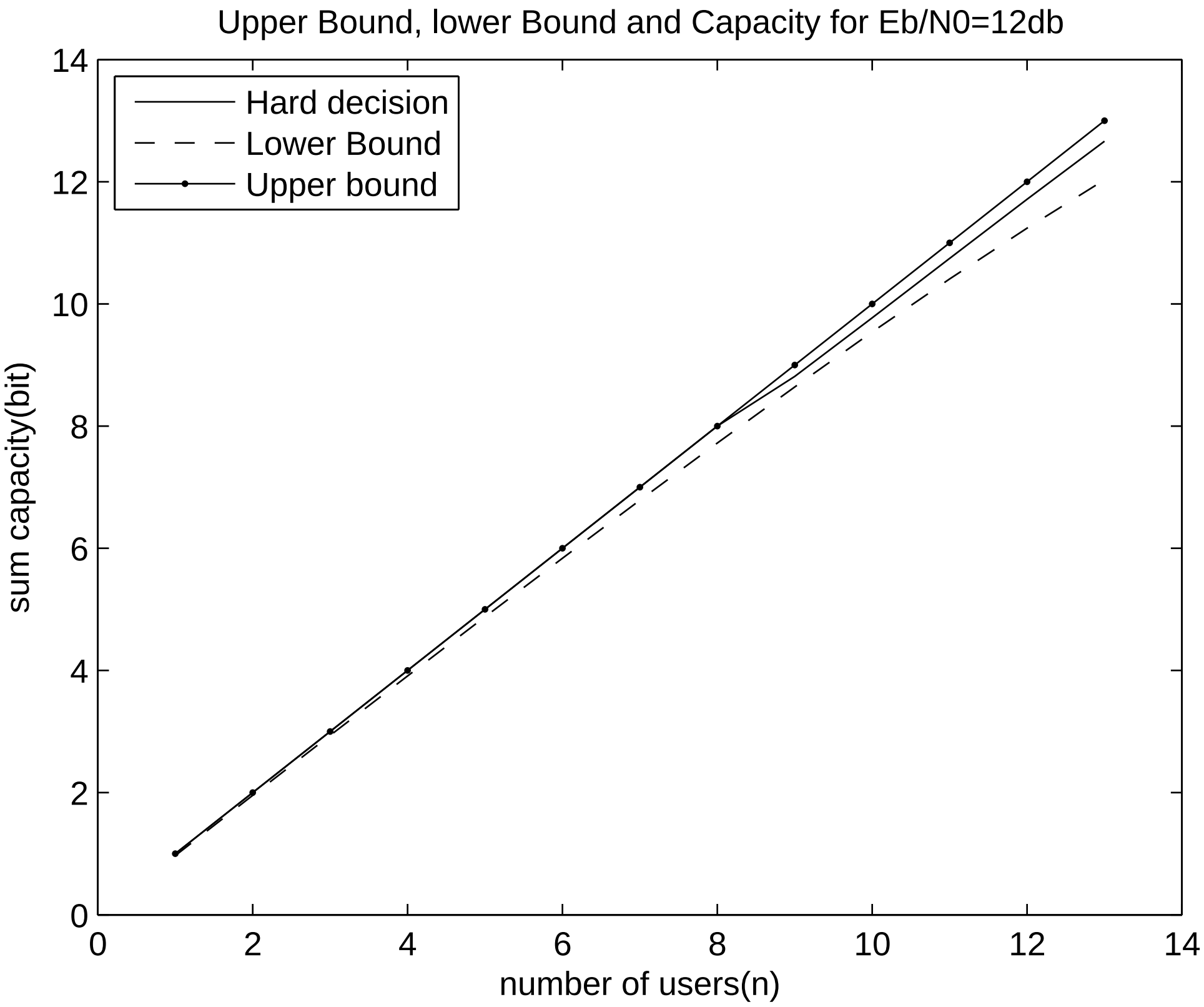

Upper Bound, lower Bound and Capacity for Eb/N0=12db
Hard decision
Lower Bound
Upper bound
sum capacity(bit)
number of users(n)
0
2
4
6
8
10
12
14

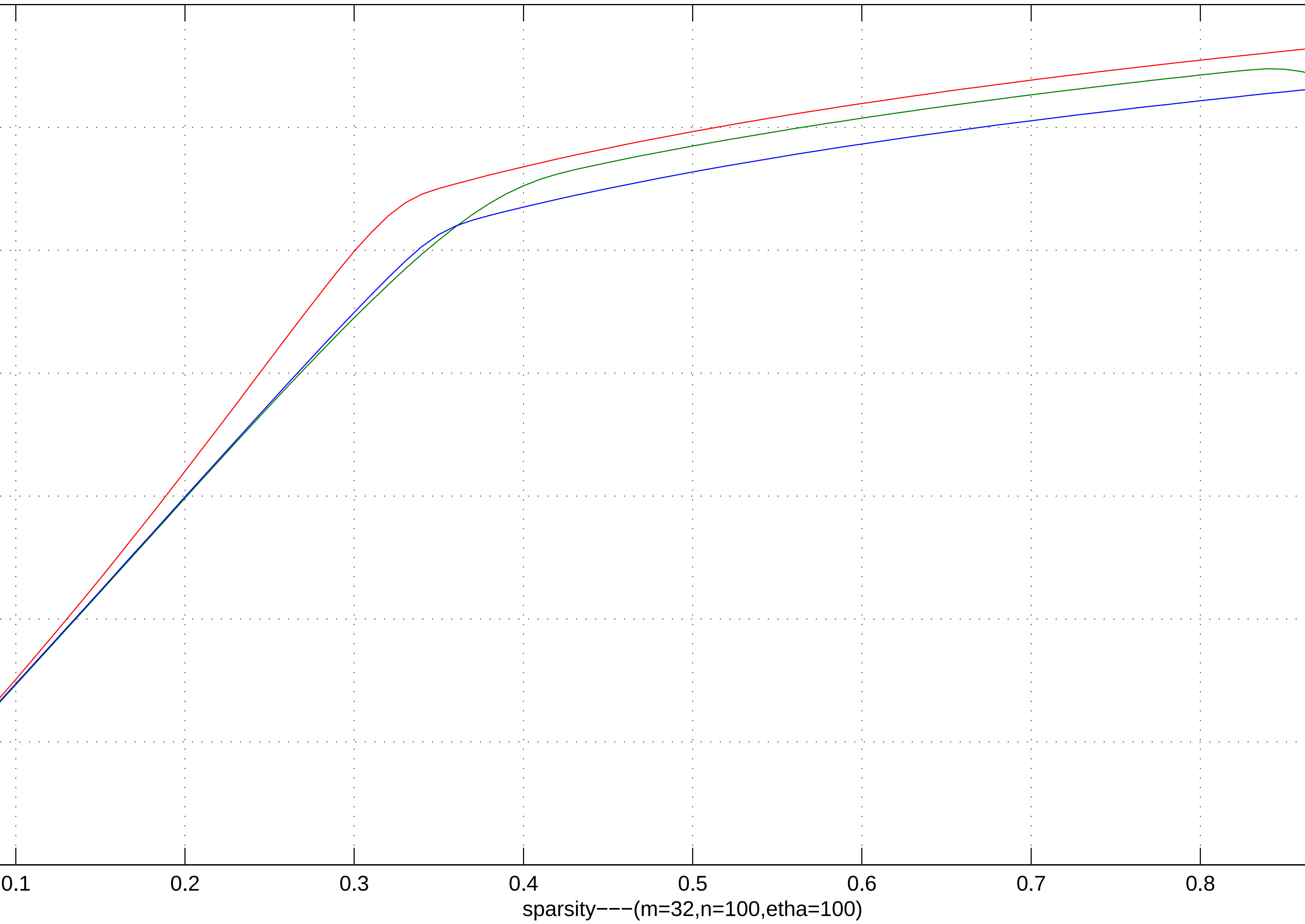

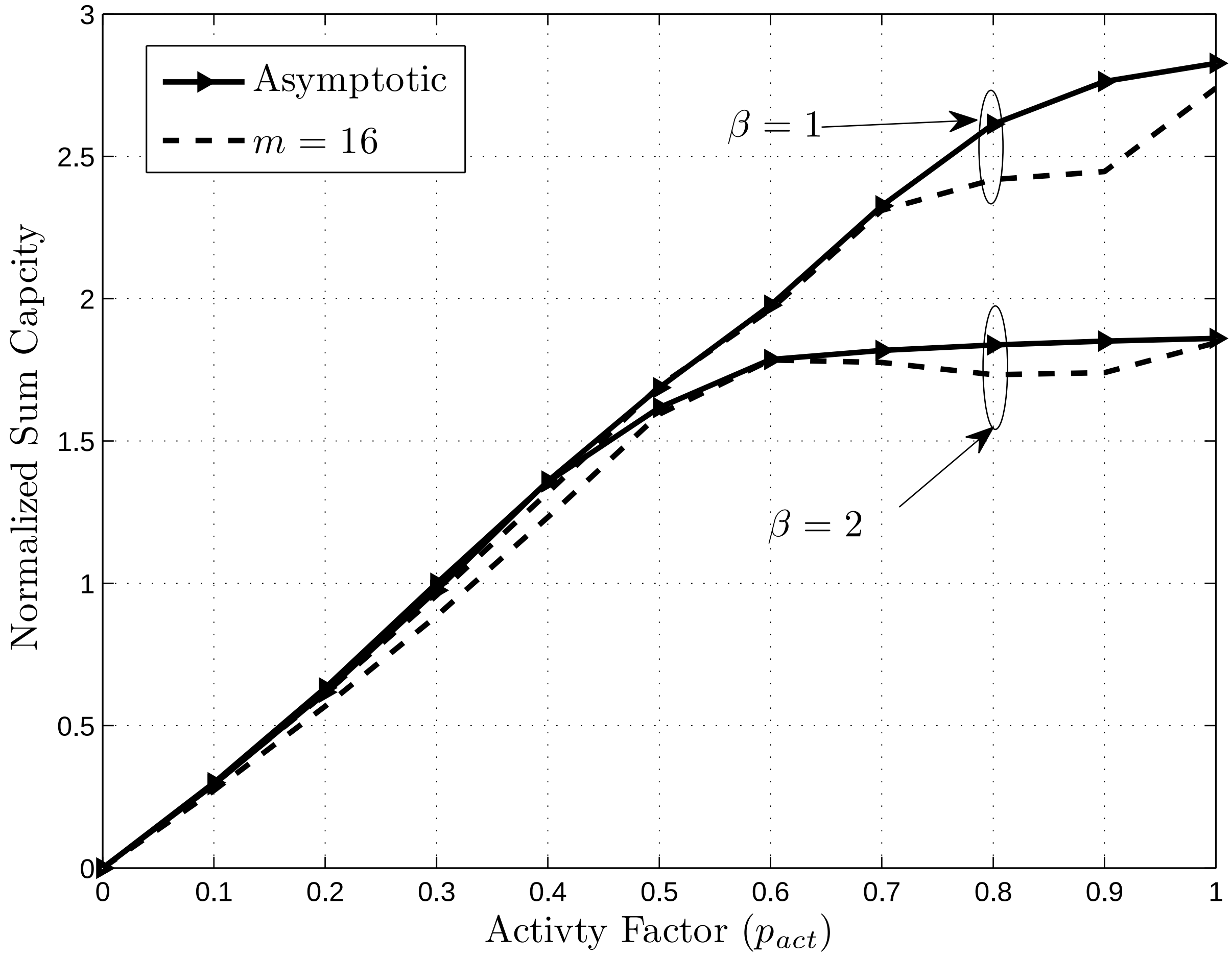

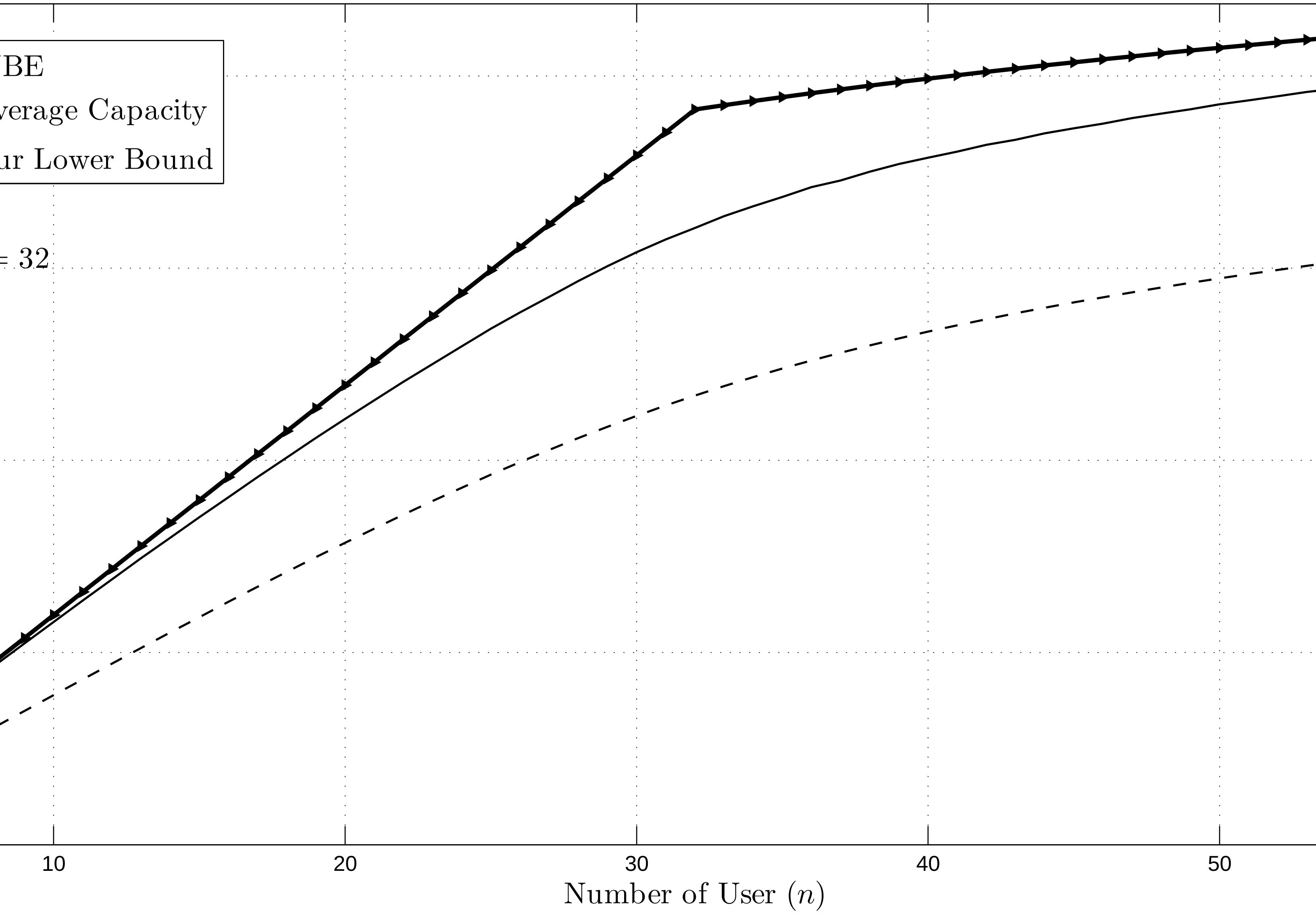

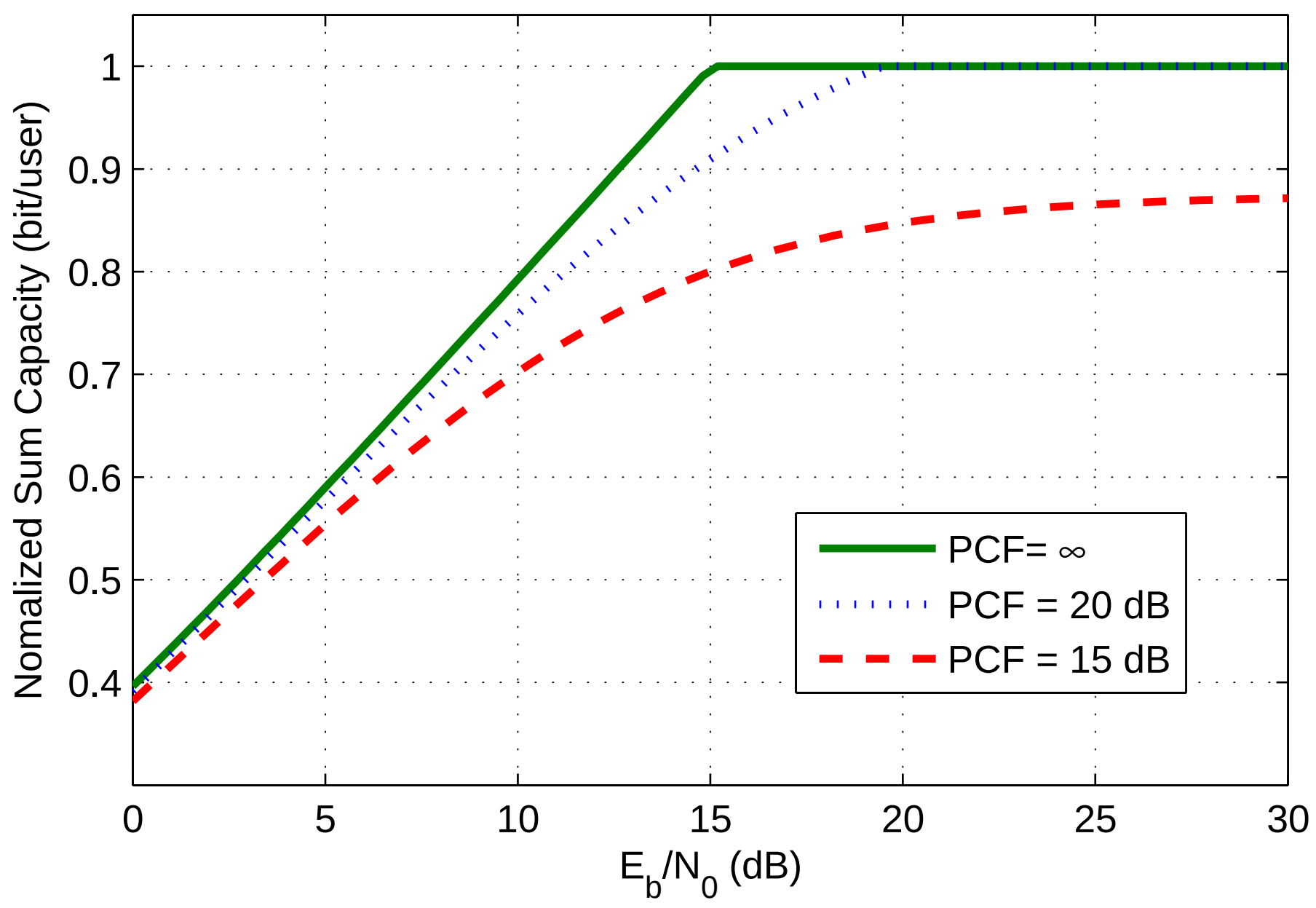